\documentclass[manuscript]{acmart}

\usepackage{booktabs, siunitx, threeparttable}
\usepackage{paralist}
\usepackage{adjustbox}
\usepackage{makecell}
\usepackage[utf8]{inputenc}
\usepackage{url}
\usepackage{colortbl}
\usepackage{balance}
\usepackage{xspace}
\usepackage{graphicx}
\usepackage{array} 
\usepackage{verbatim}
\usepackage{rotating}
\usepackage{bold-extra}
\usepackage{multirow}
\usepackage{listings}
\usepackage{boxedminipage}
\usepackage{soul}
\usepackage{enumitem}
\usepackage[normalem]{ulem}
\setlist{nolistsep,leftmargin=.5cm}
\usepackage{booktabs}
\usepackage{tabularx}
\usepackage{svg}
\usepackage{calc}
\usepackage{float}
\usepackage{multirow}
\usepackage[normalem]{ulem}
\useunder{\uline}{\ul}{}
\usepackage[most]{tcolorbox}
\definecolor{MidnightBlue}{HTML}{006895}
\definecolor{BoxesBlue}{HTML}{DEECFF}
\definecolor{BoxesYellow}{HTML}{FFF2CC}
\definecolor{StateGreen}{HTML}{91C788}
\definecolor{StateRed}{HTML}{FF8080}
\definecolor{ArrowGreen}{HTML}{61B15A}
\definecolor{ArrowViolet}{HTML}{BA94D1}
\usepackage{caption}
\usepackage[compact]{titlesec}
\usepackage{microtype} 
\usepackage{amsmath,amsfonts}
\usepackage{algorithmic}
\usepackage{textcomp}
\usepackage{xcolor}
\usepackage{hyperref}
\usepackage{ifthen}
\usepackage{wrapfig}
\usepackage{subcaption}
\usepackage{cleveref}

\crefname{listing}{listing}{listings}
\Crefname{listing}{Listing}{Listings}

\usepackage{threeparttable}

\newboolean{showcomments}
\setboolean{showcomments}{true}

\ifthenelse{\boolean{showcomments}}
{\newcommand{\nb}[2]{
		\fbox{\bfseries\sffamily\scriptsize#1}
		{\sf\small$\blacktriangleright$\textit{#2}$\blacktriangleleft$}
	}
}
{\newcommand{\nb}[2]{}
}

\newcommand\os[1]{{\color{red} \nb{OSCAR}{#1}}}

\newcommand\me[1]{{\color{blue} \nb{MEHEDI}{#1}}}
\newcommand\an[1]{{\color{purple} \nb{ANTU}{#1}}}

\newcommand{\re}{\textcolor{red}{\textbf{[REF]}\xspace}}
\newcommand{\ie}{\textit{i.e.},\xspace}
\newcommand{\eg}{\textit{e.g.},\xspace}
\newcommand{\etc}{\textit{etc.}\xspace}
\newcommand{\etal}{\textit{et al.}\xspace}

\newcommand{\goal}{\textit{Goal}\xspace}
\newcommand{\need}{\textit{Need}\xspace}
\newcommand{\benefit}{\textit{Benefits}\xspace}
\newcommand{\alternative}{\textit{Alternatives}\xspace}
\newcommand{\selectedAlternative}{\textit{Selected Alternative}\xspace}
\newcommand{\validation}{\textit{Validation}\xspace}
\newcommand{\sideEffect}{\textit{Side Effects}\xspace}
\newcommand{\maturityStage}{\textit{Maturity Stage}\xspace}
\newcommand{\constraints}{\textit{Constraints}\xspace}
\newcommand{\dependency}{\textit{Dependency}\xspace}

\newcommand{\classJavadocs}{\textit{Class Javadocs}\xspace}
\newcommand{\methodJavadocs}{\textit{Method Javadocs}\xspace}

\newcommand{\metJd}{\textit{Method Javadocs}\xspace}
\newcommand{\cc}{\textit{Code Comments}\xspace}
\newcommand{\icc}{\textit{Inline Code Comments}\xspace}
\newcommand{\cm}{\textit{Commit Messages}\xspace}
\newcommand{\issue}{\textit{Issue}\xspace}
\newcommand{\issues}{\textit{Issues}\xspace}
\newcommand{\pr}{\textit{Pull Request}\xspace}
\newcommand{\prs}{\textit{Pull Requests}\xspace}
\newcommand{\codeRev}{\textit{Code Reviews}\xspace}
\newcommand{\crc}{\textit{Code Review}\xspace}

\newcommand{\tool}{{\textsc{Argus}}\xspace}
\newcommand{\tools}{{\textsc{Argus's}}\xspace}

\newcommand{\OFourMini}{{\textit{GPT-o4-mini}}\xspace}
\newcommand{\GPTFiveTwo}{{\textit{GPT‑5.2}}\xspace}
\newcommand{\GeminiThreeFlash}{{\textit{Gemini 3 Flash}}\xspace}

\usepackage{tikz}

\definecolor{bug_red}{rgb}{.84,.23,.29}

\definecolor{nature_green}{rgb}{.47, .107, .63}

\definecolor{info-needed-color}{rgb}{1,.8,.12}

\definecolor{lightblue}{rgb}{ .753,  .902,  .961}

\definecolor{lightgreen}{rgb}{.596, 1, .596}

\definecolor{lightyellow}{rgb}{1,0.94902,0.8}

\newcounter{finding}
\renewcommand{\thefinding}{\arabic{finding}}

\newtcolorbox{findingbox}{
	enhanced,
	colback=black!5,           
	colframe=black!100,         
	boxrule=1pt,               
	arc=5pt,                   
	left=8pt,right=8pt,        
	top=5pt,bottom=5pt,        
}

\newcommand{\finding}[2][]{%
	\refstepcounter{finding}%
	\begin{findingbox}
		\textbf{RQ\thefinding \space Summary:}~#2%
		\if\relax\detokenize{#1}\relax\else\label{#1}\fi
	\end{findingbox}%
}

\newcounter{rq}

\newtcolorbox{rqbox}{
	enhanced,
	colback=black!5,
	colframe=black!100,
	boxrule=1pt,
	arc=5pt,
	left=8pt,right=8pt,
	top=5pt,bottom=5pt,
}

\newcommand{\RQ}[2][]{%
	\if\relax\detokenize{#1}\relax
	\refstepcounter{rq}%
	\else
	\setcounter{rq}{#1}%
	\fi
	\begin{rqbox}
		\textbf{}~#2%
	\end{rqbox}%
}

\newcommand{\bluelink}[2]{%
  \href{#1}{\textcolor{blue}{#2}}%
}

\newfloat{listing}{tbp}{lol}
\floatname{listing}{Listing}

\definecolor{diffheader}{RGB}{0,0,128}
\definecolor{diffhunk}{RGB}{128,0,128}
\definecolor{diffadd}{RGB}{0,128,0}
\definecolor{diffremove}{RGB}{160,0,0}

\lstdefinelanguage{gitdiff}{
  morecomment=[l][\color{diffheader}\bfseries]{diff },
  morecomment=[l][\color{diffhunk}\bfseries]{@@},
  morecomment=[l][\color{diffadd}]{+},
  morecomment=[l][\color{diffremove}]{-}
}

\lstdefinestyle{minteddiff}{
  language=gitdiff,
  basicstyle=\ttfamily\footnotesize,
  numbers=left,
  numberstyle=\tiny\color{black},
  numbersep=7pt,
  stepnumber=1,
  frame=single,
  rulecolor=\color{black},
  framesep=3pt,
  breaklines=true,
  breakatwhitespace=false,
  columns=fullflexible,
  keepspaces=true,
  showstringspaces=false,
  tabsize=2,
  xleftmargin=\dimexpr\fboxsep+\fboxrule\relax,
  xrightmargin=\dimexpr\fboxsep+\fboxrule\relax,
  framexleftmargin=0pt,
  framexrightmargin=0pt,
  resetmargins=true,
  aboveskip=2pt,
  belowskip=3pt
}

\AtBeginDocument{%
	}

\begin{document}
	
    \title{Recovering Fine-Grained Code Change Rationale from Multiple Software Artifacts}
	
	\author{Mehedi Sun}
	\affiliation{%
		\institution{William \& Mary}
		\city{Williamsburg}
		\country{USA}
	}
	\email{msun12@wm.edu}
	
\author{Antu Saha}
\affiliation{%
	\institution{William \& Mary}
	\city{Williamsburg}
	\country{USA}
}
\email{asaha02@wm.edu}

\author{Nadeeshan De Silva}
\affiliation{%
	\institution{William \& Mary}
	\city{Williamsburg}
	\country{USA}
}
\email{kgdesilva@wm.edu}

\author{Antonio Mastropaolo}
\affiliation{%
	\institution{William \& Mary}
	\city{Williamsburg}
	\country{USA}
}
\email{amastropaolo@wm.edu}

\author{Oscar Chaparro}
\affiliation{%
	\institution{William \& Mary}
	\city{Williamsburg}
	\country{USA}
}
\email{oscarch@wm.edu}

\begin{abstract}

Understanding the reasons behind past code changes is critical for various software engineering tasks, including refactoring, code reviews, and debugging. 
%
Unfortunately, locating and reconstructing this code change rationale can be difficult for developers because such information is often fragmented, inconsistently documented, and scattered across heterogeneous artifacts such as commit messages, issue reports, and pull requests.

In this paper, we address this challenge with \textit{two novel contributions}. 
\textit{First}, we conduct an empirical study that analyzes nine rationale components drawn from an established taxonomy and traces where each is documented across the artifacts (\eg issue reports, pull requests, and code reviews) associated with 63 commits from five widely used open-source Java projects. Seven of the nine components appear in practice, and rationale is highly fragmented: commit messages and pull requests primarily capture \goal, while \need and \alternative are more often found in issues and pull requests. Other components are scarce but found in artifacts other than commit messages.  Importantly, no single artifact type consistently captures all components, underscoring the need for cross-document reasoning and synthesis.

\textit{Second}, we introduce \tool, a novel LLM-based approach that identifies sentences expressing \goal, \need, and \alternative across a commit's artifacts and synthesizes these sentences into concise rationale summaries to support code comprehension and maintenance tasks. We developed and evaluated \tool on the 63 commits and compared its performance against baseline variants. The best-performing version achieved 51.4\% overall precision and 93.2\% recall for rationale identification, outperforming an alternative approach, while generating rationale summaries rated as accurate and correct relative to reference summaries.  Our experiments with different LLMs for both rationale identification and generation further show that although the rationale identification accuracy varies across models, \tool can synthesize rationale summaries consistently rated as accurate. A user study with 12 Java programmers, who engaged in understanding unfamiliar code changes, further showed that these summaries were found useful for code change understanding and could support tasks such as code review, documentation, and debugging.

Overall, our results show that fine-grained rationale components are distributed across heterogeneous software artifacts and cannot be recovered from commit messages alone. By automatically identifying and synthesizing this fragmented rationale information, \tool provides developers with concise rationale summaries that support code comprehension, code review, debugging, and long-term software maintenance.
\end{abstract}
 

\maketitle


\section{Introduction}
\label{sec:intro}

Understanding the \textit{rationale behind past code changes} is one of the most frequent and important activities that software developers perform when developing and maintaining software systems~\cite{al2022developers,Rastkar:ICSE13,Ebert:SANER19,pascarella2018information,Codoban:ICSME15,Tao:FSE12,LaToza:EUPLT10,Ko:ICSE07,Burge2008}. 
This \textbf{code change rationale} provides essential context about the \textit{reasoning and justification} underlying a code change, beyond describing \textit{what} was modified in the code (\ie the code changes themselves)~\cite{Tao:FSE12,Rastkar:ICSE13,al2022developers,liang2023qualitative}. Such rationale includes the motivations for the change, the goals it seeks to achieve, the alternatives considered, the implementation decisions made, and the evidence supporting the correctness or effectiveness of the resulting solution~\cite{al2022developers}.
Prior studies underline that understanding code change rationale is a critical requirement for developers to perform various software engineering tasks~\cite{al2022developers,LaToza:EUPLT10,saha2025automatically}, including refactoring and code review, diagnosing and fixing bugs, documenting design and functionality, and reusing code for implementing new features. Code change rationale allows developers to interpret the structure and behavior of the codebase, understand trade-offs considered during development, and make informed decisions when correcting defects, implementing enhancements, or designing new features~\cite{al2022developers,Rastkar:ICSE13,Burge2008,Tao:FSE12}. Without this context, developers risk introducing regressions, violating architectural constraints, or duplicating previous efforts, especially when they are unfamiliar with the project's history or were not involved in earlier development phases~\cite{al2022developers,Rastkar:ICSE13,saha2025decoding,Burge2008}.

Despite its importance, collecting and understanding rationale can be difficult for developers~\cite{al2022developers,Tao:FSE12,LaToza:EUPLT10,Ko:ICSE07}. 
Rather than being formally documented, rationale is frequently confined to ephemeral media.
Even when recorded digitally, it is often documented informally, expressed inconsistently, and distributed across multiple software artifacts~\cite{Rastkar:ICSE13,Sillito:TSE08}. Junior developers and newcomers, who might rely on senior colleagues for guidance, often lack easily accessible sources of rationale and are forced to reconstruct it from heterogeneous and unstructured artifacts~\cite{Rastkar:ICSE13,Sillito:TSE08,bacchelli2013expectations,cortes2014automatically,li2023commit}, including issue reports, pull requests, commit messages, code reviews, code comments, chat conversations, and meeting logs. These artifacts are created by different stakeholders, at different stages of the development process, and for different purposes. They vary widely in style, content, and length and often use inconsistent terminology to express similar ideas~\cite{Rastkar:ICSE13,Tian:ICSE2022,chaparro2017detecting,zimmermann2010makes}. Consequently, locating and reconstructing the rationale behind a code change often requires developers to navigate, interpret, and connect information distributed across multiple artifacts, which can be tedious, error-prone, and time-consuming~\cite{al2022developers,Rastkar:ICSE13,Sillito:TSE08,roehm2012professional}. 

{Although prior research has explored methods for identifying rationale in software artifacts~\cite{Mouna2025MSR, Alkadhi:SANER18, Rani2021JSSCOMMENTTYPE, Di2019TSE, Jiuang2024ASE, Lopez2012, Rogers2012, RogersSpringer2015, Sharma:ICSE21, pascarella2019classifying}, existing approaches fall short in several key dimensions. First, they focus on detecting rationale content within a single artifact type~\cite{Mouna2025MSR, Alkadhi:SANER18, pascarella2019classifying, Jiuang2024ASE, Sharma:ICSE21}, such as commit messages or issue reports, limiting their applicability in real-world projects where documentation is typically fragmented~\cite{Hou2008APSEC}. 
Second, they lack mechanisms to connect and integrate rationale fragments across documents, making it difficult for developers to obtain a coherent understanding of the reasons behind a change~\cite{Hou2008APSEC, Rath2018ICSE}. Third, they are unable to identify fine-grained rationale components (\eg a change's goal, underlying need, or alternative solutions considered)\; that developers often require and look for when making sense of software changes~\cite{al2022developers}. Finally, they do not support cross-artifact generation of concise rationale summaries that developers can easily consume to facilitate software comprehension and maintenance tasks. 


To address these challenges, this paper makes \textbf{two overall contributions}. \textbf{First}, we conduct a comprehensive empirical study of how code change rationale components are expressed across multiple software artifacts (issue reports, pull requests, commit messages, \etc). We manually analyze the artifacts associated with 63 real-world commits from five widely used Java open-source systems.
These commits span a wide range of change types and sizes, contributed by different developers, and are linked to diverse artifacts that capture various development workflows used to implement and validate the changes.
%
%
Using multi-coder qualitative analysis guided by an established rationale taxonomy~\cite{al2022developers}, we identify seven fine-grained rationale components. These include the \goal of the change, the \need motivating it, \textit{Alternative solutions} considered, and the \textit{Selected Alternative}. 
Our analysis reveals that rationale information is highly fragmented and inconsistently distributed across artifacts. Commit messages and pull requests primarily document a change's \goal, while \need is mostly captured in issues and pull requests. Other components (\eg \alternative and \selectedAlternative) appear less frequently and are scattered across different artifacts. Pull requests and issues stand out as the most diverse sources of rationale information. Importantly, most rationale components beyond \goal are documented outside commit messages. In essence, no single artifact type consistently captures all rationale components, highlighting the need for techniques that can identify, retrieve, and synthesize rationale information across multiple artifacts.
\looseness=-1

\textbf{Second}, building on these insights, we design and evaluate \tool, a novel LLM-based approach for fine-grained, multi-document rationale extraction and generation. \tool identifies the software artifacts associated with a code change (\eg issue reports, pull requests, and code reviews) and performs two main tasks: (1) it identifies sentences that express the three most common rationale components from the study (\goal, \need, and \alternative), and (2) it generates concise, coherent summaries of these components that integrate rationale information from the artifacts. We experiment with multiple prompting strategies for \tool to guide the model in both rationale extraction and generation using the 63 studied commits. Employing a data-driven methodology using a development set of 13 commits, we identify the most effective prompting strategy (namely, task decomposition and few-shot exemplars with explanations) and evaluate its effectiveness on a held-out test set of 50 commits. Our analysis reveals that \tool outperforms prompting baselines, achieving 51.4\% precision and 93.2\% recall for rationale identification, while generating rationale summaries rated as accurate and correct relative to reference summaries.
We also conducted a cross-LLM analysis of rationale identification and generation. The results show that model choice has an effect on rationale identification accuracy, whereas rationale generation is much less sensitive to the choice of LLM when the same rationale sentences are provided as input.

We further evaluated the perceived usefulness of \tool in a user study with 12 Java developers, who assessed the generated rationale components for 18 commits. Participants reported that the synthesized rationale helped them understand why changes were made across different types of commits. They also indicated that the rationale could be useful in comprehending code changes and can support maintenance tasks such as code review, documentation, and debugging.
\looseness=-1

In summary, our work makes the following contributions:
\begin{enumerate}
	\item An empirical study that found that rationale components for a code change, such as the \goal, \need, and \alternative, are fragmented and inconsistently expressed in various software artifacts (issues,  pull requests, code reviews, \etc). This highlights the need for cross-document rationale identification and synthesis to support developers in understanding the reasons behind code changes. To the best of our knowledge, we are the first to study how code change rationale components are documented in software artifacts. 
	\item A novel multi-document LLM-based approach, \tool, that identifies fine-grained rationale component information across diverse artifact kinds and synthesizes summaries of each component. 
    \item A data-driven tool development and evaluation methodology for rationale identification and generation to identify the most effective prompting strategies for \tool. Combining reasoning-based few-shot exemplars with task-decomposition prompting leads to accurate and correct rationale summaries ready to be consumed by developers. 
    \item A cross-model sensitivity analysis examining how different LLMs affect the rationale identification and generation modules of \tool.
	\item A user study that assessed the usefulness of \tool's generated rationale summaries, showing that developers find them useful for understanding code changes and supporting various tasks such as code review, debugging, and documentation.
	\item A novel dataset of fine-grained, multi-document code change rationale, with rationale content annotated in artifacts associated with 63 commits of five popular and active open source projects. This dataset is part of a publicly available replication package that provides source code and infrastructure to support verification and reproduction of work, as well as further research in this area~\cite{package}. 
\end{enumerate}


\section{Background and Motivating Example}
\label{sec:background}

\subsection{Code Change Rationale Components}
A \textit{code change} is a set of modifications,  additions, or removals made to one or more code files in a software system, recorded as a \textit{commit} in a version control system.
%
The \textit{rationale of a code change} captures the reasoning and justification underlying that change, including its motivations, goals, alternatives, decisions, trade-offs, and supporting evidence.
A code change's rationale can be decomposed into \textit{rationale components}, each representing a distinct aspect of the reasoning behind the change, such as the \textit{Need} motivating it, the \textit{Goal} it seeks to achieve, the \textit{Alternatives} considered, or the \textit{Validation} supporting the chosen solution~\cite{al2022developers}. 

Our work is based on the taxonomy of code-change rationale by Al Safwan \etal~\cite{al2022developers}, developed through a rigorous mixed-methods study, which included a literature review and surveys and interviews with practitioners. The taxonomy defines 15 rationale components that capture different aspects of the information developers seek when understanding software evolution.

The components in the taxonomy are categorized into four broad themes: the code change's \textit{objective}, \textit{design}, \textit{execution}, and \textit{evaluation}.  
\textit{Objective} rationale components include the code change's \goal (what the developers aimed to achieve) and \need (the reasons or motivations for the change). \textit{Design} components include  \textit{Constraints} (limitations or restrictions for implementing the change) and \alternative (potential solutions discussed by developers but not implemented in the final
code change).  \textit{Execution} components include \textit{Time} (timing-related reasons for implementing the change) and \textit{Modifications} (reasons related to code modifications).  \textit{Evaluation} components include \sideEffect (reasons related to mitigating a change's side effects) and \maturityStage (reasons related to how mature the change is).
The components overlap by design as some can apply in certain development contexts while others can apply in other contexts. We refer the reader to Al Safwan \etal's paper~\cite{al2022developers} for the full taxonomy.

While the original taxonomy contains 15 rationale components, our study focuses on the subset of components that require interpretive reasoning and are expressed by different stakeholders within software artifacts. As discussed in \Cref{sec:rationale_study} this results in seven rationale components that were observed and analyzed in our dataset.


\begin{figure}[t]
	\caption{A Motivating Example from the \textit{OkHttp} Project~\cite{okhttp_commit_4c86085}}
	\centering \includegraphics[width=\columnwidth]{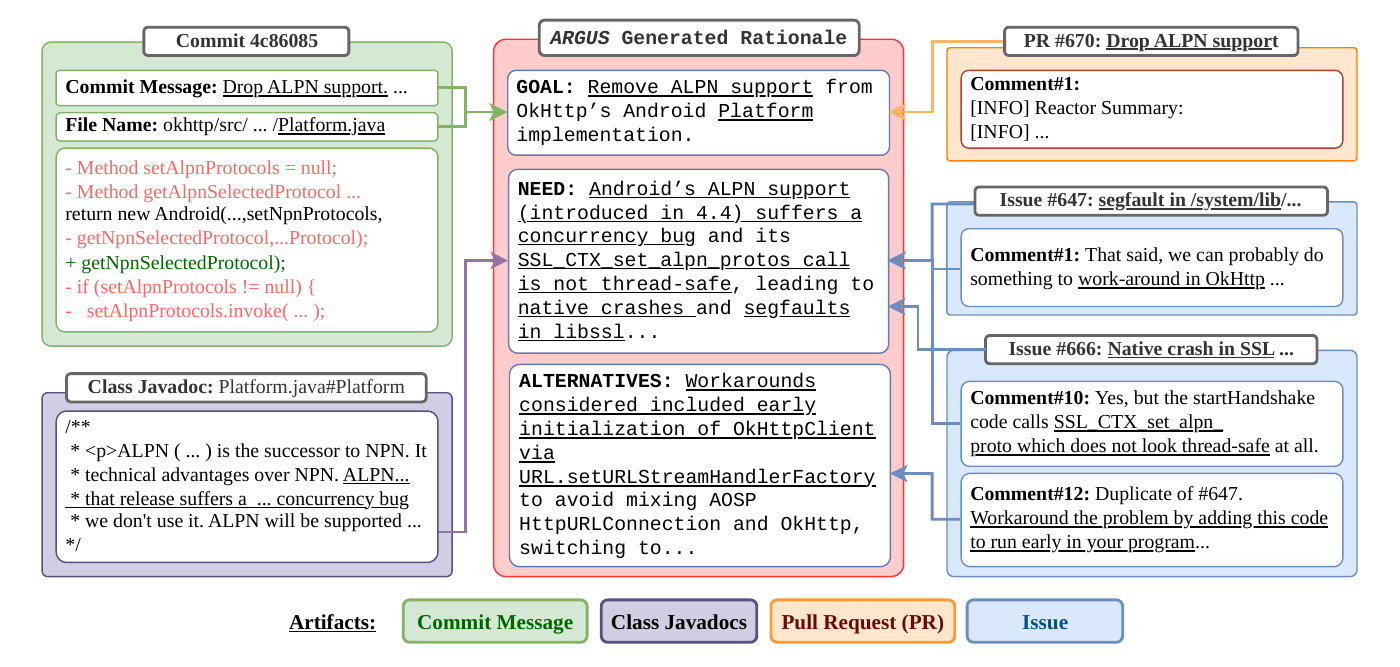}
	\label{fig:example}
\end{figure}

\subsection{Problem and Motivating Example}
\Cref{fig:example} illustrates a code change and the related rationale fragments scattered across multiple artifacts that explain why the change was made. The example shows a commit\footnote{https://github.com/square/okhttp/commit/4c86085429edbeef0a383941936ee7b64cc3805e} from OkHttp\footnote{https://square.github.io/okhttp/}, a Java HTTP client for the Java Virtual Machine (JVM), Android, and the Graal Virtual Machine (GraalVM). The commit removed logic in \texttt{Platform.java} that supported Application-Layer Protocol Negotiation (ALPN), a Transport Layer Security (TLS) extension enabling clients to inform servers which application-layer protocols they support. While the commit message states the change's \goal (“Dropping ALPN support”), it does not explain why this change was necessary or what alternatives were considered.

Artifacts linked to the commit, shown in the figure, provide richer context that helps explain the change. For instance, the \textit{Class Javadoc} of \texttt{Platform.java} noted that ``\textit{Android 4.4 … suffers from a concurrency bug}'' while Issues \#666\footnote{https://github.com/square/okhttp/issues/666} and \#647\footnote{https://github.com/square/okhttp/issues/647} include detailed comments about the problem: ``\textit{the SSL\_CTX\_set\_alpn\_protos call is not thread safe}” and causes a ``\textit{segfault in … libssl.so.}'' Issue \#666 also records alternative solutions that developers considered, including small code changes as workarounds until a permanent fix in Android became widely available.

Importantly, no individual artifact contains enough information to reconstruct a complete view of the commit's rationale, including its \goal, \need, and \alternative. A developer seeking a comprehensive understanding of the change would therefore need to manually locate, examine, and synthesize information scattered across multiple artifacts, particularly when the original developers are unavailable, as is often the case in open-source projects. This process can be tedious and time-consuming, particularly for junior developers or newcomers unfamiliar with the project’s code and history. Our goal in this paper is to automate this process by extracting and synthesizing rationale information from multiple artifacts into coherent rationale summaries---such as those shown in \Cref{fig:example} (see \tool generated rationale) and produced by our tool, \tool. These summaries provide developers with a concise view of a change's rationale and can support software maintenance and evolution tasks, such as investigating regressions, reviewing code changes, and implementing new features.



\section{Investigating Code Change Rationale Components in Software Artifacts}
\label{sec:rationale_study}

We investigate how software artifacts (\eg issue reports, pull requests, and code reviews) related to a commit capture different fine-grained rationale components (\goal, \need, \etc) Guided by the taxonomy of Al Safwan \etal~\cite{al2022developers} (see \Cref{sec:background}), our study examines the extent to which these artifacts provide rationale information that can be extracted, integrated, and presented to developers when understanding code changes.

The study addresses the following research questions (RQs):

\begin{enumerate}[label=\textbf{RQ$_\arabic*$:}, ref=\textbf{RQ$_\arabic*$}
	,itemindent=0cm,leftmargin=1cm]

    \item \label{rq:components}{\textit{How frequently are rationale components documented in software artifacts?}}

    \item \label{rq:comp_prevalence}{\textit{Where are rationale components documented across software artifact types?}}

\end{enumerate}

\ref{rq:components} provides insight into the prevalence of different rationale components, \ie which are more or less commonly documented.  \ref{rq:comp_prevalence} examines how rationale components are distributed across artifact types, identifying the primary sources of rationale.

\subsection{Commit Collection}

Our study builds on the dataset curated by Tian \etal~\cite{Tian:ICSE2022}, which consists of 1,649 manually selected commits from five large, widely used Java open-source software (OSS) projects: \textit{Spring Boot}~\cite{springboot}, a Java-based framework for building production-grade applications; \textit{Apache Dubbo}~\cite{dubbo}, an RPC and microservice framework; \textit{OkHttp}~\cite{okhttp}, an HTTP client for the JVM, Android, and GraalVM; \textit{JUnit4}~\cite{junit4}, a unit testing framework; and \textit{Retrofit}~\cite{retrofit}, an HTTP client built on top of \textit{OkHttp}. These commits were sampled from a larger population of 41.8K commits collected from these projects' development up to 2021. Tian \etal carefully curated this dataset to capture diverse commit message characteristics, including variations in how developers describe the change (\textit{what}), the rationale (\textit{why}), or both. We refer the reader to Tian \etal's paper~\cite{Tian:ICSE2022} for a detailed description of their data collection process.

To focus our analysis on commits where rationale is meaningful and attributable, we apply several filtering criteria to the 1,649 commits:

\begin{itemize}
	\item \textbf{Language Filtering:} We retained only commits that modify at least one Java file, as our analysis focuses on code-level rationale. This resulted in excluding \num{510} commits without Java file changes. 
	
	\item \textbf{Atomicity:} We excluded \num{52} non-atomic commits that bundle multiple unrelated changes, as such commits hinder the attribution of rationale to a single change. Non-atomic commits were already identified by Tian \etal~\cite{Tian:ICSE2022}. 
	
	\item \textbf{Modification Size Filtering:} We filtered commits based on modification size (the number of modified files and changed lines of code) to remove extremely large changes as they often combine multiple concerns~\cite{kirinuki2014hey}. For example, commit~\cite{junit4_commit_44e7458} has almost 400 changed lines and includes several unrelated changes \eg `Moved InitializationError to ParentRunner', `Updated version in docs to 4.5' or `included docs about junit-dep jar'.  Using the interquartile range (IQR), we removed \num{166} commits that were outliers in the number of modified files and \num{226} commits that were outliers in the number of changed lines of code.
\end{itemize}

Applying these criteria reduced the dataset to 830 candidate commits. From these, we randomly selected 63 commits, stratified across projects (12–13 commits per project, shown in \Cref{tab:data_sample_stats}), to ensure coverage across different systems while keeping the manual artifact annotation, based on multi-coder qualitative analysis, tractable (see \Cref{subsec:annotation}). 

The sampled commits cover a wide range of change characteristics (\Cref{fig:com_diversity}). 
For example, commits range from one to five modified Java files and from one to 206 changed lines, 
while being associated with one to 25 artifacts spanning up to seven artifact types (see \Cref{subsec:artifact_collection} for the process of identifying associated artifacts). 
This diversity provides a rich set of scenarios for analyzing how rationale is expressed across artifacts. While the filtering reduces the size of the dataset, it allows us to focus on commits where rationale is more likely to be present, attributable, and interpretable, which is essential for our qualitative analysis.

\begin{figure}[t]
	\centering
    \includegraphics[width=\linewidth]{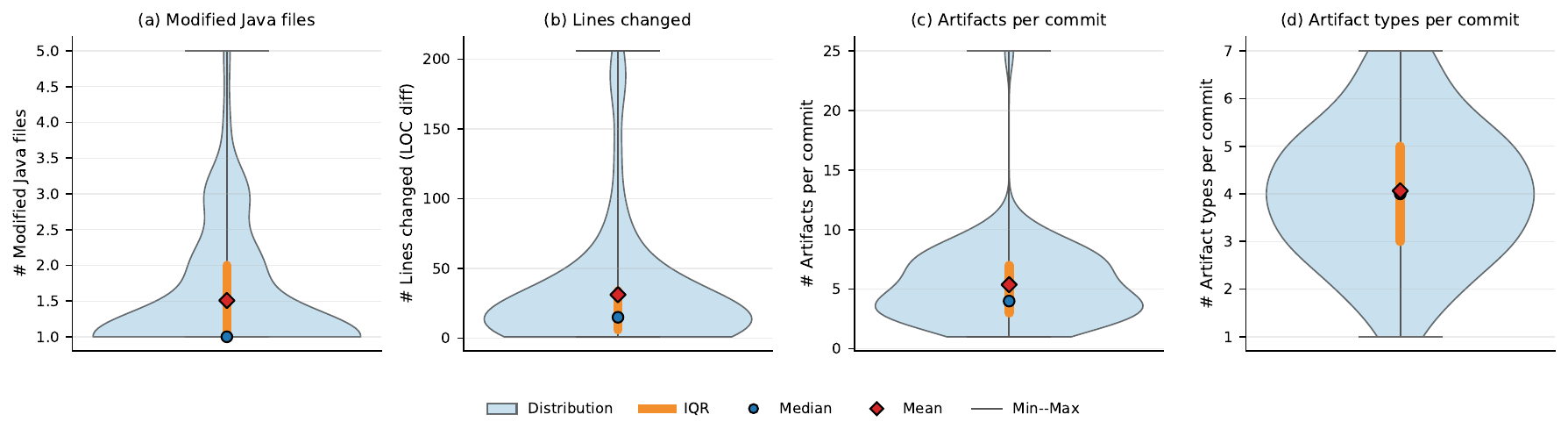}
    \caption{Distribution of sampled commit characteristics}
    \label{fig:com_diversity}
\end{figure}

In our online replication package~\cite{package}, we provide the full list of sampled commits along with the applied filtering criteria and stratification details for transparency and reproducibility.

\subsection{Software Artifact Collection}
\label{subsec:artifact_collection}

To study how the rationale information is distributed across different artifacts, we first collected artifacts associated with each sampled commit. This subsection describes which artifact types we considered, why we selected them, and how we retrieved them.

\subsubsection{Types of Software Artifacts and Motivation\\} Understanding the rationale behind code changes requires examining information that is distributed across multiple software artifacts associated with those changes. Prior research on rationale sources has shown that developers document rationale information in a wide range of artifacts, including commit messages~\cite{dhaouadi2024ICPCRationaleDataset}, issue reports~\cite{saha2025decoding,Rastkar:ICSE13, Jiuang2024ASE}, emails~\cite{Di2019TSE}, and code comments~\cite{pascarella2019classifying, Rani2021JSSCOMMENTTYPE}. Informal communication channels such as mailing lists~\cite{Sharma:ICSE21, bi2021JSSarchitecture}, discussions on forums and Q\&A sites~\cite{zhou2025using} and chat discussions~\cite{Alkadhi:SANER18}  have also been shown to contain rationale, including arguments and design decisions, although such information is often unstructured and difficult to systematically extract and link to specific code changes.

Despite this diversity of sources, most existing work focuses on individual artifact types in isolation, leaving limited understanding of how rationale is distributed, complemented, and reconstructed across artifacts. To address this gap, we adopt a multi-artifact perspective and systematically collect artifacts that capture different aspects of rationale information, \ie rationale components such as a change's \goal, \need, and \alternative.

We collected six types of artifacts linked to the 63 sampled commits: \issues, \prs, \crc, \textit{Javadoc Comments}, \icc, and \cm. We focus on these artifacts because they are consistently available and publicly accessible across all five projects, as they are managed within GitHub, and they can be reliably linked to specific commits through explicit references or repository structure. Other potential sources of rationale, such as mailing lists, chat logs (\eg Slack or IRC), and external discussions (online meeting transcripts), were not included in this study due to their inconsistent availability across the selected projects. For example, while mailing list archives are available for some projects (\eg Apache Dubbo~\cite{dubbo_mailing_list}), similar resources were not found for others (\eg Spring Boot). We also did not identify publicly accessible chat channels for any of the studied projects. Although GitHub Discussions~\cite{githubdiscussions} are available, they are not consistently used across projects, and it is often difficult to reliably link discussion threads to specific commits or issues. As a result, incorporating these artifacts would require substantial manual effort and would hinder the consistency and reproducibility of the dataset. Focusing on the six types of artifacts mentioned above therefore allows us to construct a dataset that is systematically linked to commits and suitable for reliable analysis.

We distinguish between class-level Javadocs, method-level Javadocs, and inline code comments (all these representing code documentation) because they convey rationale at different levels of granularity and serve distinct documentation purposes. Class-level Javadocs typically describe high-level design intent, architectural roles, and overall responsibilities of a class. Method-level Javadocs provide more localized explanations, such as the behavior of individual methods, parameter semantics, and implementation constraints. Inline code comments are embedded directly within the code and can capture fine-grained, implementation-specific rationale. Separating these three forms of code documentation enables us to analyze how rationale is expressed across different abstraction levels, from high-level design decisions to low-level implementation details.

\subsubsection{Defining Individual Artifacts\\} 
We defined each software artifact as a self-contained unit of textual content for analysis. An \textit{Issue Report} consists of the issue title, description body, and all comments within the issue thread. A \textit{Pull Request} (PR) consists of the PR title, description body, and all comments within the PR discussion thread. For source code comments, a \textit{Class Javadoc} comprises all documentation comments immediately preceding the declarations of modified classes in a given Java file (as reported in the commit), while a \textit{Method Javadoc} comprises all documentation comments immediately preceding the declarations of modified methods in that file. An \textit{(Inline) Code Comment} artifact consists of all inline comments contained within modified classes and modified methods, excluding comments located in unchanged methods of the same file. 
For \textit{Class Javadocs}, \textit{Method Javadocs}, and \textit{(Inline) Code Comments}, we focus on comments associated with modified code elements, as they are more likely to capture the rationale for the changes introduced in the commit than comments attached to unchanged code.
A \textit{Code Review} artifact consists of all review comments attached to changed lines in the commit diff of a single PR, potentially spanning multiple code files. Finally, a \textit{Commit Message} consists of the commit subject line and the message body. 

\subsubsection{Process of Collecting the Artifacts\\}
All five projects in the dataset are hosted on GitHub, which provides unified access to issues, pull requests, and code reviews. To establish accurate commit–artifact links, we combined heuristics, regular expressions, GitHub's API requests, and manual validation. Specifically, we (i) searched commit messages for explicit references to issue or pull request identifiers using common patterns (\eg ``\#\num{}'', ``pull/\num{}'') and issue or pull request (PR) URLs, (ii) queried the GitHub API to identify mentions of commit hashes in issue and pull request discussions, and (iii) performed cross-artifact searches (\eg identifying issues or pull requests referenced in commits, comments, titles, or discussion threads) to recover additional links, accounting for cases where the referenced issue/PR is not explicitly mentioned in the commit message but appears elsewhere in related artifacts.

After identifying the related issues and pull requests, we used the GitHub API to systematically retrieve their detailed metadata, including descriptions and discussion threads. Javadocs and inline comments were extracted from modified Java files using the \textit{comment-parser} library~\cite{commentparser} with pattern-based heuristics, while code review comments were obtained from pull request discussions via the GitHub API. A single commit may be associated with multiple artifacts of the same type (\eg multiple issues or pull requests), and artifacts may also reference each other.

In total, we identified and constructed \textbf{339} (5.4 artifacts per commit) artifacts across the 63 commits, providing a diverse and multi-granular dataset for subsequent annotation and analysis.

\subsection{Artifact Validation and Annotation Process}
\label{subsec:annotation}

We conducted an iterative multi-coder qualitative analysis to (1) validate that the identified issue reports and pull requests were truly associated with the commits, and (2) annotate the sentences in the artifacts according to the rationale components they convey. Our process follows a consensus-based coding approach guided by the taxonomy of Al Safwan \etal~\cite{al2022developers}. We first parsed the textual content of all collected artifacts into sentences using SpaCy’s English transformer pipeline (\texttt{en\_core\_web\_trf})\footnote{We parsed sentences from each code comment individually, rather than from all comments concatenated within a Java file, to reduce sentence-segmentation errors, such as incorrectly identifying a sentence that spans two adjacent comments.}~\cite{spacymodel}, resulting in a total of \textbf{3,088 sentences} in the \textbf{339 artifacts} associated with the \textbf{63 commits}.

\subsubsection{Artifact Relevance Validation\\}
\label{artifact_validation}

Our goal was to analyze only artifacts that were genuinely associated with the target code change. Because artifacts were collected automatically, some retrieved artifacts could be false positives: although linked to the target commit, they primarily discussed a different code change and therefore did not provide rationale for the target commit.

Two annotators independently assessed the relevance of all extracted \issues and \prs. We did not perform this validation for \cc, \methodJavadocs, or \classJavadocs because these artifacts were extracted directly from modified Java files in the target commit. Similarly, \crc comments were obtained from the validated \prs. An \issue or \pr was considered relevant if it discussed the target code change or provided contextual information necessary to understand it.

To judge relevancy, the annotators first reviewed the target commit's diff and commit message, then examined the associated \issue or \pr, including its title, description, and discussion comments. An artifact was judged as irrelevant if it was linked to the target commit but primarily focused on a different code change. For example, a retrieved \pr might implement a separate change while merely referencing the target commit, or an \issue might mention multiple commits, including the target one, while centering on another modification. Otherwise, it was judged as relevant.

From the 339 artifacts collected across 63 commits, both annotators agreed that 44 artifacts were irrelevant, while 7 artifacts were flagged as irrelevant by only one of the annotators. All disagreements were subsequently resolved through consensus during discussion meetings. In total, 291 of 339 (85.8\%) artifacts were considered relevant after reconciliation, and 48 artifacts were deemed irrelevant and excluded from further analysis. These excluded artifacts contained 486 sentences, which were not considered for qualitative analysis. The remaining 2,602 sentences, drawn from relevant artifacts, were retained for rationale annotation. 

For the automated rationale identification and generation experiments described in \Cref{sec:prompt_dev}, we intentionally included irrelevant content to evaluate how our approach performs under realistic conditions where such artifacts may be retrieved for analysis. 

\subsubsection{Rationale Annotation Process\\}

Our goal is to identify sentences that express \emph{code change rationale} and assign them to fine-grained components from Al Safwan \etal's taxonomy~\cite{al2022developers} (\eg \goal, \need, \alternative, \validation \etc). We annotate at the sentence level, where a sentence may express multiple rationale components. 
We conducted annotation over six iterative rounds, each involving 10–13 commits. Two annotators independently labeled the sentences in each round as follows:
\begin{enumerate}[leftmargin=2em]
	\item The annotators labeled each sentence with one or more rationale components. Sentences that did not express rationale were left unlabeled.
	\item After each round, the annotators reviewed all conflicts and resolved them through discussion to reach consensus.
	\item Following agreement, we summarized the identified rationale components for each commit into concise natural-language statements, which serve as ground truth for evaluating automated rationale generation.
	\item This process was repeated across all six rounds until all commits were annotated.
\end{enumerate}

The annotation and ground-truth construction required substantial manual effort. Across the six annotation iterations, we spent approximately 52.8 hours on independent coding, 10.1 hours on reconciliation, and 20 hours on ground-truth rationale creation, for a total of about 82.9 human hours.

\underline{Codebook Development.}
A shared codebook guided the annotation process. It was initially derived from the definitions and examples provided in Al Safwan \etal's taxonomy~\cite{al2022developers}.  For example, following Al Safwan \etal's definitions, we defined `\goal' as statements describing what the developer aims to achieve in a commit, including both high- and low-level descriptions of the code change (\eg "Add a new overload of \texttt{Response.success()} that accepts a custom HTTP status code" --- example for the Retrofit commit \texttt{329a9fd}~\cite{retrofit_commit_329a9f}, detailed in \Cref{subsec:illustrative_example}); the `\need' includes statements conveying reasoning or motivation for why the change is required or why it was implemented in a particular way (\eg "Tests currently cannot verify behavior for successful responses other than 200 OK"); `\validation' includes statements that provide evidence or justification that the implemented changes achieve their intended objective, thus supporting the correctness or effectiveness of the chosen solution (\eg "The newly added test cases confirm that responses with status code 204 are handled correctly").
Following each iteration of the annotation process, the codebook was refined to incorporate additional rules for identifying rationale components, particularly in response to annotator disagreements and sources of ambiguity.  To minimize possible sources of confusion when interpreting the content across multiple components, we introduced explicit rules; for instance, a sentence that merely describes the situation in which a problem occurs is not labeled as `\need' unless it also conveys the underlying motivation for the change.
After each round, the annotators collaboratively updated the codebook by clarifying definitions, adding illustrative examples, and refining annotation guidelines to reduce subjectivity. This iterative refinement process improved consistency and minimized ambiguity across annotation rounds. 
The final version of the codebook containing annotation rules for all the components is included in our replication package~\cite{package}. 

\underline{Contextual Grounding and Bias Mitigation.}
To ensure that sentences were accurately judged as reflecting rationale, and to minimize interpretation biases stemming from limited familiarity with the project's code, the annotators followed a structured preparation and contextualization process. Before coding a commit's artifacts, they first examined the code change itself, along with relevant surrounding code in the modified files, to establish the technical context of the modification. Next, they carefully reviewed all information contained in the associated artifacts (\eg the full discussion and metadata in the issue reports) to capture the broader context of the commit. Only after this contextual review did the annotators proceed with sentence-level annotation of rationale. When necessary, they also consulted official project documentation, including web API references and code-level usages of modified APIs, to better understand their intended usage.

This process ensured that annotations were not based on isolated sentences but grounded in the full technical and organizational context. It allowed annotators to determine (1) whether the artifacts were directly associated with the commit (rather than serving merely as cross-references, general discussions, or historical notes), and (2) whether a given sentence expressed rationale and, if so, which components of the taxonomy it represented. Ambiguous cases were resolved during reconciliation meetings. 

\underline{Rationale Component Selection.}
\label{comp_select}
We initially conducted a pilot study on a stratified sample of 10 commits. During this phase, we observed that certain execution-related components (\eg Committer, Time, Location, Modifications, and Explanation of Modifications) are directly obtainable from version-control metadata and therefore do not require interpretive annotation. This observation is also consistent with findings by Al Safwan \etal's study~\cite{al2022developers}, whose participants identified these components as the easiest to locate. Additionally, the \benefit component was excluded due to its strong semantic overlap with \need, which made consistent distinction of the two components difficult during annotation. From the 15 rationale components defined by Al Safwan \etal~\cite{al2022developers}, we therefore excluded these six components. Our analysis focuses on the remaining nine components that capture more interpretive, decision-oriented aspects of developer rationale.

\underline{Inter-Annotator Agreement.}
We measured inter-annotator agreement using Cohen’s $\kappa$~\cite{cohen1960}. For artifact relevance identification, agreement remained consistently high across annotation rounds, reaching $\kappa = 0.93$ in the final iteration, indicating almost perfect agreement. For rationale component labeling, agreement improved over successive rounds from $\kappa = 0.58$ to $\kappa = 0.79$, reflecting increased alignment between annotators. As shown in \Cref{fig:agreement_trend}, both tasks exhibit stable and high agreement, with rationale labeling achieving substantial agreement in later rounds. Disagreements between annotators mainly arose from (1) implicit rationale sentences \ie rationale spreading across sentences; (2) component overlap \ie a single sentence expressing information about multiple rationale components; and  (3) differences in technical knowledge and subjective interpretation. 
Detailed component-wise agreement rates are available in our replication package~\cite{package}.

\begin{figure}[h]
\centering
\includegraphics[width=0.5\linewidth]{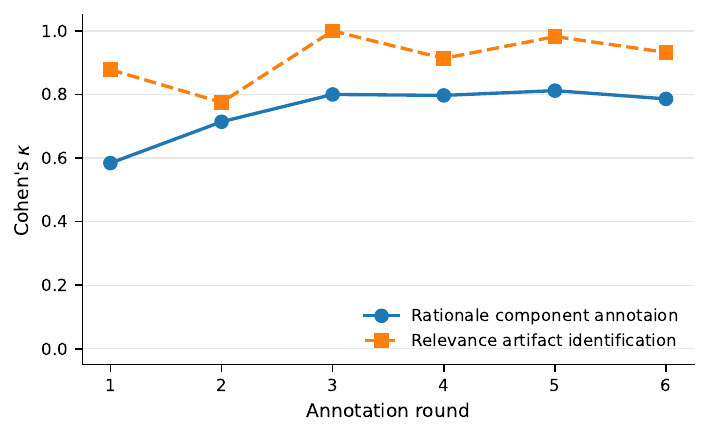}
\caption{Inter-annotator agreement (Cohen’s $\kappa$) across annotation rounds for rationale component labeling and relevance artifact identification.}
\label{fig:agreement_trend}
\end{figure}

The final outcome of this process is a sentence-level, multi-label dataset of rationale components grounded in validated artifacts. This dataset serves as the foundation for our empirical analysis and the evaluation of our automated rationale extraction approach.

\subsection{Results and Analysis}
\label{sec:results}

We analyzed 63 sampled commits comprising 127 modified code files (2 files per commit on average). For these commits, we initially collected 77 \prs and 37 \issues. After filtering out artifacts not directly related to the corresponding commits, 49 \prs and 24 \issues were retained for analysis.

Across all artifacts, we identified a total of 3,088 sentences. Following the artifact relevance validation process described in ~\Cref{artifact_validation}, 48 artifacts (486 sentences) were deemed irrelevant and excluded. From the retained artifacts, we collected multiple sources of rationale. Specifically, from the modified files, we extracted 57 \classJavadocs, 16 \methodJavadocs , and 68 \textit{Inline} \cc (\cc hereon). Among the 49 \prs, 14 included \codeRev. Overall, this resulted in 291 artifacts (4.6 per commit on average) used in our analysis. The number of artifacts per project is presented in \Cref{tab:data_sample_stats}.

\begin{table}[t]
\small
\centering
\caption{Number of artifacts associated with the 63 sampled commits.}
\label{tab:data_sample_stats}
\renewcommand{\arraystretch}{1.1}
\small		
\begin{tabular}{c|c|c|c|c|c|c|c|c|c}
    \toprule
    \textbf{Project}     & \textbf{Commits} & \makecell{\textbf{Changed}\\\textbf{Code Files}} & \makecell{\textbf{Code}\\\textbf{Comments}} & \makecell{\textbf{Class}\\\textbf{Javadocs}} & \makecell{\textbf{Pull}\\\textbf{Requests}} & \textbf{Issues} & \makecell{\textbf{Method}\\\textbf{Javadocs}} & \makecell{\textbf{Code}\\\textbf{Reviews}} & \textbf{Total}  \\ \midrule
    Dubbo       & 13                                                           & 18 (1.4)                    & 16 (1.3)               & 11 (1.4)                 & 15 (1.5)                & 6 (1.2)                 & 4 (1.0)                   & 1 (1.0)                & 66 (5.1)           \\ 
    JUnit4      & 13                                                           & 18 (1.4)                    & 7 (1.0)                & 10 (1.1)                 & 9 (1.0)                 & 3 (1.0)                 & 2 (1.0)                   & 3 (1.0)                & 47 (3.6)           \\ 
    OkHttp      & 12                                                           & 40 (3.3)                    & 21 (1.8)               & 13 (1.9)                 & 11 (1.0)                & 4 (1.3)                 & 5 (1.3)                   & 2 (1.0)                & 68 (5.7)           \\ 
    Retrofit    & 13                                                           & 23 (1.8)                    & 17 (1.5)               & 8 (1.0)                  & 11 (1.1)                & 4 (1.0)                 & 5 (1.0)                   & 7 (1.0)                & 65 (5.0)           \\ 
    Spring-Boot & 12                                                           & 28 (2.3)                    & 7 (1.2)                & 15 (1.4)                 & 3 (1.0)                 & 7 (1.2)                 & 0 (0.0)                    & 1 (1.0)                & 45 (3.8)           \\ \midrule
    Total     & 63                                                  & 127 (2.0)          & 68 (1.4)      & 57 (1.3)        & 49 (1.1)       & 24 (1.1)       & 16 (1.1)         & 14 (1.0)      & 291 (4.6) \\ \bottomrule
\end{tabular}
\footnotesize{Each commit includes a commit message which is also considered an artifact. For individual artifact types, values in parentheses indicate the average among commits containing at least one artifact of that type}

\end{table}

\subsubsection{\ref{rq:components}: How Frequently are Rationale Components Documented in Software Artifacts?\\}

We identified 7 rationale components in the analyzed artifacts, spanning 253 sentences and 62 of the 63 commits (see \Cref{tab:component-distribution}). Among these, \goal\ (what the change achieves), \need\ (why the change was necessary), and \alternative\ (considered alternatives) were the most frequently documented. Specifically, 146 sentences (57.7\%) expressed \goal, 86 (34.0\%) expressed \need, and 20 (7.9\%) expressed \alternative. In contrast, the remaining components were rare: \selectedAlternative, \maturityStage, \validation, and \sideEffect\ appeared in only 1--6 sentences overall. We found no artifact content expressing \constraints or \dependency.

\Cref{tab:component-distribution} reveals that, at the commit level, artifacts associated with 61 of the 63 commits document the change’s \goal. The \need is documented for 34 commits, indicating that nearly half of the commits lack an explicit statement of motivation in the analyzed artifacts. The remaining components (\alternative, \selectedAlternative, \maturityStage, \validation, and \sideEffect) appear in artifacts associated with 1-9 commits only, which means that most of the commits (54 or more) do not have any of these rationale components documented.

\begin{table}[t]
	\caption{Number of commits, artifacts, and sentences containing each rationale component}
	\label{tab:component-distribution}
		\small
		\begin{tabular}{l|c|c|c}
			\toprule
			\textbf{Component} & \textbf{Commit}       & \textbf{Artifact}      & \textbf{Sentence}      \\ \midrule
			\goal                                       & 61 (98.4\%)  & 114 (84.4\%)  & 146 (57.7\%)  \\ 
			\need                                       & 34 (54.8\%)  & 49 (36.3\%)   & 86 (34.0\%)   \\ 
			\alternative                               & 9 (14.5\%)   & 9 (6.7\%)     & 20 (7.9\%)    \\ 
			\selectedAlternative                     & 4 (6.5\%)    & 4 (3.0\%)     & 6 (2.4\%)     \\ 
			\maturityStage                            & 1 (1.6\%)    & 1 (0.7\%)     & 1 (0.4\%)     \\ 
			\validation                                 & 2 (3.2\%)    & 2 (1.5\%)     & 4 (1.6\%)     \\ 
			\sideEffect                              & 2 (3.2\%)    & 2 (1.5\%)     & 2 (0.8\%)     \\ \midrule
			Total                                    & 62 & 135  & 253  \\ \bottomrule
		\end{tabular}%
\end{table}

To examine whether the aggregate trends reported above are consistent across projects, we also analyzed the number of commits containing rationale components for each project (see \Cref{tab:system_level_components}). Overall, the same trend holds across all five repositories. In every project, \goal is the most frequently documented component, and it is documented for nearly all sampled commits. \need is the second most common component, as it is documented in artifacts for about half of the commits in each project. All remaining components are rarely found across projects. For example, \alternative and \selectedAlternative are documented for 1 to 4 commits only in JUnit4, OkHttp, and Spring-Boot, while \maturityStage and \sideEffect are observed only once. Overall, these results suggest that our aggregated findings are not driven by a single project; rather, the trend for developers to document some rationale components, especially \goal and \need, much more often than others is consistent across systems.

\begin{table}[t]
	\caption{Number of commits with artifacts containing each rationale component}
	\label{tab:system_level_components}
	\small
	\begin{tabular}{l|c|c|c|c|c}
		\toprule
		\textbf{Component} & \textbf{JUnit4} & \textbf{Dubbo} & \textbf{Retrofit} & \textbf{OkHttp} & \textbf{Spring-Boot} \\ \midrule
		\goal                  & 11 & 13 & 13 & 12 & 12 \\
		\need                  &  7 &  7 &  7 &  6 &  7 \\
		\alternative           &  1 &  0 &  0 &  4 &  4 \\
		\selectedAlternative   &  1 &  0 &  0 &  2 &  1 \\
		\validation            &  1 &  1 &  0 &  0 &  0 \\
		\maturityStage         &  0 &  0 &  0 &  0 &  1 \\
		\sideEffect           &  0 &  0 &  0 &  1 &  1 \\
		\midrule
		{Total}    & 12 & 13 & 13 & 12 & 12 \\
		\bottomrule
	\end{tabular}
\end{table}

These findings partially align with Al Safwan \etal~\cite{al2022developers}. Their study, based on developers’ reported experiences (via surveys/interviews), found that \sideEffect and \alternative are among the most difficult rationale components to find, and that \alternative, \selectedAlternative, \constraints, and \maturityStage are among the least frequently recorded. Our artifact-based results confirm this pattern: \alternative and \selectedAlternative are rarely documented, \maturityStage is nearly absent, and \constraints do not appear at all. We also found \sideEffect to be extremely uncommon. Taken together, these results suggest that the components developers report as hard to find are indeed sparsely documented in commit-related artifacts.

The only sampled commit with no identifiable rationale component was \textit{JUnit4} commit \texttt{cec4a6b}~\cite{junit4_commit_cec4a6b}. This was a change affecting a single code file and adding only TODO-style comments, with an empty commit message and no associated discussion artifacts. 

We believe \goal, \need, and \alternative are more frequently found because they are the components most naturally expressed in developer communication, such as commit messages, pull request descriptions, and issue discussions. In contrast, components such as \constraints, \dependency, \maturityStage, and \sideEffect are often left implicit unless they become especially salient during implementation or review. The JUnit4 commit~\cite{junit4_commit_cec4a6b} with no identifiable rationale illustrates this pattern: because the change was minimal and accompanied by no descriptive message or discussion, even basic rationale components such as \goal were absent.



\RQ{ \textbf{\ref{rq:components} Findings}: Seven of nine rationale components were found in the software artifacts associated with the studied 63 commits. The most frequently documented components are the change's \goal, \need, and \alternative, found in the artifacts of 61, 34, and 9 commits, respectively. The remaining components,  \textit{Selected Alternative}, \textit{Validation}, \textit{Side Effects}, and \textit{Maturity Stage} are less frequently documented (found for 1-4 commits). Two components, \constraints and \dependency, were not found in associated artifacts. }

\subsubsection{\ref{rq:comp_prevalence}: Where are Rationale Components Documented Across Software Artifact Types?\\}

We examined how rationale is distributed across artifact types by analyzing the frequency with which artifacts contain rationale information overall and for individual rationale components, the presence of rationale in commit messages compared to other artifacts, and the proportion of content within individual artifacts (\eg a single \issue or \pr) that expresses rationale.

\begin{table}[t]
    \small
    \centering
    \caption{Rationale presence across artifact types and \\ rationale density in rationale-containing individual artifacts}
    \label{tab:individual_rationale}
    \setlength{\tabcolsep}{4pt}
    \begin{threeparttable}
        \begin{tabular}{l|c|c|c|c}
        \toprule
        \textbf{Artifacts} &
        \makecell{\textbf{Number of}\\\textbf{Artifacts ($A_T$)}} &
        \makecell{\textbf{Rationale}\\\textbf{Containing}\\\textbf{Artifacts ($A_R$)}$^*$} &
        \makecell{\textbf{Avg. Sentences }\\\textbf{from $A_R$}$^+$} &
        \makecell{\textbf{Avg. Proportion}\\\textbf{of Sentences}\\\textbf{from $A_R$}$^+$} \\
        \midrule
        \cc         & 68  & 3 (4.4\%)    & 5.0 (5)     & 60.0\% (60.0\%) \\
        \cm       & 63  & 61 (96.8\%)  & 1.7 (1)   & 88.0\% (100.0\%) \\
        \classJavadocs       & 57  & 3 (5.3\%)    & 7.0 (5)     & 48.9\% (40.0\%) \\
        \prs         & 49  & 42 (85.7\%)  & 5.5 (4)   & 60.6\% (53.6\%) \\
        \issues                 & 24  & 18 (75.0\%)  & 36.1 (21)   & 19.5\% (12.0\%) \\
        \methodJavadocs      & 16  & 6 (37.5\%)   & 4.2 (3)     & 53.3\% (37.5\%) \\
        \codeRev & 14  & 2 (14.3\%)   & 28.0 (2)     & 50.4\% (50.0\%) \\
        \midrule
        Total           & 291 & 135 (46.4\%) & 8.3 (2) & 66.7\% (75.0\%) \\
        \bottomrule
        \end{tabular}%
        \begin{tablenotes}[flushleft]
        \footnotesize
        \item[] 
        \textit{Note:} `*' indicates the proportional value of all artifacts is placed in parentheses; `+' indicates the median value of all rationale-containing artifacts is placed in parentheses.


        \end{tablenotes}
    \end{threeparttable}
\end{table}

\underline{Rationale Presence Across Artifacts.}
Overall, rationale components were identified in 135 of the 291 analyzed artifacts (46.4\%) --- see \Cref{tab:individual_rationale}. Rationale information is distributed across all artifact types, although its prevalence varies. \cm, \prs, and \issues are the most common sources of rationale, containing rationale in 61 of 63 (96.8\%), 42 of 49 (85.7\%), and 18 of 24 (75.0\%) artifacts, respectively. In contrast, rationale was identified in 6 of 16 \metJd (37.5\%), 2 of 14 \codeRev (14.3\%), 3 of 57 \classJavadocs (5.3\%), and 3 of 68 \cc (4.4\%). This distribution likely reflects the different purposes of these artifacts: discussion-oriented artifacts, such as \issues and \prs, are commonly used to capture the reasoning behind changes, whereas code-centric artifacts, such as \textit{Javadocs}, primarily document code behavior and implementation details. Below, we examine more closely the content of \cm, as it is a code change-centric artifact.

\begin{table}[t]
	\centering
	\caption{Number of rationale sentences found for each artifact (with column-wise proportions)}
	
	\label{tab:component-distribution-artifacts}
	\resizebox{\columnwidth}{!}{%
		\begin{tabular}{l|r|r|r|r|r|r|r|r}
			\toprule
			\textbf{Artifacts} & \multicolumn{1}{c|}{\textbf{\goal}} & \multicolumn{1}{c|}{\textbf{\need}} & \multicolumn{1}{c|}{\textbf{\alternative}} & \multicolumn{1}{c|}{\textbf{\textit{Sel. Altern.}}} & \multicolumn{1}{c|}{\textbf{\validation}} & \multicolumn{1}{c|}{\textbf{\sideEffect}} & \multicolumn{1}{c|}{\textbf{\textit{Maturity}}} & \multicolumn{1}{c}{\textbf{Total}} \\
			\midrule
			\classJavadocs & 3 (2.0\%) & 1 (1.2\%) &           &           &           &           &           & 4 (1.6\%) \\
			\cc            & 2 (1.4\%) & 1 (1.2\%) &           &           &           &           &           & 3 (1.2\%) \\
			\codeRev       &           &           & 3 (15.0\%) & 3 (50.0\%) &           & 1 (50.0\%) &           & 6 (2.4\%) \\
			\cm            & 68 (46.6\%) & 14 (16.3\%) &           &           &           &           &           & 80 (31.6\%) \\
			\issues        & 7 (4.8\%) & 39 (45.4\%) & 12 (60.0\%) & 1 (16.7\%) &           &           & 1 (100.0\%) & 60 (23.7\%) \\
			\metJd         & 5 (3.4\%) & 3 (3.5\%) &           &           &           &           &           & 8 (3.2\%) \\
			\prs           & 61 (41.8\%) & 28 (32.6\%) & 5 (25.0\%) & 2 (33.3\%) & 4 (100.0\%) & 1 (50.0\%) &           & 92 (36.4\%) \\
			\midrule
			\textbf{Total} & \textbf{146 (100.0\%)} & \textbf{86 (100.0\%)} & \textbf{20 (100.0\%)} & \textbf{6 (100.0\%)} & \textbf{4 (100.0\%)} & \textbf{2 (100.0\%)} & \textbf{1 (100.0\%)} & \textbf{253 (100.0\%)} \\
			\bottomrule 
	\end{tabular}}
   \footnotesize{\textit{Note:} For each artifact, the total number of rationale sentences may be smaller than the sum of the component counts because a single sentence can express multiple rationale components.}
\end{table}


\underline{Presence of Rationale Components in Artifacts.}
\Cref{tab:component-distribution-artifacts} shows that \cm, \issues, and \prs are the richest sources of rationale, together accounting for 91.7\% (232 of 253) of all rationale sentences. Information about what a code change accomplishes (\goal) is primarily documented in \cm (46.6\% of \goal sentences) and \prs (41.8\%). In contrast, information about the motivation behind code changes (\need) is most often recorded in \issues and \prs, with only 16.3\% of \need sentences appearing in \cm. The components \alternative and \selectedAlternative occur exclusively in \issues, \prs, and \codeRev, which also capture some of the less frequent components (\validation, \sideEffect, and \maturityStage).

\prs capture 36.4\% of all rationale sentences and cover nearly all (6 of 7) components, making them the most diverse artifact for documenting rationale. \issues and \codeRev follow as the second (5 of 7) and third (3 of 7) most diverse sources, respectively. Overall, these results indicate that developers and other stakeholders primarily rely on \prs, \cm, and \issues to document rationale across the 63 studied commits—a pattern consistently observed in all five software projects.

Interestingly, all artifacts except \codeRev include \goal and \need. \codeRev instead primarily contain information about \alternative, \selectedAlternative, and \sideEffect. This suggests that developers use \codeRev mainly to document implementation-level rationale, focusing on decision-making and potential consequences, rather than high-level goals or motivations.

\underline{Rationale Components in \cm vs. Other Artifacts.} We compare the presence of rationale components in \cm versus other sources, which may not be as readily accessible as commit messages for developers when understanding code changes. \Cref{tab:cm-vs-others} shows the number of commits where rationale components are found exclusively in commit messages, exclusively in artifacts different from commit messages, or in commit messages and other artifacts. 
As shown in the table, the \goal is captured in nearly all commit messages (a total of 60 of 63, \ie for 19 + 41 commits, regardless if the \goal appears or not in other artifacts). Only three commits do not have any \goal documented in the commit message. However, \need is far more often documented outside commit messages (24 commits) than inside them (10 commits). The remaining components are found exclusively in sources beyond commit messages.

\begin{table}[t]
    \small
	\centering
	\caption{Number of commits where rationale components are found in commit messages only, in other artifacts only, or in both commit messages and other artifacts}
	\label{tab:cm-vs-others}
    \small
		\begin{tabular}{l|c|c|c|c|c|c|c|c}
			\toprule
			\makecell{\textbf{Artifacts}} & \textbf{\goal} & \textbf{\need} & \textbf{\alternative} & \textbf{\textit{Sel. Alt.}} & \textbf{\validation} & \textbf{\sideEffect} & \textbf{\textit{Maturity}} & \textbf{\textit{Total}} \\
			\midrule
            
            \makecell{Only in\\commit msgs.}     & 19 (30.2\%) & 2 (3.2\%)   & 0 (0.0\%)   & 0 (0.0\%)  & 0 (0.0\%) & 0 (0.0\%) & 0 (0.0\%) & 15 (23.8\%)\\
            
			  \makecell{Only in\\other artifacts} & 1 (1.6\%)   & 24 (38.1\%) & 9 (14.3\%)  & 4 (6.3\%)  & 2 (3.2\%) & 2 (3.2\%) & 1 (1.6\%) & 1 (1.6\%) \\
			
			\makecell{In both\\artifact groups}      & 41 (65.1\%) & 8 (12.7\%)  & 0 (0.0\%)   & 0 (0.0\%)  & 0 (0.0\%) & 0 (0.0\%) & 0 (0.0\%) & 46 (73.0\%) \\
			\midrule
			\textbf{Total} & \textbf{61 (96.8\%)} & \textbf{34 (54.0\%)} & \textbf{9 (14.3\%)} & \textbf{4 (6.3\%)} & \textbf{2 (3.2\%)} & \textbf{2 (3.2\%)} & \textbf{1 (1.6\%)} & \textbf{62 (98.4\%)} \\
			\bottomrule
	\end{tabular}
    \footnotesize{\textbf{Note}: Component columns count commits where that component appears only in the indicated source. Total counts commits where all identified rationale appears only in that source}
\end{table}

In essence, commit messages primarily capture \goal, while the remaining six components are typically documented elsewhere. This means that developers inspecting only commits often miss information about \need and other components, and must search additional artifacts to retrieve it. This result motivates our work on automated cross-artifact identification and generation of rationale components.
\looseness=-1

\underline{Rationale Components in Individual Artifacts.} 
\Cref{tab:component-distribution-artifacts} reports the density of rationale within individual artifacts (\eg single \issues or \prs) and  reveals additional differences across artifact types. Among artifacts that contain rationale, \cm exhibit the highest concentration, with approximately 88\% of their sentences expressing one or more rationale components on average. This is unsurprising given that \cm are short (1.7 sentences on average) and are expected to communicate what a change accomplishes, resulting primarily in \goal-related rationale. In contrast, \issues contain rationale in only about 20\% of their sentences on average, indicating that much of the discussion focuses on topics other than the rationale of the associated code change. The remaining artifact types fall between these patterns, with roughly half of their sentences (49--61\% on average) conveying rationale, suggesting that when rationale is documented in these artifacts, it tends to constitute a substantial portion of their content.

\RQ{\textbf{\ref{rq:comp_prevalence} Findings}: \cm and \prs predominantly document the \goal of code changes, while \need is most frequently captured in \prs and \issues. \prs and \issues are the most diverse sources of rationale, documenting 5--7 rationale components. More generally, rationale is most commonly documented in \cm, \prs, and \issues, whereas \classJavadocs, \metJd, \codeRev, and \textit{Inline} \cc contain rationale less frequently. Most rationale components beyond \goal are documented in artifacts other than commit messages, highlighting the need for automated identification and synthesis of rationale across multiple sources.}

\section{\tool: Automated Extraction and Generation of Code Change Rationale}
\label{sec:approach}

Our qualitative study in \Cref{sec:rationale_study} showed that rationale information is fragmented across software artifacts associated with a commit, yet these are rich sources of fine-grained rationale components. While commit messages frequently document a change's \goal, other rationale components, such as \need and \alternative, are often found in issues, pull requests, code reviews, and code comments. These findings motivate automated support for identifying and synthesizing rationale information across multiple sources.

We introduce \tool, an LLM-powered approach for fine-grained \textbf{A}utomatic \textbf{R}ationale extraction and \textbf{G}eneration from m\textbf{U}lti-document \textbf{S}ources. Given a commit, \tool reconstructs its rationale by retrieving related software artifacts, identifying rationale content, and synthesizing concise rationale summaries.

As illustrated in \Cref{fig:approach}, \tool consists of three modules executed in sequence:
\begin{enumerate}
    \item The \textbf{Artifact Retriever} first collects artifacts associated with a commit (\eg commit messages, issues, pull requests, code reviews, and code comments) and parses their textual content into sentences;
    
    \item Then, the \textbf{Rationale Component Extractor} identifies sentences expressing the rationale components \goal, \need, and \alternative in the collected artifacts; and
    
    \item Finally, the \textbf{Rationale Component Generator} synthesizes the extracted rationale information into concise summaries for each rationale component.

\end{enumerate}
The following subsections describe each module in detail.

\begin{figure}[t]
	\centering \includegraphics[width=\textwidth]{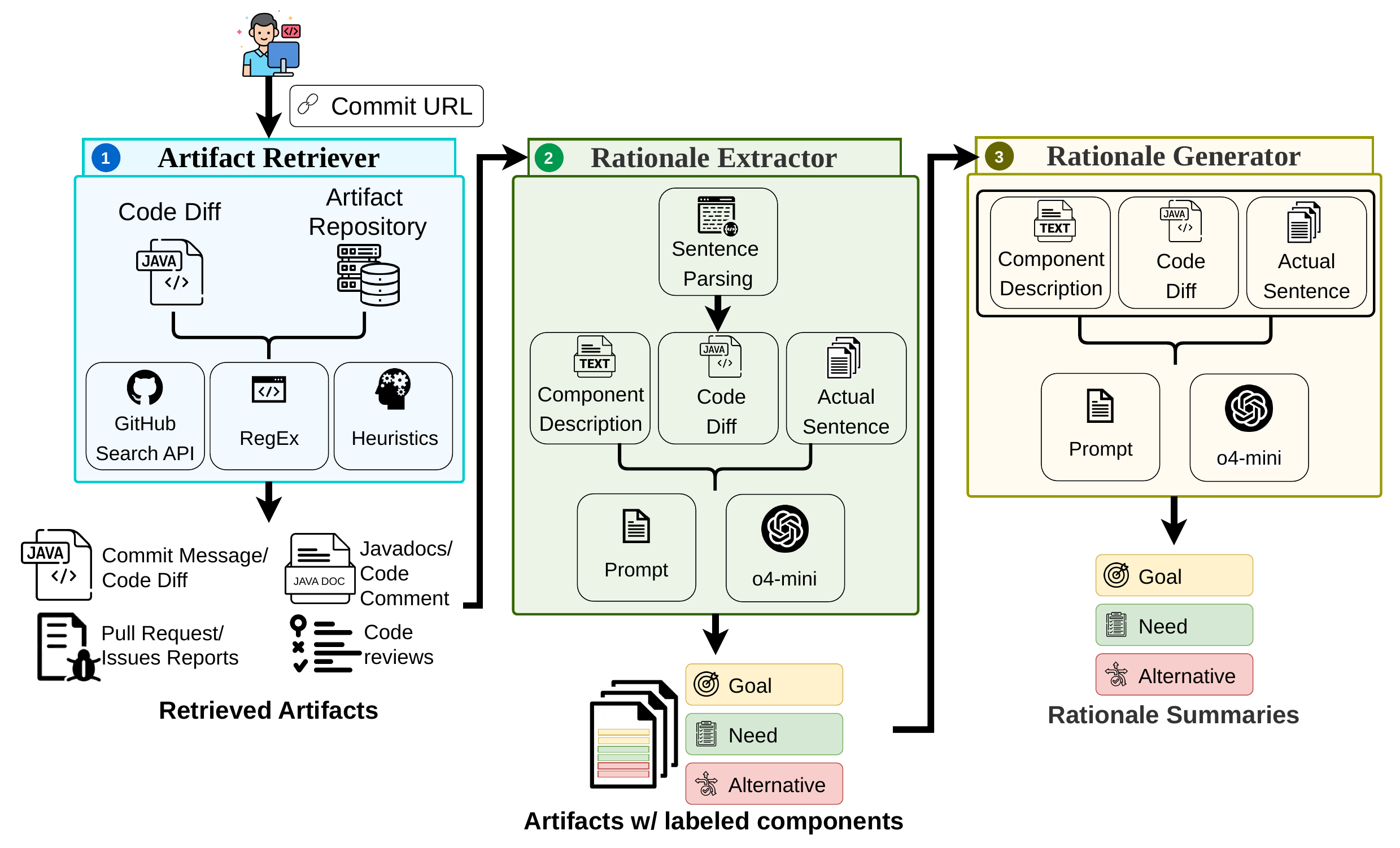}
	\caption{\tools Architecture}
	\label{fig:approach}
\end{figure}

\subsection{Artifact Retriever}

The goal of this module is to gather the software artifacts most likely to contain rationale information for a target commit.
This module takes as input a project commit (\ie the commit URL) and outputs a collection of associated artifacts, including their textual content parsed into sentences. 
To construct this collection, this module retrieves the commit message and diff, as well as linked issue reports, pull requests, and code reviews. As described in \Cref{subsec:artifact_collection}, artifacts are identified through a combination of regular expressions, GitHub API queries, and heuristic-based text matching, which detect artifact IDs in commit messages or commit references in the artifacts' text. The retriever also parses commit data to extract Javadocs and inline comments from modified classes and methods, using the \textit{comment-parser} library~\cite{commentparser}. Finally, all textual content is segmented into sentences using SpaCy’s English transformer pipeline (\texttt{en\_core\_web\_trf})~\cite{spacymodel}.

This retrieval step is essential because, as shown by our qualitative study (\Cref{sec:rationale_study}), rationale information is distributed across multiple artifacts. By collecting information from the artifacts most likely associated with a commit, \tool provides subsequent phases with a comprehensive set of potentially rationale content.

Because the retriever relies on explicit references and heuristic matching, it may retrieve artifacts that are only incidentally related to the target commit. We therefore analyze the downstream effect of such retrieval noise in Section~\ref{sec:prompt_dev}.

\subsection{Rationale Component Extractor}

The goal of this module is to identify sentences that express rationale components.
The module takes as input the sentences produced by the \textit{Artifact Retriever} and outputs those sentences annotated with one or more rationale component labels.
\tool focuses on the rationale components \goal, \need, and \alternative because our empirical study found them to be the most frequently documented across software artifacts.

To perform the labeling, this module prompts the LLM to reason about the meaning of each sentence and assign one or more component labels, or \textit{none} if the sentence does not express rationale. The prompt follows a task decomposition approach~\cite{khot2022arXiv}, incorporating component definitions, sentences grouped by specific artifact, and few-shot exemplars with ground-truth labels and explanations. This design guides the LLM to interpret sentence semantics and apply fine-grained classification. \Cref{sec:prompt_dev} provides details of our data-driven methodology, which experimented with different prompting strategies to determine the most accurate approach to guide the LLM for this task.



\subsection{Rationale Component Generator}

The goal of this module is to generate concise and human-readable summaries of the identified rationale components.
This module takes the sentences labeled by the \textit{Rationale Component Extractor} and outputs a summary for each rationale component. 
To generate these summaries, this module prompts the LLM to reason over information collected from multiple artifacts, connect related pieces of information, and synthesize them into coherent descriptions that capture the essence of each component.
Similar to the extractor, the prompt follows a task-decomposition approach~\cite{khot2022arXiv}, incorporating component definitions, labeled sentences grouped by artifact, and few-shot exemplars with ground-truth labels. As detailed in \Cref{sec:prompt_dev}, we adopted a data-driven methodology, experimenting with different prompting strategies and contexts to guide the LLM toward producing accurate and correct rationale summaries.


\subsection{\tool's Design Rationale}

\tool adopts a decomposition-based design in which rationale identification and rationale generation are performed as separate LLM tasks. We chose this design for three reasons.
First, rationale information is often distributed across multiple artifacts and interleaved with large amounts of irrelevant content. Processing all retrieved artifacts in a single prompt may therefore reduce the model's ability to focus on rationale-relevant information, particularly as input length increases~\cite{Levy2024,ling2025longreason,kuratov2024babilong,zhang2025attention}.
Second, prior work has shown that decomposing complex reasoning tasks into smaller sub-tasks can improve LLM performance~\cite{khot2022arXiv}. Similar strategies have proven effective in software-engineering tasks such as automated program repair~\cite{YinISSTA2024} and code generation~\cite{LeiACL2025}.
Third, decomposition improves interpretability and debugging. The intermediate rationale-identification output can be independently inspected and evaluated, making it easier to understand and diagnose errors. This design also enables future extensions, such as alternative rationale-generation strategies tailored to specific developers or software-engineering tasks.

\subsection{LLM Selection}

\tool uses \OFourMini as its underlying language model. We selected \OFourMini because it was one of the strongest reasoning-oriented models available when this project began (mid-2025) and had demonstrated to be highly effective across software-engineering tasks, including fault localization~\cite{YeoarXiv2025} and code generation~\cite{Li2025,RontogiannisAAAI2026}. We therefore used this model throughout prompt development, repeated executions, and user evaluation.
To assess whether \tool's effectiveness depends on the underlying LLM, we conducted a cross-model analysis using more recent models, including \GPTFiveTwo and \GeminiThreeFlash. The results of this analysis are presented in \Cref{argus:methodology}.


\section{\tools Development and Evaluation}
\label{sec:prompt_dev}

We adopted a data-driven methodology to develop prompt templates for \tool's two main modules: rationale component identification and generation. Our goal was to determine the most effective prompting strategies to guide \OFourMini in accomplishing both tasks. To this end, we used a \textit{development set} consisting of a subset of the 63 commits and their corresponding ground-truth data (\ie manually annotated artifact sentences). The best-performing prompting strategies were then incorporated into \tool and evaluated against baseline prompting approaches on a held-out \textit{evaluation set} comprising the remaining commits and their associated ground-truth data. This evaluation allowed us to assess the accuracy of rationale component identification and the correctness of the rationale summaries generated by \tool.

We experimented with five prompt templates that implement well-known prompting strategies across \tools two modules: rationale component identification (three templates) and generation (two templates). The strategies include zero-shot, few-shot, and reasoning-based few-shot prompting, all combined with task decomposition prompting~\cite{khot2022arXiv}. 

Importantly, in all experiments of this section, we included both artifacts relevant and irrelevant to the commits. This choice reflects real-world deployment of \tool, where the artifact retriever is expected to return both types. 
\looseness=-1


With this in mind, we addressed the following research questions (RQs):

\begin{enumerate}[start=3, label=\textbf{RQ$_\arabic*$:}, ref=\textbf{RQ$_\arabic*$}, itemindent=0cm,leftmargin=1cm]
	\item \label{rq:prompt-dev-identification}{\textit{What prompting strategies are most effective to identify rationale components in software artifacts?}}
	\item \label{rq:prompt-dev-generation}{\textit{What prompting strategies are most effective to generate rationale component summaries?}}
	\item  \label{rq:tool-eval-ident}{\textit{How accurate is \tool's rationale identification module?}}
	\item  \label{rq:tool-eval-gen}{\textit{How accurate are \tool's generated rationale summaries?}}
\end{enumerate}

\ref{rq:prompt-dev-identification} and \ref{rq:prompt-dev-generation} focus on developing \tool's prompts by comparing different prompting strategies for rationale identification and generation on the development set. The best-performing strategies are then integrated into \tool. \ref{rq:tool-eval-ident} and \ref{rq:tool-eval-gen} evaluate the resulting system on the held-out evaluation set, measuring the accuracy of rationale identification and the correctness of the generated rationale summaries.

\Cref{subsec:dev_data} describes the process we followed to construct the dataset to answer our RQs. \Cref{sec:prompt_dev_methodology,sec:prompt_dev_results} describe the methodology and results of our data-driven approach for \tools prompt development (\ref{rq:prompt-dev-identification} and \ref{rq:prompt-dev-generation}), while \Cref{argus:methodology,subsec:argus_eval_results} present \tool's evaluation methodology and results (\ref{rq:tool-eval-ident} and \ref{rq:tool-eval-gen}).


\subsection{Dataset Construction}
\label{subsec:dev_data}

Prompt development and evaluation relied on two types of ground truth data:
(1) manually labeled artifact sentences tagged with the rationale components they convey (for identification), and
(2) manually written descriptions of the three rationale components \tool targets: \goal, \need, and \alternative (for generation).

The annotated sentences were obtained from the dataset built in our qualitative study reported in \Cref{sec:rationale_study}. The ground truth component descriptions were created by the same two annotators who participated in that study. One annotator synthesized clear and concise summaries for each commit by reviewing the commit diff, message, and artifact sentences annotated with \goal, \need, and \alternative. The second annotator then reviewed the summaries for clarity, correctness,  and accuracy compared to the labeled sentences. Disagreements were resolved by discussion until consensus was reached. This iterative, multi-coder approach was repeated in six rounds of 10–13 commits each, producing concise and reliable summaries while mitigating subjectivity and errors.

We applied this methodology to all 63 commits collected in the qualitative study, and divided them into two subsets: 

\begin{itemize}
    
    \item The \textit{development set}, consisting of 13 commits, is used to answer \ref{rq:prompt-dev-identification} and \ref{rq:prompt-dev-generation}. This dataset is exclusively used for prompt formulation and refinement and was randomly selected to include a diverse set of commits across projects, rationale components, and artifact types. It covers all five projects and includes 13 commits with all rationale components represented (13 with \goal, 9 with \need, and 4 with \alternative), as well as 76 artifacts: 18 \prs, 18 \cc, 13 \cm, 12 \classJavadocs, 7 \issues, 6 \codeRev, and 2 \methodJavadocs.
    
    \item The \textit{evaluation set}, containing the remaining 50 commits, is used to answer \ref{rq:tool-eval-ident} and \ref{rq:tool-eval-gen}. This dataset contains 48 commits with \goal, 25 with \need, and 5 with \alternative, and comprises 263 artifacts: 59 \prs, 50 \cc, 50 \cm, 45 \classJavadocs, 30 \issues, 15 \codeRev, and 14 \methodJavadocs.
\end{itemize}

About 12\% of sentences in the development set and 17\% in the evaluation set came from irrelevant artifacts, helping us refine and assess prompts under realistic settings. Together, these two sets provide coverage across all five projects, multiple rationale components, and diverse artifact sources, allowing us to develop and assess \tool across varied rationale-extraction scenarios.

\begin{figure}[t]
	\centering
	\begin{subfigure}[t]{0.49\textwidth}
		\centering
		\vspace{0pt}
		\includegraphics[height=5.5cm]
		{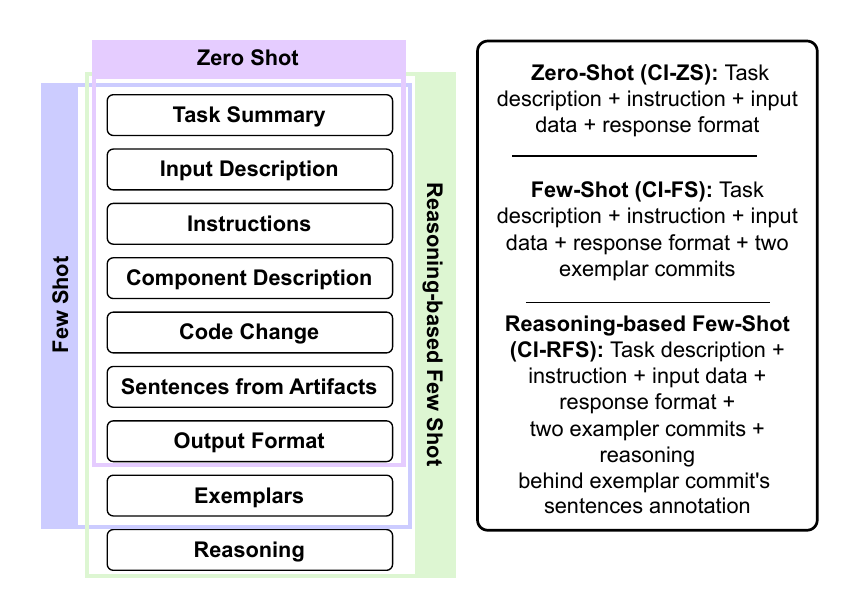}
		\caption{Rationale component identification}
		\label{fig:ident_prompts}
	\end{subfigure}
	\hfill
	\begin{subfigure}[t]{0.49\textwidth}
		\centering
		\vspace{0pt}
		\includegraphics[height=5.2cm]
		{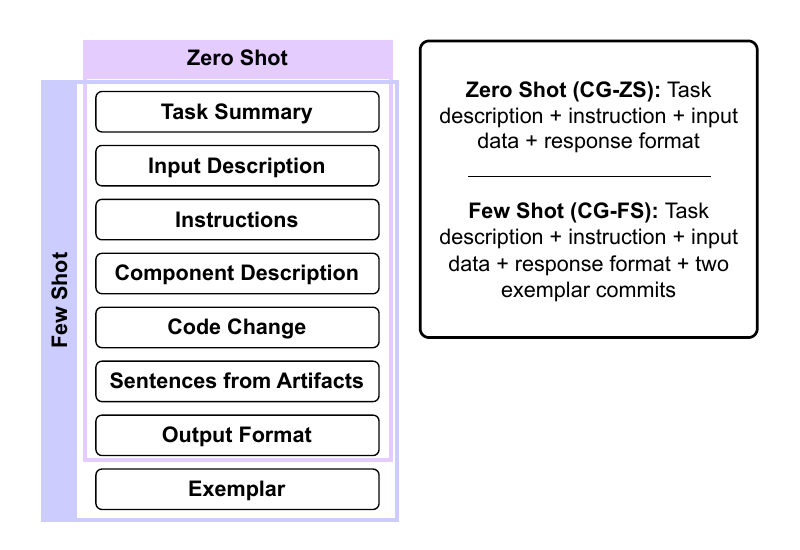}
		\par\vspace{0.3cm}
		\caption{Rationale component generation}
		\label{fig:gen_prompts}
	\end{subfigure}

	\caption{Structure of the developed prompts for both of \tool's main tasks.}
	\label{fig:prompts}
\end{figure}


\subsection{Prompt Development Methodology}
\label{sec:prompt_dev_methodology}

Our prompt development process explored three strategies: zero-shot, few-shot, and reasoning-based few-shot prompting, all combined with task decomposition prompting~\cite{khot2022arXiv}. For rationale identification, we created three templates (one per strategy), and for rationale generation, we designed two templates (zero-shot and few-shot).  This section describes how the prompts were developed, the prompting strategies that were evaluated, and the methodology used to assess their effectiveness.

\newcolumntype{M}[1]{>{\raggedright\arraybackslash}m{#1}}
\begin{table}[t]
\centering
\caption{Prompt snippets used for rationale component identification and generation. The component description element is common for both tasks}
\label{tab:prompt_snippets}
\small

\begin{tabular}{M{0.13\columnwidth}|M{0.40\columnwidth}|M{0.40\columnwidth}}
\hline

\centering\textbf{Prompt Element} &
\centering\textbf{Snippet for Component Identification} &
\centering\textbf{Snippet for Component Generation}
\tabularnewline
\hline

\textbf{Task Summary}
&
You are an experienced Java programmer on the \texttt{<project\_name>} project. Your task is to analyze a set of sentences, identify whether they express the rationale behind a target code commit ...
&
You are an experienced Java programmer on the
\texttt{<project\_name>} project. Your task is to generate concise and accurate rationale components that explain why a specific code change was made.
\tabularnewline
\hline

\textbf{Input Description}
&
You will be provided with:\newline
-- A list of rationale components, each one describing a different aspect that explains why the code was changed the way it was.\newline
-- The diff of the target commit, which shows how one or more code files were changed.
&
You will be provided with:\newline
-- A list of rationale component types...\newline
-- A code diff...\newline
-- A set of annotated sentences related to the commit... labeled with rationale components... extracted from various software artifacts...
\tabularnewline
\hline

\textbf{Instructions}
&
-- Analyze the content of the sentences, identify the rationale components that they describe, and annotate them with those components.\newline
-- Annotate the sentences ... using one or more of the ... rationale components only if the sentence contains information about those components... &
-- Analyze the code diff to understand the technical nature of the code change.\newline
-- Review the annotated sentences to find relevant information about the rationale behind the change...  
-- For each rationale component, identify sentences labeled with that component and discard irrelevant...
\tabularnewline
\hline

\textbf{Component Description}
&
GOAL: Describes what changes have been made by the developer in the code change... indicating what the developer aims to achieve with the code change ... \newline NEED: Describes the specific reason, motivation, or benefit of the code change. This indicates why the developer wants to make the code change. &
GOAL: Describes what changes have been made by the developer in the code change... indicating what the developer aims to achieve with the code change ... \newline NEED: Describes the specific reason, motivation, or benefit of the code change. This indicates why the developer wants to make the code change.
\tabularnewline
\hline

\textbf{Artifact Sentences}
&
The following sentences come from the commit message;\newline 
Id: <Sentence ID>;\newline
Source: COMMIT\_MESSAGE;\newline 
Sentence: Upate TestName to make the name field volatile.    &
The following sentences come from the commit message;\newline  Id: <Sentence ID>;\newline 
Source: COMMIT\_MESSAGE;\newline 
Sentence: Unnecessary array in varargs in AnnotatedBuildrer...\newline 
Labels: GOAL, NEED 
\tabularnewline
\hline

\textbf{Output Format}
&
\# Response Format (CSV):\newline 
sentence\_id, labels: <ID1>, "<LABEL1>, <LABEL2>..." &
\# Response Format:\newline
GOAL: <brief\_description>;\newline
NEED: <brief\_description>;\newline
ALTERNATIVES: <brief\_description> 
\tabularnewline
\hline

\textbf{Reasoning}
&
\textit{Sentence}: ``Adds {@code matcher} to the list of requirements for the cause of any thrown exception.''
\textit{Label}: ``GOAL''\newline
\textit{Reason}: ``This sentence describes what the expectCause() function does. The method adds a matcher requirement for the cause, which aligns with the commit’s goal, i.e., match causes easily using a Matcher instead of an explicit Throwable. Hence, it was labeled as GOAL. It doesn’t describe why the code change was made, i.e., the reason, motivation, or benefit of the code change. Moreover, it does not describe any other potential solution that could be implemented instead of the current implementation. Hence, it is neither labeled as NEED nor ALTERNATIVES.''

&
(Not defined)
\tabularnewline
\hline

\end{tabular}
\end{table}

\subsubsection{Prompt Design\\}

Two of the 13 commits from the development set were selected as exemplars for the few-shot templates chosen by the annotators. 
We intentionally limited the number of exemplars to two commits to provide sufficient guidance while avoiding unnecessarily long prompts. Since each exemplar includes multiple artifact sentences, rationale labels, and explanations, additional exemplars would substantially increase the amount of context the model must process, potentially distracting it from reasoning about the target commit.
These exemplar commits were drawn from two different projects, and collectively contain 52 sentences from 10 software artifacts and 6 different rationale components, providing diverse coverage of rationale components across artifacts.

Each prompt was initialized from a common template containing standard elements, such as a task description, instructions, exemplars, input data, and the expected output format, following established prompt-engineering guidelines~\cite{promptingguide}. To illustrate the prompt design, \Cref{tab:prompt_snippets} presents representative excerpts from the prompts used for rationale identification and generation. The complete prompt templates are available in our replication package~\cite{package}.

We iteratively refined the prompts using the development set of 11 commits. After each round of prompt execution, we analyzed the model's outputs, identified common error patterns, and revised the prompts accordingly. For rationale identification, errors were categorized as false positives and false negatives by comparing the model's predictions against the ground-truth annotations. For rationale generation, errors were categorized according to whether the generated summaries omitted relevant rationale information or introduced information not present in the ground truth. Two annotators, both authors of this paper, reviewed these errors and refined the prompts through meta-prompting and consensus-based discussion, deriving additional decision rules and prompt adjustments.

\subsubsection{Prompt Variants for Rationale Identification\\}
\label{subsec:prompt_dev_identification}

Rationale component identification (CI) is formulated as a multi-label classification task where artifact sentences are tagged with one or more components (\goal, \need, \alternative) or \textit{none}. We explored three prompting strategies, all using task decomposition~\cite{khot2022arXiv} (see \Cref{fig:ident_prompts}):

\begin{enumerate}
    \item \textbf{Zero-Shot (CI-ZS)}: This prompt includes a task summary, an input data description (the definition of the rationale components, the commit diff, and the sentences grouped by individual artifact), specific task instructions, the inputs, and the output format.
  
    \item \textbf{Few-Shot (CI-FS)}: the prompt extended CI-ZS with two exemplar commits containing labeled and a few unlabeled artifact sentences.
    
    \item \textbf{Few-Shot-Reasoning (CI-RFS)}: the prompt further extended the CI-FS prompt with explanations for why the exemplar sentences were tagged with specific rationale components. We developed this prompt in two steps. First, we used explanations behind each sentence of the example commits to better guide the model. Then, we analyzed the failed cases in that prompt and derived a set of rules to improve the prompt.
\end{enumerate}

\subsubsection{Prompt Variants for Rationale Generation\\}
\label{sec:prompt_dev_rat_gen}

For rationale generation (CG), \tool produces concise rationale descriptions, given sentences previously labeled as \goal, \need, and/or \alternative. We designed two templates using task decomposition  prompting~\cite{khot2022arXiv} (see \Cref{fig:gen_prompts}):

\begin{enumerate}
    \item \textbf{Zero-Shot (CG-ZS)}: the prompt included the task summary, component definitions, specific task instructions, the inputs, and the output format.  The inputs included the commit diff and the labeled sentences grouped by artifact.
    \item \textbf{Few-Shot (CG-FS)}: The prompt extends the CG-ZS prompt with two exemplar commits, including their labeled sentences and ground truth rationale summaries.
\end{enumerate}
		
We did not employ reasoning-based few-shot prompting for rationale generation because this task operates on rationale sentences that have already been identified and categorized as \goal, \need, or \alternative. Unlike rationale identification, which requires the model to distinguish rationale from non-rationale content and assign fine-grained labels, rationale generation focuses on synthesizing and summarizing already-selected information. Consequently, we limited our investigation to zero-shot and few-shot prompting strategies for this task.


\subsubsection{Evaluation Methodology of Rationale Identification\\}
\label{subsec:prompt_dev_identification}

We evaluated rationale identification performance by comparing predicted labels against the ground-truth annotations. For each rationale component, we computed:

\begin{itemize}
\item \textbf{True Positives (TP):} sentences correctly labeled by the model as expressing the rationale component;
\item \textbf{False Positives (FP):} sentences labeled by the model as expressing the rationale component yet the ground truth does not assign that label;
\item \textbf{True Negatives (TN):} sentences correctly identified as not expressing the rationale component; and
\item \textbf{False Negatives (FN):} sentences that express the rationale component according to the ground truth but were not labeled as such by the model.
\end{itemize}

Based on these values, we calculated precision, recall, and F2-score:
\[ \pmb{Precision} = \frac{TP}{TP + FP}\qquad 
\pmb{Recall} = \frac{TP}{TP + FN}\qquad 
\pmb{F2} = \frac{(1+2^2) \cdot Precision \cdot Recall}{(2^2 \cdot Precision) + Recall} \]

Precision measures the proportion of predicted rationale labels that are correct, while recall measures the proportion of ground-truth rationale labels that are successfully identified. Overall metrics were computed by aggregating TP, FP, TN, and FN across all rationale components. We use the F2-score as the primary measure because rationale identification serves as a candidate-retrieval stage for subsequent generation. A false negative doesn't include relevant rationale information from the pipeline, which prevents the generation module from using it. In contrast, a false positive retains the relevant rationale information and introduces additional noise that can be reviewed and disregarded later. We therefore prioritize recall while still accounting for precision, and report precision, recall, and F2 score separately.

To account for the non-deterministic behavior of the LLM, we executed each prompt three times, measured its performance using the metrics described above, and assessed the consistency of the resulting predictions across runs using Krippendorff's $\alpha$~\cite{Krippendorff2018}. We also included 44 sentences from 4 irrelevant artifacts for 3 commits in the development set to assess the effect of these irrelevant artifacts in our approach. More details are provided in \Cref{sec:prompt_dev_results}.

\subsubsection{Evaluation Methodology of Rationale Generation\\}
\label{sec:prompt_dev_rat_gen}

To evaluate summary quality, two researchers independently compared generated outputs with ground truth summaries at a semantic level, focusing on two dimensions:
\begin{itemize}
    \item \textbf{Information Coverage (IC)}: the extent to which ground-truth rationale content is captured by the generated summary, where higher coverage is more desirable.
    
    \item \textbf{Extra Information (EI)}: the extent to which a generated summary includes information beyond the ground-truth rationale, where lower amounts of extra information are more desirable.
    
\end{itemize}

Both dimensions were rated on 5-point Likert scales, and \textit{throughout this paper both scales are oriented so that higher is better}. For Information Coverage (IC), 5 means every piece of ground-truth rationale is present in the generated summary and 1 means none of it is. For Extra Information (EI), 5 means the summary introduces no content beyond the ground truth and 1 means essentially all of its content is unsupported. Raters recorded EI on the natural ``amount of extra information'' scale and we inverted it once, at data-entry time, so that every EI value reported below follows the higher-is-better convention.

Annotators evaluated rationale summaries at the semantic level rather than based on lexical similarity. For each rationale component, they first decomposed the ground-truth rationale into its distinct pieces of information and then assessed the extent to which those pieces were captured by the generated summary. For example, consider the ground-truth \need: \textit{``The protocol register registers a protocol, even when the register attribute is set to false. Besides, the log message is incorrect.''} This rationale contains two distinct pieces of information: (1) the protocol is registered even when registration is disabled, and (2) the log message is incorrect. A generated rationale would receive a high Information Coverage score (4--5) if it captured both pieces of information, a moderate score (2--3) if it captured only one of them or omitted important details, and a low score (1) if it did not capture either. Within these ranges, annotators assigned higher scores when the generated rationale captured the information more completely and accurately. 

We used explicit guidelines to conduct this evaluation process. For example, assigning lowest score for both IC (Information Coverage) and EI (Extra Info) if a rationale component was missing from either the generated output or the ground truth. If a ground-truth rationale component was not generated, we assigned an IC score of 1 because none of the information associated with that component was generated, and we assigned an EI score of 5 because no unsupported information was introduced to compare with ground truth. Conversely, if the model generated a rationale component for which no corresponding component was present in the ground truth, we assigned an IC score of 1 because the generated content did not cover any ground-truth rationale information, and an EI score of 1 because all of the generated information was considered extra.

Extra Information (EI) was evaluated independently from Information Coverage (IC). Annotators assessed whether the generated summary introduced information that was not supported by the ground-truth rationale. Additional details were considered acceptable when they elaborated on or rephrased information already present in the ground truth. In contrast, unsupported claims, interpretations, or facts not grounded in the reference rationale were penalized through lower EI scores. 
This separate evaluation of IC and EI allowed summaries to receive high coverage scores while still being penalized for introducing unsupported information, or vice versa. Disagreements between the two annotators were resolved through discussion until consensus was reached.

\underline{Evaluation Metrics.} We report the \textbf{average Information Coverage (IC) and Extra Information (EI) ratings} across evaluated commits.
To obtain an overall measure of rationale generation performance that accounts for IC and EI, we computed the \textbf{F2-score}~\cite{Rijsbergen1979} using the average IC and EI ratings. We use the F2-score because it weights information coverage twice as heavily as extra information. We consider omissions of important rationale information more detrimental than the inclusion of some additional information, as incomplete summaries may fail to communicate the goal, motivations, or alternatives underlying the change.

{{\underline{Inter-rater Agreement}.}}
Overall inter-rater agreement, measured using \textit{weighted Cohen’s Kappa ($\kappa$)}~\cite{cohenweighted1968}, was 0.88, indicating high annotator reliability. We used this index because the Likert scales are ordinal, and disagreements between distant scores (\eg 1 vs. 5) are more severe than those between adjacent scores (\eg 3 vs. 4), which this index captures. 
Disagreements were mostly due to subjective interpretation of the summaries, which were resolved through annotators' discussion and consensus.

{\underline{Prompt Execution and LLM Consistency.}}
We executed each generation prompt using as input the sentences labeled by the best-performing rationale identification strategy, namely reasoning-based few-shot prompting (see \Cref{subsec:prompt-dev-identification-results}). This setup allowed us to evaluate rationale generation under realistic conditions, where the input may contain both correctly identified rationale sentences and sentences incorrectly labeled as rationale, including those originating from relevant and irrelevant artifacts. It also captures cases in which the rationale identification step fails to label sentences that express rationale, thus propagating missing information to the generation stage.

Although each prompt was executed three times, we assessed the consistency of the generated summaries across runs. To this end, we measured semantic similarity using embedding-based cosine similarity computed with OpenAI's \texttt{text-embedding-3-large} model~\cite{textembedding3large}. For each commit, we formed the three possible pairs of summaries generated across the three executions and computed the cosine similarity for each pair. We repeated this process for all commits and for both the CG-ZS and CG-FS prompts, yielding a total of 66 similarity scores (2 prompts $\times$ 11 commits $\times$ 3 summary pairs).

Across these 66 pairs, the average cosine similarity was 0.92, with a median of 0.94 and a first quartile of 0.90, indicating that the generated summaries were highly consistent across executions despite potential variations in wording. Similar results were observed for each prompt individually; the complete results are available in our replication package~\cite{package}. Given this high degree of consistency, we evaluated only the summaries generated in the first execution using the human-evaluation methodology and metrics described above.

\subsection{Prompt Development Results}
\label{sec:prompt_dev_results}


\subsubsection{\ref{rq:prompt-dev-identification}: Prompt Development Results for Rationale Identification\\}
\label{subsec:prompt-dev-identification-results}

We executed all three identification prompts three times using \OFourMini~\cite{openaio4mini}. To assess the consistency of predictions, we calculated Krippendorff's $\alpha$~\cite{Krippendorff2018} to compute the agreement of the labeling produced by the model across the three runs. We observed moderate-to-high consistency across runs, with Krippendorff's $\alpha$ ranging from 0.64 to 0.88 across rationale components: $\alpha_{\text{alternative}} = 0.64$, $\alpha_{\text{need}} = 0.72$, and $\alpha_{\text{goal}} = 0.88$. To address this variation, we adopted a majority-voting strategy to decide the final label for a sentence: a sentence is assigned a label set $X$ (\eg \goal and \need) only if the model produced the same outcome in at least two of the three runs. The intuition is that agreement across multiple executions increases confidence in the correctness of the prediction. Experiments on the development set (available in our replication package~\cite{package}) confirmed that the voting approach outperformed the average performance of the individual runs, yielding relative improvements of 7\% in precision, 3\% in recall, and 4\% in F2-Score. Based on these results, we implemented the voting approach for all three rationale identification prompts described in \Cref{subsec:prompt_dev_identification} and report their performance in this section.

\begin{table}[t]
	\centering
	\caption{Prompt development results for rationale identification (CI)}
	\begin{subtable}[b]{\textwidth}
    \centering
		\caption{Results by component and prompt}
		\label{tab:rat-iden-dev}
        \small
			\begin{tabular}{c|c|rrrr|ccc}
				\toprule
				\textbf{Component} & \textbf{Prompt} & \textbf{TP} & \textbf{FP} & \textbf{TN} & \textbf{FN} & \textbf{Precision} & \textbf{Recall} & \textbf{F2} \\ \midrule
				                     & Zero-Shot (CI-ZS)            & 28 & 5  & 668 & 3 & 84.8\%  & 90.3\%  & 89.2\%  \\
				\goal                & Few-Shot (CI-FS)             & 27 & 4  & 669 & 4 & 87.1\%  & 87.1\%  & 87.1\%  \\
				                     & \textbf{Few-Shot-Reasoning (CI-RFS)} & \textbf{30} & \textbf{1} & \textbf{672} & \textbf{1} & \textbf{96.8\%} & \textbf{96.8\%} & \textbf{96.8\%} \\ \midrule

				                     & Zero-Shot (CI-ZS)            & 19 & 15 & 665 & 5 & 55.9\%  & 79.2\%  & 73.1\%  \\
				\need                & Few-Shot (CI-FS)             & 21 & 17 & 663 & 3 & 55.3\%  & 87.5\%  & 78.4\%  \\
				                     & \textbf{Few-Shot-Reasoning (CI-RFS)} & \textbf{22} & \textbf{15} & \textbf{665} & \textbf{2} & \textbf{59.5\%} & \textbf{91.7\%} & \textbf{82.7\%} \\ \midrule

				                     & Zero-Shot (CI-ZS)            & 7 & 13 & 683 & 1 & 35.0\%  & 87.5\%  & 67.3\%  \\
				\alternative         & Few-Shot (CI-FS)             & 8 & 13 & 683 & 0 & 38.1\%  & 100.0\% & 75.5\%  \\
				                     & \textbf{Few-Shot-Reasoning (CI-RFS)} & \textbf{7} & \textbf{7} & \textbf{689} & \textbf{1} & \textbf{50.0\%} & \textbf{87.5\%} & \textbf{76.1\%} \\ \midrule

				                     & Zero-Shot (CI-ZS)            & 54 & 33 & 2016 & 9 & 62.1\%  & 85.7\%  & 79.6\%  \\
				Overall              & Few-Shot (CI-FS)             & 56 & 34 & 2015 & 7 & 62.2\%  & 88.9\%  & 81.9\%  \\
				                     & \textbf{Few-Shot-Reasoning (CI-RFS)} & \textbf{59} & \textbf{23} & \textbf{2026} & \textbf{4} & \textbf{72.0\%} & \textbf{93.7\%} & \textbf{88.3\%} \\ \bottomrule
			\end{tabular}%
	\end{subtable}
	\begin{subtable}[b]{\textwidth}
		\centering
        \small
        \vspace{0.3cm}
		\caption{Relative Improvement (RI) between prompts}
		\label{tab:rat-iden-rel-inc}
        \small
			\begin{tabular}{c|c|c|c|c}
				\toprule
				\textbf{Comparison} & \textbf{Component}  & \textbf{Precision}       & \textbf{Recall}          & \textbf{F2}                              \\ \midrule
				\multirow{4}{*}{\makecell{Few-Shot (CI-FS)\\vs\\Zero-Shot (CI-ZS)}}      &  \goal       &  3\%          & -4\% & -2\% \\
                                                     &	\need       & -1\%          & 11\% & 7\% \\
                                        			&	\alternative& 9\%           & 14\% & 12\% \\
                                                    \cline{2-5}
				                                    &    \textbf{Overall}     & \textbf{0\% } & \textbf{4\%}  & \textbf{3\%} \\ \midrule
				\multirow{4}{*}{\makecell{Few-Shot-Reasoning (CI-RFS)\\vs\\Few-Shot (CI-FS)}}    &   \goal &  11\% & 11\% & 11\% \\
				&   \need           & 8\%   & 5\%    & 6\% \\
				&   \alternative    & 31\%  & -13\%  & 1\% \\
                \cline{2-5}
				&   \textbf{Overall}         & \textbf{16\%}     & \textbf{5\%}      & \textbf{8\%} \\ \bottomrule
			\end{tabular}%
	\end{subtable}
\end{table}

\Cref{tab:rat-iden-dev} reports the performance of the three prompting strategies across components. Overall, reasoning-based few-shot (CI-RFS) achieves the highest accuracy, with 72.0\% precision, 93.7\% recall, and 88.3\% F2-score. It substantially outperforms both the few-shot (CI-FS) and zero-shot (CI-ZS) strategies, which rank second and third, respectively. In practical terms, this means that few-shot-reasoning (CI-RFS) correctly identifies about nine out of ten rationale sentences, and that roughly seven out of ten sentences it identifies as rationale indeed convey rationale. These results indicate that providing exemplars with explanations enables the LLM to better learn how to recognize rationale. In fact, when examining the relative improvement (RI) between strategies (\Cref{tab:rat-iden-rel-inc}), we observe that the presence of explanations for the exemplars is the key factor driving few-shot-reasoning (CI-RFS)’s superior performance, yielding gains over CI-FS of 16\% in precision, 5\% in recall, and 8\% in F2-score (compared to few-shot (CI-FS) vs. zero shot (CI-ZS) improvements of 0\%, 4\%, and 3\%, respectively).

Examining the performance achieved for each rationale component in \Cref{tab:rat-iden-dev}, we observe that \goal\ is the ``\textit{easiest}'' to identify, with Few-Shot-Reasoning (CI-RFS) achieving very high precision and recall (96\%+). For \need\ and \alternative, Few-Shot-Reasoning (CI-RFS) attains high recall but lower precision (59.5\% and 50\%, respectively). As rationale identification is an intermediate retrieval step for rationale generation, we prioritized recall during prompt development. False negatives are especially harmful because missed rationale sentences cannot be used by the generator in the next step. False positives, however, can introduce unsupported information into the rationale generation input and the generation step, or a developer can still filter out or ignore that unsupported information. Therefore, we do not treat low precision as harmless as missing rationale sentences. Besides, we evaluate the downstream effect of FPs indirectly through the rationale generation results, where the Extra Information (EI) metric captures unsupported content in the generated summaries (see \Cref{sec:res-rat-gen}). From this perspective, we consider CI-RFS’s performance acceptable for the task. Notably, irrelevant artifacts did not affect rationale identification during prompt development. As shown in \Cref{tab:irrelevant_dev}, none of the 44 sentences from irrelevant artifacts in the development set were identified as rationale by any prompting strategy (see the "From Irrel." column in the table).

\begin{table*}[t]
\centering
\caption{Identified rationale sentences from irrelevant artifacts by each prompt or approach.}
\label{tab:irrelevant_sentence_effect}

\begin{subtable}[t]{0.45\columnwidth}
\centering
\caption{Development set}
\label{tab:irrelevant_dev}
\small
\setlength{\tabcolsep}{5pt}
\begin{tabular}{@{}lrrrr@{}}
\toprule
& \multicolumn{2}{c}{\textbf{Input Sentences}}
& \multicolumn{2}{c}{\textbf{Identified Rationale}} \\
\cmidrule(lr){2-3}
\cmidrule(l){4-5}
\textbf{Prompt}
& \textbf{Rel.}
& \textbf{Irrel.}
& \textbf{\#Rat. Sent.}
& \textbf{From Irrel.} \\
\midrule
CI-ZS  & 704 & 44 & 84 & 0 \\
CI-FS  & 704 & 44 & 90 & 0 \\
CI-RFS & 704 & 44 & 81 & 0 \\
\bottomrule
\end{tabular}
\end{subtable}
\hfill
\begin{subtable}[t]{0.45\columnwidth}
\centering
\caption{Evaluation set}
\label{tab:irrelevant_test}
\small
\setlength{\tabcolsep}{5pt}
\begin{tabular}{@{}lrrrr@{}}
\toprule
& \multicolumn{2}{c}{\textbf{Input Sentences}}
& \multicolumn{2}{c}{\textbf{Identified Rationale}} \\
\cmidrule(lr){2-3}
\cmidrule(l){4-5}
\textbf{Prompt}
& \textbf{Rel.}
& \textbf{Irrel.}
& \textbf{\#Rat. Sent.}
& \textbf{From Irrel.} \\
\midrule
\tool & 1,890 & 390 & 311 & 12 \\
CI-FS & 1,890 & 390 & 329 & 14 \\
\bottomrule
\end{tabular}
\end{subtable}
\end{table*}

\begin{table}[t]
	\centering
	\caption{Prompt development results for rationale generation (CG). F2-score weights Information Coverage (IC) twice as much as Extra Information (EI). Higher EI values indicate less extra information}
	\label{tab:prompt_dev_gen_results}
    
    \begin{subtable}[b]{0.72\textwidth}
    	\centering
    	\caption{Results by component and prompt}
    	\label{tab:rat-gen-dev}
    	\resizebox{\columnwidth}{!}{%
    	\small
    	\begin{tabular}{c|c|c|c|c}
    		\toprule
    		\makecell[c]{\textbf{Component}}
    		&
    		\makecell[c]{\textbf{Prompt}}
    		&
    		\makecell[c]{\textbf{Information}\\\textbf{Coverage (IC)}}
    		&
    		\makecell[c]{\textbf{Extra}\\\textbf{Information (EI)}}
    		&
    		\makecell[c]{\textbf{F2}}
    		\\ \midrule
    
    		\multirow{2}{*}{\goal}
    			& \textbf{Zero-Shot (CG-ZS)} & \textbf{4.7} & \textbf{3.6} & \textbf{4.5} \\
    			& Few-Shot (CG-FS)  & 4.3 & 3.5 & 4.1 \\ \midrule
    
    		\multirow{2}{*}{\need}
    			& Zero-Shot (CG-ZS) & 4.0 & 3.6 & 3.9 \\
    			& \textbf{Few-Shot (CG-FS)} & \textbf{4.7} & \textbf{4.0} & \textbf{4.6} \\ \midrule
    
    		\multirow{2}{*}{\alternative}
    			& Zero-Shot (CG-ZS) & \textbf{2.7} & 2.3 & 2.6 \\
    			& Few-Shot (CG-FS)  & \textbf{2.7} & \textbf{2.5} & \textbf{2.6} \\ \midrule
    
    		\multirow{2}{*}{Overall}
    			& Zero-Shot (CG-ZS) & \textbf{4.0} & 3.3 & 3.8 \\
    			& \textbf{Few-Shot (CG-FS)} & \textbf{4.0} & \textbf{3.4} & \textbf{3.9} \\ \bottomrule
    	\end{tabular}
    	}
    \end{subtable}
    
    \begin{subtable}[b]{0.72\textwidth}
    	\centering
    	\vspace{0.3cm}
    	\caption{Relative Improvement (RI) between prompts}
    	\label{tab:rat-gen-rel-inc}
    	\resizebox{\columnwidth}{!}{%
    	\small
    	\begin{tabular}{c|c|c|c|c}
    		\toprule
    		\makecell[c]{\textbf{Comparison}}
    		&
    		\makecell[c]{\textbf{Component}}
    		&
    		\makecell[c]{\textbf{Information}\\\textbf{Coverage (IC)}}
    		&
    		\makecell[c]{\textbf{Extra}\\\textbf{Information (EI)}}
    		&
    		\makecell[c]{\textbf{F2}}
    		\\ \midrule
    
    		\multirow{4}{*}{\makecell{Few-Shot (CG-FS)\\vs\\Zero-Shot (CG-ZS)}}
    			& \goal        & -10.6\% & -5.3\% & -9.3\% \\
    			& \need        & 15.2\%  & 9.9\%  & 13.9\% \\
    			& \alternative & 0.0\%   & 6.7\%  & 1.5\% \\ \cline{2-5}
    			& \textbf{Overall} & \textbf{0.0\%} & \textbf{1.7\%} & \textbf{0.4\%} \\ \bottomrule
    	\end{tabular}
    	}
    \end{subtable}
\end{table}

\subsubsection{\ref{rq:prompt-dev-generation}: Prompt Development Results for Rationale Generation\\}
\label{sec:res-rat-gen}

\Cref{tab:prompt_dev_gen_results} reports the rationale generation results for the two prompting strategies: Zero-Shot (CG-ZS) and Few-Shot (CG-FS). The reported values correspond to the average \textit{Information Coverage} (IC) and \textit{Extra Information} (EI) ratings assigned by the evaluators across the 11 development commits. As described in \Cref{sec:prompt_dev_rat_gen}, both metrics are measured on 5-point Likert scales, where higher values indicate better performance. Specifically, higher IC scores indicate that a larger proportion of the ground-truth rationale is captured in the generated summary, while higher EI scores indicate that the summary introduces less information beyond the ground truth.

Overall, both prompts achieve an IC score of 4.0, which indicates that the generated summaries typically capture most of the information contained in the ground-truth rationale. In addition, the prompts achieve an EI score of 3.3/3.4, suggesting that the generated summaries generally introduce limited amounts of additional information. Since the prompts use sentences identified by the rationale identification module, their performance is affected by incorrect identifications of the module. Across the development commits, the reasoning-based few-shot identification prompt (CI-RFS) produced 23 false positive and 4 false negative sentences. False positives cause the generated rationale to include extra information, and false negatives remove relevant input information. For example, the sentence, ``\textit{Network calls can throw checked `IOException`s so you need to declare `@Throws(IOException::class)` on every single function on your interface, or it will blow up at runtime.}''\footnote{Project: Retrofit;\quad Source: \bluelink{https://github.com/square/retrofit/issues/3128\#issuecomment-507266667}{Issue \#3128};\quad Commit: \bluelink{https://github.com/square/retrofit/commit/9d683b75b111e60d30da8a03335d2ce71f5e8585}{9d683b7}} was labeled as \alternative by the LLM. Commit \bluelink{https://github.com/square/retrofit/commit/9d683b75b111e60d30da8a03335d2ce71f5e8585}{9d683b7} did not have any alternative content in the related artifacts as determined by the annotators. Since the sentence was mislabeled as \alternative, the generated rationale also included an \alternative summary. This resulted in reduced information coverage (IC) and extra information (EI).
\looseness=-1

For \goal, the zero-shot generation prompt (CG-ZS) performed 10.6\% and  5.3\% better than the zero-shot prompt (CG-FS) on IC and EI, respectively. 
%
In contrast, for \need, the few-shot prompt (CG-FS) achieves 15.2\% higher IC (4.7 vs. 4.0) and 9.9\% higher EI (4.0 vs. 3.6). For \alternative, the two prompts achieve similar performance on IC, while the few-shot prompt (CG-FS) achieves a 6.7\% higher EI.  For \need and \alternative, providing prompt exemplars helped the model produce more accurate summaries (\ie covering more of the ground truth while including less extra information).
However, the exemplars were not as effective for generating \goal.

Overall, both prompting strategies are comparable. However, few-shot prompting (CG-FS) offers slightly more benefits for rationale generation, as it offers overall gains in performance compared to zero-shot prompting, particularly for generating \need and \alternative. For these reasons, we selected few-shot prompting (CG-FS) as our prompting strategy for \tool.


\RQ{\textbf{\ref{rq:prompt-dev-identification} and \ref{rq:prompt-dev-generation} Findings}: For rationale identification, combining reasoning-based few-shot exemplars with task-decomposition prompting outperformed both zero-shot and few-shot prompting, achieving the highest precision, recall, and F2-score across all rationale components. 

For rationale generation, zero-shot and few-shot prompting produced similarly accurate summaries overall, although few-shot prompting provided modest improvements for generating \need and \alternative summaries. 

Overall, rationale identification benefits substantially from exemplar explanations, whereas rationale generation is less sensitive to the choice of prompting strategy once rationale information has been identified.
}

\subsection{\tools Evaluation Methodology}
\label{argus:methodology}

To evaluate \tool’s two main modules (rationale identification and rationale generation), we applied the best prompting strategy identified in the previous section (reasoning-based few-shot with task decomposition). The evaluation followed the same methodology as prompt development, including execution procedures and metrics, with one key difference: we used 50 commits with corresponding ground-truth data from the evaluation dataset (see \Cref{subsec:dev_data}).

Our goal was to assess the accuracy and generalizability of \tool's prompts, including the quality of the generated rationale summaries. As a baseline for rationale identification, we used the second-best prompt from the development phase, since it provides a direct comparison against an alternative prompting strategy for the same task. For rationale generation, since both candidate prompts performed similarly, we report only the accuracy of \tool's generated summaries. 
To the best of our knowledge, no existing approach from the literature generates fine-grained rationale components for code changes by integrating information from multiple software artifacts, as \tool does---see \Cref{sec:related_work} for more details. Consequently, there is no directly comparable technique that could serve as a baseline technique for \tool.
\looseness=-1

To understand the impact of the underlying LLM on \tool's performance, we evaluated \tool with two more recent models: \GPTFiveTwo~\cite{openai2025gpt52} and \GeminiThreeFlash~\cite{google2025gemini3flash}. These models (1) have strong code understanding capabilities, (2) can effectively reason about natural language and source code (3) have been consistently ranked among the top-performing models~\cite{openai2025gpt52, google2025gemini3flash} in well-known software engineering and code reasoning benchmarks, such as SWE-Bench Verified~\cite{Jimenez2024SWE}. This benchmark requires models to reason over natural-language issue descriptions and source code, making them relevant to our task. Our goal was not to benchmark all available LLMs, but to assess the extent to which \tools performance and behavior vary across different models.

\begin{table}[t]
\centering
\caption{\tool's rationale component identification performance on the evaluation dataset}
\label{tab:evaluation_results_identification}

\begin{subtable}[b]{\textwidth}
	\centering
	\caption{Results by component and prompt}
	\small
		\begin{tabular}{c|c|rrrr|ccc}
			\toprule
			\textbf{Component}
			& \textbf{Prompt}
			& \textbf{TP}
			& \textbf{FP}
			& \textbf{TN}
			& \textbf{FN}
			& \textbf{Precision}
			& \textbf{Recall}
			& \textbf{F2}
			\\ \midrule

			\multirow{2}{*}{\goal}
				& \tool             & 105 & 32 & 2138 & 5 & 76.6\% & 95.5\% & 91.0\% \\
				& Few-Shot (CI-FS)  & 103 & 40 & 2130 & 7 & 72.0\% & 93.6\% & 88.3\% \\ \midrule

			\multirow{2}{*}{\need}
				& \tool             & 51 & 79 & 2145 & 5 & 39.2\% & 91.1\% & 72.0\% \\
				& Few-Shot (CI-FS)  & 49 & 83 & 2141 & 7 & 37.1\% & 87.5\% & 68.8\% \\ \midrule

			\multirow{2}{*}{\alternative}
				& \tool             & 8 & 44 & 2226 & 2 & 15.4\% & 80.0\% & 43.5\% \\
				& Few-Shot (CI-FS)  & 9 & 53 & 2217 & 1 & 14.5\% & 90.0\% & 44.1\% \\ \midrule

			\multirow{2}{*}{Overall}
				& \tool             & 164 & 155 & 6509 & 12 & 51.4\% & 93.2\% & 80.2\% \\
				& Few-Shot (CI-FS)  & 161 & 176 & 6488 & 15 & 47.8\% & 91.5\% & 77.3\% \\ \bottomrule
		\end{tabular}%
\end{subtable}

\begin{subtable}[b]{\textwidth}
	\centering
	\caption{Relative Improvement (RI) between prompts}
	\small
		\begin{tabular}{c|c|c|c|c}
			\toprule
			\textbf{Comparison}
			& \textbf{Component}
			& \textbf{Precision}
			& \textbf{Recall}
			& \textbf{F2}
			\\ \midrule

			\multirow{4}{*}{\makecell{\tool\\vs\\Few-Shot (CI-FS)}}
				& \goal         & 6.4\%  & 1.9\%   & 3.0\%  \\
				& \need         & 5.7\%  & 4.1\%   & 4.7\%  \\
				& \alternative  & 6.0\%  & -11.1\% & -1.4\% \\ \cline{2-5}
				& \textbf{Overall}
				& \textbf{7.6\%}
				& \textbf{1.9\%}
				& \textbf{3.7\%}
				\\ \bottomrule
		\end{tabular}%
\end{subtable}

\end{table}

\subsection{\tools Evaluation Results}
\label{subsec:argus_eval_results}

\subsubsection{\ref{rq:tool-eval-ident}: Performance of \tool's Rationale Identification\\}

\Cref{tab:evaluation_results_identification} presents the performance of \tool's rationale identification module on the evaluation dataset. Overall, \tool achieves strong recall (93.2\%) across rationale components. This indicates that it is generally effective at retrieving rationale sentences from the collected artifacts. Its performance for \goal is 91.0\% F2, where both precision~(76.6\%) and recall (95.5\%) are high. In contrast, identifying rationale sentences for \need (72.0\% F2) and \alternative (43.5\% F2) remains challenging, as \tool (and the baseline) produces more false positives than true positives.

\Cref{tab:evaluation_results_identification} also shows that \tool's rationale identification module outperforms the few-shot identification prompt (CI-FS) in terms of precision, recall, and F2-score, with overall relative improvements of 7.6\%, 1.9\%, and 3.7\%. This trend holds for both \goal and \need, but not for \alternative: \tool improves precision at the expense of recall. Still, in terms of F2-score, \tool performs on par with the baseline for \alternative. We do not consider the recall drop particularly problematic, since it is explained by \tool missing only one \alternative sentence out of ten. More importantly, compared to the baseline, \tool reduces false positives for \alternative from 53 to 44. Taken together, these results indicate that \tool is more accurate than the baseline at identifying rationale sentences in artifacts.

Comparing these evaluation results with the prompt development outcomes in \Cref{tab:rat-iden-dev}, we observe a consistent trend: \tool achieves somewhat lower recall but substantially higher precision across components, particularly for \alternative. The baseline shows similar behavior for \need and \alternative. Two main factors explain these findings. First, \tool misclassified non-\alternative sentences as \alternative because they contain phrasing suggestive of alternatives. Sentences such as ``\textit{Workaround [for] the problem by adding this code to run early in your program: <CODE\_SNIPPET> It's caused by mixing both AOSP's built-in HttpURLConnection and OkHttp in the same process.}''\footnote{Project: OkHttp;\quad Source: \bluelink{https://github.com/square/okhttp/issues/666\#issuecomment-38999763}{Issue \#666};\quad Commit: \bluelink{https://github.com/square/okhttp/commit/4c86085429edbeef0a383941936ee7b64cc3805e}{4c86085}} may be coded as \alternative because it explicitly uses the term \textit{workaround}. However, a follow-up discussion in this Issue \bluelink{https://github.com/square/okhttp/issues/666\#issuecomment-38999763}{\#666} states that this workaround was already present in the application and that this workaround is not related to the actual problem. Thus, it was not a candidate solution considered for the target change. Second, the few-shot exemplars may not capture the diverse ways the \alternative component is expressed, such as rejected solutions, suggestions, comparisons, and workarounds. With only two exemplar commits, the model may incorrectly label descriptions of the implemented solution as \alternative. For example, the sentence \textit{``Another option would be to keep it `@Transactional` and add a note in the Javadoc that the user can annotate a given test class with `@Transactional(propagation = NOT\_SUPPORTED)` to disable transactional test support.\footnote{Project: Spring-boot;\quad Source: \bluelink{https://github.com/spring-projects/spring-boot/issues/23630\#issuecomment-707627777}{Issue \#23630};\quad Commit: \bluelink{https://github.com/spring-projects/spring-boot/commit/a5b27789c0bcc7ee65e798be68dbb44ba0e2d05b}{a5b2778}}''} describes a possible implementation. However, after analyzing the associated code change, we found that this solution is the \selectedAlternative. In our taxonomy, \alternative captures candidate solutions that were discussed but not used, whereas the implemented solution is captured as \selectedAlternative. Thus, this false positive suggests that \tool
did not sufficiently relate the sentence to the code change to determine whether the proposed solution was \alternative or \selectedAlternative. Future versions of \tool could also use a larger or retrieval-based exemplar set so that the model is exposed to a broader range of \alternative expressions, including rejected options, tentative suggestions, workarounds, and explicit trade-off discussions.

We also investigated whether the inclusion of 390 sentences from 39 irrelevant artifacts for 11 commits in the evaluation dataset increased the likelihood of misclassifications (see \Cref{tab:irrelevant_sentence_effect}). However, we found that only 12 of the 311 sentences~(3.9\%) identified as rationale by \tool originated from irrelevant artifacts. This indicates that \tool is largely robust to noise introduced by the Artifact Retriever.


\subsubsection{\ref{rq:tool-eval-gen}: Performance of \tool's Rationale Generation\\}

\Cref{tab:test-results-rat-gen} shows the quality of the rationale summaries generated by \tool on the evaluation dataset. Across the 97 evaluated rationale-component instances, \tool achieved an average Information Coverage (IC) score of 3.8, an Extra Information (EI) score of 3.2, and an F2 score of 3.7. Both IC and EI are measured on 5-point scales, where higher values indicate better performance. Thus, the overall IC score indicates that the summaries captured a substantial portion of the ground-truth rationale, while the EI score indicates that they introduced some extra information.

\tool's performance varied considerably across rationale components. \goal components achieved the highest scores, with an IC of 4.5, an EI of 3.7, and an F2 of 4.3. These results indicate that \tool generally captured nearly all of the ground-truth \goal information while introducing relatively little unsupported content. For \need, the IC score of 3.5 indicates substantial but incomplete coverage, while the EI score of 3.2 suggests that the generated summaries contained some additional information not supported by the ground truth. \alternative was the most challenging component: its IC score of 2.1 indicates limited coverage of the ground-truth rationale, and its EI score of 1.5 indicates that much of the generated information was extra. 

These results are affected by errors propagated from the rationale identification module, whose predicted sentences are used as input to rationale generation. Specifically, \tool generated rationale components for commits that did not contain ground truth rationale of certain types (\eg some commits had no \goal, others had no \need, and others had no \alternative), because the identification module incorrectly labeled sentences as containing rationale. As a result, \tool generated rationale where none was expected. This occurred in 2 of the 50 commits where \tool incorrectly generated a \goal, 9 of the 33 commits for \need, and 9 of the 14 commits for \alternative. These cases received the lowest possible IC and EI scores as a penalty, substantially reducing the overall scores, particularly for \alternative. These findings show that false positives during rationale identification have an important impact on rationale generation.
\looseness=-1

False negatives from rationale identification can also affect the rationale generation module. When sentences expressing a ground truth component are not identified, the generation module may receive incomplete input or produce no summary for that component. Such cases reduce IC because relevant rationale information cannot be recovered, although they do not introduce extra information. 

For the generated components that contain irrelevant information, the main impact was on Extra Information (EI). This is reflected in the lower overall EI score of 3.0, compared with the overall IC score of 3.8. The \goal and \need summaries still captured most or all of the ground-truth rationale (IC = 5.0 and 4.0), but their EI scores of 3.5 indicate that some extra information was included. The effect was stronger for \alternative, where the generated summary received an EI score of 1.0. These results suggest that irrelevant information primarily reduces EI by a small margin (6.7\%).
\looseness=-1

Overall, the results show that \tool produces accurate \goal summaries and reasonably informative \need summaries, whereas generating accurate \alternative summaries remains difficult, primarily because this component is the most sensitive to errors propagated from rationale identification.

\begin{table}[t]
	\centering
    \caption{\tools generated rationale quality (IC = Information Coverage; EI=Extra Information; \ie components generated by \tool but absent from the ground truth because of false-positive sentences)}
	\label{tab:test-results-rat-gen}

    \small
    \begin{tabular}{c|c|c|c|c}
        \toprule
        \makecell[c]{\textbf{Component}}
        &
        \makecell[c]{\textbf{Prompt}}
        &
        \makecell[c]{\textbf{Information}\\\textbf{Coverage (IC)}}
        &
        \makecell[c]{\textbf{Extra}\\\textbf{Information (EI)}}
        &
        \makecell[c]{\textbf{F2}}
        \\ \midrule

        \goal
            & Few-Shot (CG-FS) & 4.5 & 3.7 & 4.3 \\

        \need
            & Few-Shot (CG-FS) & 3.5 & 3.2 & 3.4 \\

        \alternative
            & Few-Shot (CG-FS) & 2.1 & 1.5 & 1.9 \\ \midrule

        \textbf{Overall}
            & \textbf{Few-Shot (CG-FS)}
            & \textbf{3.8}
            & \textbf{3.2}
            & \textbf{3.7}
            \\ \bottomrule
    \end{tabular}%
\end{table}

\RQ{\textbf{\ref{rq:tool-eval-ident} and \ref{rq:tool-eval-gen} Findings}: \tool identifies \goal sentences in artifacts with high accuracy (76.6\% precision and 95.5\% recall), while its performance is lower for \need and \alternative sentences. Nevertheless, \tool outperforms the baseline second-best prompting strategy by 7.6\% in precision, 1.9\% in recall, and overall 3.7\% in F2-score, and remains robust to irrelevant artifact content. For rationale generation, \tool produces accurate \goal summaries that capture nearly all ground-truth information, whereas \need and \alternative summaries are less accurate, primarily due to false positives from the rationale identification module.}



\subsubsection{Illustrative Example of \tool's Rationale Identification and Synthesis\\}
\label{subsec:illustrative_example}

To illustrate how \tool reconstructs rationale from multiple artifacts, we discuss Retrofit's commit \texttt{329a9fd} shown in \Cref{lst:retrofit-response-success}. This commit introduced a new overload of \texttt{Response.success()} that allows callers to specify an HTTP status code when creating a successful response.

The commit message, \textit{``Ability to add HTTP status code to Response.success,''} provides a brief description of the change and conveys its high-level \goal. However, it does not explain why the change was needed or what problem motivated the implementation. Additional rationale information appears in the associated pull request (PR) discussion (PR \#2715~\cite{retrofit_pull_2715}), where developers explain that tests were previously limited to responses with status code \texttt{200 OK}. Consequently, developers could not easily verify behavior for other successful HTTP status codes, such as \texttt{204 No Content}.

Both artifacts provide complementary pieces of rationale. The commit message captures \textit{what} was added, while the pull request explains \textit{why} the change was introduced. Reconstructing the rationale for this commit therefore requires integrating information from both sources rather than relying on the commit message alone.

In contrast, \tool synthesizes this information into concise rationale summaries. The generated \goal explains the new API behavior introduced by the change, while the generated \need explicitly captures the testing motivation behind it. By combining rationale information distributed across the commit message and pull request, \tool produces a coherent explanation ready for developers to consume when understanding the change.

More broadly, this example reflects a pattern observed throughout our empirical study described in \Cref{sec:rationale_study}: rationale information is often distributed across multiple artifacts, requiring developers to connect complementary sources to reconstruct a change's rationale. This observation motivated the design of \tool and its focus on cross-artifact rationale identification and synthesis.
\looseness=-1

\begin{listing}[t]
\centering

\noindent\fbox{%
\begin{minipage}{\dimexpr\linewidth-2\fboxsep-2\fboxrule\relax}
\footnotesize
\textbf{Commit message}: Ability to add HTTP status code to
\texttt{Response.success} (\#2715)
\end{minipage}%
}

\begin{lstlisting}[style=minteddiff]
diff --git a/retrofit/src/main/java/retrofit2/Response.java b/retrofit/src/main/java/retrofit2/Response.java
@@ -35,6 +35,22 @@ public final class Response<T> {

+  public static <T> Response<T> success(int code, @Nullable T body) {
+    if (code < 200 || code >= 300) {
+      throw new IllegalArgumentException("code < 200 or >= 300: " + code);
+    }
+    return success(body, new okhttp3.Response.Builder()
+        .code(code)
+        .message("Response.success()")
+        .protocol(Protocol.HTTP_1_1)
+        .request(new Request.Builder().url("http://localhost/").build())
+        .build());
+  }
\end{lstlisting}

\noindent\fbox{%
\begin{minipage}{\dimexpr\linewidth-2\fboxsep-2\fboxrule\relax}
\footnotesize
\textbf{Associated Pull Request}

\smallskip
\smallskip

\noindent\fbox{%
\begin{minipage}{\dimexpr\linewidth-2\fboxsep-2\fboxrule\relax}
\footnotesize
\textbf{PR \#2715}: Ability to add HTTP status code to
\texttt{Response.success}

\textbf{Comment \#1}: ``When writing tests using Retrofit, it would be
great if there were a way to set the `success' HTTP status code in
order to verify the functionality of other status codes, such as
\texttt{204 No Content}.''
\end{minipage}%
}
\end{minipage}%
}

\vspace{0.1cm}

\noindent\fbox{%
\begin{minipage}{\dimexpr\linewidth-2\fboxsep-2\fboxrule\relax}
\footnotesize

\textbf{\tool-Generated Rationale}

\smallskip
\smallskip

\noindent\fbox{%
\begin{minipage}{\dimexpr\linewidth-2\fboxsep-2\fboxrule\relax}
\footnotesize
\textbf{\goal:} Add a new overload of
\texttt{Response.success(int code, T body)} that creates a synthetic
successful response with the specified HTTP status code and
deserialized body.

\textbf{\need:} Enable tests to set the ``success'' HTTP status code,
for example \texttt{204 No Content}, when using
\texttt{Response.success}, so behavior for different success codes can
be verified.
\end{minipage}%
}
\end{minipage}%
}

\caption{Generated rationale from the Retrofit commit
\texttt{329a9fd}~\cite{retrofit_commit_329a9f}. Although the commit
message and pull request only state the goal briefly, \tool's generated
rationale explains both the implemented API behavior and the testing
need behind the change.}
\label{lst:retrofit-response-success}
\end{listing}

\subsubsection{Failure Analysis\\}
\label{sec:failure-analysis}

To better understand the limitations of \tool, we qualitatively analyzed the sentences that were incorrectly classified during rationale component identification and generation. Specifically, we inspected false positive and false negative predictions for each rationale component and grouped the errors according to recurring patterns. 

\vspace{0.1cm}

\noindent\underline{\textbf{Rationale Identification Failures.}}
Across the three rationale components, the LLM produced 155 false positive
and 12 false negative component predictions. These consisted of 32 false
positives and 5 false negatives for \goal, 79 and 5 for \need, and 44 and 2
for \alternative, respectively.

\textbf{\goal.} We inspected these false positives and false negatives produced by the LLM for the \goal component and identified several common error patterns. First, the LLM sometimes labeled \textbf{low-level implementation details as \goal}. For example, sentences such as \textit{``I also removed the use of \texttt{Pattern}\footnote{Project: JUnit4;\quad Source: \bluelink{https://github.com/junit-team/junit4/pull/1402\#issuecomment-270190682}{PR \#1402};\quad Commit: \bluelink{https://github.com/junit-team/junit4/commit/c85c6147ada5ad1afdd3be10769670d309730132}{c85c614}}''} and \textit{``Changed to use \texttt{mkdirs()} and added a test for paths that contain a forward slash\footnote{Project: JUnit4;\quad Source: Code review comment from \bluelink{https://github.com/junit-team/junit4/pull/1402\#discussion_r94349230}{PR \#1402};\quad Commit: \bluelink{https://github.com/junit-team/junit4/commit/c85c6147ada5ad1afdd3be10769670d309730132}{c85c614}}''} describe low-level implementation actions rather than the overall objective of the change. Similarly, the LLM incorrectly labels these low-level implementation details in \classJavadocs and \methodJavadocs with the change's \goal. For instance, the method Javadoc sentence: \textit{``Returns an input stream containing one or more certificate PEM files\footnote{Project: OKHttp;\quad Source: \bluelink{https://github.com/square/okhttp/commit/2e07d308d2189e9ffef4fa958fa20f467ffdcfe5}{\methodJavadocs from CustomTrust.java};\quad Commit: \bluelink{https://github.com/square/okhttp/commit/2e07d308d2189e9ffef4fa958fa20f467ffdcfe5}{2e07d30}}''} describe a helper method in a newly added \texttt{CustomTrust} recipe that reads PEM certificates before constructing an \texttt{SSLContext}. Thus, the sentence describes local method-level details rather than the overall objective of the commit: to create a custom trust recipe that demonstrates how to convert PEM certificates into an \texttt{SSLContext}. Although the method contributes to this objective, the LLM failed to distinguish the implementation details from the broader commit-level \goal.

Second, the LLM sometimes labeled \textbf{sentences describing the cause of a
failure as \goal}. For example, \textit{``The underlying cause of the failure is a NullPointerException\footnote{Project: Spring-Boot;\quad Source: \bluelink{https://github.com/spring-projects/spring-boot/issues/4826\#issuecomment-172505410}{\issue \#4826};\quad Commit: \bluelink{https://github.com/spring-projects/spring-boot/commit/c4f756daee43f89e0ba832ceac17bac216fc899b}{c4f756d}}''} identifies the cause of the reported failure but does not express the intention of the code change. The \goal of this commit is \textit{Update MavenSettings ... to fix NullPointerException} but \tool considered this sentence as the \goal, instead of as a \need. According to the participants in Safwan \etal's study~\cite{Safwan:FSE19}, some components can overlap: ``\textit{In terms of overlap among components, one participant thought that goal and need can be the same most of the time and preferred to merge them together}''. 

Finally, some false positives were \textbf{context dependent generic statements} which do not provide enough information, such as \textit{``Here's my proposed fix for \#380\footnote{Project: JUnit4;\quad Source: \bluelink{https://github.com/junit-team/junit4/pull/752}{PR\#752};\quad Commit: \bluelink{https://github.com/junit-team/junit4/commit/445ea85dd748e4e83cf7be5940b41df95b3ffa8d}{445ea85}}.''} In this case, additional contextual information is needed, particularly from pull request \#380, yet the LLM considered them as the \goal. 

For \goal false negatives, the LLM sometimes missed \textbf{\goal information
phrased as a potential solution}. For example, \textit{``Another option would be to keep it \texttt{@Transactional} and add a note in the Javadoc...\footnote{Project: Spring-Boot;\quad Source: \bluelink{https://github.com/spring-projects/spring-boot/issues/23630\#issuecomment-707627777}{\issue \#23630};\quad Commit: \bluelink{https://github.com/spring-projects/spring-boot/commit/a5b27789c0bcc7ee65e798be68dbb44ba0e2d05b}{a5b2778}}''} and \textit{``One possible solution is to subclass the \texttt{DefaultConfiguration} and set register shutdown hook to \texttt{false}\footnote{Project: Spring-Boot;\quad Source: \bluelink{https://github.com/spring-projects/spring-boot/issues/11360\#issuecomment-384619974}{\issue \#11360};\quad Commit: \bluelink{https://github.com/spring-projects/spring-boot/commit/8a0f0354dfed2fc6d4cb1b046ec03fe5b7ff1146}{8a0f035}}''} express possible implementation goals, but their alternative-like phrasing may have led the model to avoid labeling them as \goal.

\textbf{\need.} The LLM identified 130 sentences as containing \need information. Of these, 79 were false positives. In addition, the LLM missed five \need sentences, counted as false negatives. Most \need false positives occurred when the LLM treated \textbf{problem-related sentences as the motivation for the target change}. For example, sentences such as \textit{``Debug mode is not logging web and sql related loggers\footnote{Project: Spring-Boot;\quad Source: \bluelink{https://github.com/spring-projects/spring-boot/issues/15986}{\issue \#15986};\quad Commit: \bluelink{https://github.com/spring-projects/spring-boot/commit/978f80122b9138f5606bd9e33597807b90129d61}{978f801}}''} describe observed failures or problems. However, the commit is about polishing the existing codebase, and the \need is to ``\textit{Remove unnecessary package}'', yet the LLM failed to consider the broader perspective of the commit and misclassified the problem description as \need. 

Second, the LLM labeled \textbf{technical information or background context as \need}. For instance, \textit{``RpcContext may already have been changed when callback happens\footnote{Project: Apache Dubbo;\quad Source: \cc from \bluelink{https://github.com/apache/dubbo/commit/b5408b752631644d5934971f4ba11d165c776536}{AsyncRpcResult.java};\quad Commit: \bluelink{https://github.com/apache/dubbo/commit/b5408b752631644d5934971f4ba11d165c776536}{ b5408b7}}''} explains why the \texttt{RpcContext} must be restored around an asynchronous callback. However, the target commit modifies callback chaining so that exceptions thrown by \texttt{onResponse} or \texttt{onError} can be propagated through the returned asynchronous result; it does not introduce or modify the \texttt{RpcContext} restoration behavior. So, the sentence provides relevant background about the surrounding code but does not state the motivation for the target change. We observed this behavior in our prompt development process and attempted to address the issue by including this rule in the prompt: ``If the discussed information ... is not related to the actual code change, it will not be labeled as any component.'' However, more guidance is needed for the LLM to avoid these cases.

Third, the LLM sometimes misclassified \textbf{logs, stack traces, and reproduction evidence as \need}, such as long exception traces showing a \texttt{NullPointerException} or a failing test. These sentences are useful evidence, but they mainly support the existence of a problem rather than directly stating the problem or rationale for the change.

For \need false negatives, we observed that the LLM missed \need sentences when the \textbf{motivation was expressed as a design principle rather than as an explicit problem or failure}. For example, consider the sentences from Spring Boot's commit \texttt{8f23ee4}, \textit{``Support property based MeterFilters\footnote{Project: Spring-Boot;\quad Source: \bluelink{https://github.com/spring-projects/spring-boot/issues/11800}{\issue \#11800};\quad Commit: \bluelink{https://github.com/spring-projects/spring-boot/commit/8f23ee4e58e8d78dae7f6dd15441fb4b5ec86933}{ 8f23ee4}}''} and \textit{``Our recommendation for library authors is always just instrument with plain timers and let application developers decide at the last minute which summary statistics they want via filters\footnote{Project: Spring-Boot;\quad Source: \bluelink{https://github.com/spring-projects/spring-boot/issues/11800\#issuecomment-360893146}{\issue \#11800};\quad Commit: \bluelink{https://github.com/spring-projects/spring-boot/commit/8f23ee4e58e8d78dae7f6dd15441fb4b5ec86933}{ 8f23ee4}}''}. The \need only becomes clear when these two sentences are interpreted together: property-based \texttt{MeterFilter}s are needed to let developers configure metric statistics, introducing flexibility. The LLM likely missed this case because it may not be able to consider the broad context of the change and considered these sentences in isolation. During prompt development, we observed the same problem and introduced this rule in the prompt: ``\textit{Do not annotate in isolation. Consider the broader context provided by surrounding sentences and related artifacts.}'' However, the LLM requires additional context or guidance for avoiding these issues.

\textbf{\alternative.}
For \alternative, the LLM identified 52 sentences as describing \alternative. Of these, 44 were false positives. Additionally, the LLM missed two \alternative sentences, counted as false negatives. Most \alternative false positives occurred when the LLM treated \textbf{suggestions as evidence of an alternative solution}, even when the sentence did not describe a concrete implementation option or did not consider broader context for the target change. For example, consider this sentence, \textit{``Workaround [of] the problem [is] by adding this code to run early in your program: OkHttpClient okHttpClient = new OkHttpClient(); URL.setURLStreamHandlerFactory(okHttpClient)\footnote{Project: OKHttp;\quad Source: \bluelink{https://github.com/square/okhttp/issues/666\#issuecomment-38999763}{\issue \#666};\quad Commit: \bluelink{https://github.com/square/okhttp/commit/4c86085429edbeef0a383941936ee7b64cc3805e}{4c86085}}''}. In isolation, this sentence appears to propose an alternative solution. However, the subsequent sentence, ``\textit{We already have that workaround in, first line in Application's oncreate}\footnote{Project: OKHttp;\quad Source: \bluelink{https://github.com/square/okhttp/issues/666\#issuecomment-38999863}{\issue \#666};\quad Commit: \bluelink{https://github.com/square/okhttp/commit/4c86085429edbeef0a383941936ee7b64cc3805e}{4c86085}}'' indicates that the workaround had already been implemented and was therefore not related to the problem or it is not a candidate solution for the target change. We also observed similar cases during prompt development and added the following rule: CommitID\_SourceID\_TextID\_LocalSentenceID ... If consecutive sentences share the same CommitID\_SourceID\_TextID, it means they are logically grouped within the artifacts' text. For labeling the sentences, consider these groups as context (as they may give more information about the sentences).'' Nevertheless, these errors suggest that further guidance is needed to help the LLM use surrounding context consistently.

For \alternative false negatives, the LLM may miss \textbf{alternatives expressed as non-technical workarounds}. For example, the sentence about \textit{``downgrad[ing] to 1.3.0 and used the app for a while without issues... I'm downgrading okhttp in this branch now, and I'll close this issue if I'm ever able to reproduce it on 1.3.x.\footnote{Project: OKHttp;\quad Source: \bluelink{https://github.com/square/okhttp/issues/666}{\issue \#666};\quad Commit: \bluelink{https://github.com/square/okhttp/commit/4c86085429edbeef0a383941936ee7b64cc3805e}{ 4c86085}}''} describes a workaround or alternative path in a bug reproduction context from an \issue. As a result, the LLM may have treated it as a debugging step rather than \alternative content.

\noindent\underline{\textbf{Rationale Generation Failures.}}
For rationale generation, we observed the lower-rated Extra Information (EI) and Information Coverage (IC) scored rationales (less than 2). We found 7 such rationales, and most of these failures were caused by \textbf{errors propagated from the rationale identification module}. In particular, false-positive sentences identified as \goal, \need, or \alternative were passed to the generator as input, which sometimes caused the LLM to generate summaries for components that were absent from the ground truth. This explains the lower EI scores, because the generated summaries included the extra information. This issue was especially visible for \need and \alternative, where the LLM sometimes generated rationale from problem descriptions, technical background, or suggestion-like sentences that were incorrectly labeled during identification. Conversely, false negatives in the identification step removed relevant rationale evidence before generation, causing the generated summaries to miss important ground-truth information and lowering IC. For example, when the LLM missed the \alternative sentence describing \textit{``downgrad[ing] to 1.3.0 and used the app for a while without issues... I'm downgrading okhttp in this branch now, and I'll close this issue if I'm ever able to reproduce it on 1.3.x.\footnote{Project: OKHttp;\quad Source: \bluelink{https://github.com/square/okhttp/issues/666}{\issue \#666};\quad Commit: \bluelink{https://github.com/square/okhttp/commit/4c86085429edbeef0a383941936ee7b64cc3805e}{ 4c86085}}''}, the generator had less evidence to recover that workaround in the final \alternative summary. Similarly, when the LLM missed \need sentences such as \textit{``Support property based MeterFilters\footnote{Project: Spring-Boot;\quad Source: \bluelink{https://github.com/spring-projects/spring-boot/issues/11800}{\issue \#11800};\quad Commit: \bluelink{https://github.com/spring-projects/spring-boot/commit/8f23ee4e58e8d78dae7f6dd15441fb4b5ec86933}{ 8f23ee4}}''} and the corresponding design recommendation about allowing application developers to configure summary statistics through filters, the generated \need summary could not fully capture the motivation for the change. These observations suggest that generation failures are not independent generation errors; rather, they are downstream effects of sentence-level rationale identification.

\subsubsection{Impact of the LLM on \tools Performance\\}

We evaluated the best-performing prompts with two additional models, \GPTFiveTwo and \GeminiThreeFlash~\cite{gemini3flashmodelcard,openai2025gpt52}, to assess how \tools prompting and pipeline design perform beyond \OFourMini.

For rationale identification, we re-ran the best-performing prompting strategy, CI-RFS with voting across three runs, on the evaluation set using all three models. Because the CI-RFS prompt was developed using \OFourMini, this analysis examines how the performance changes when applied to \GPTFiveTwo and \GeminiThreeFlash.

\begin{table*}[t]
	\centering
	\caption{Rationale identification results for three models (using the CI-RFS prompt, developed for \OFourMini, and voting with three prompt executions); RI = relative improvement of \OFourMini compared to \GeminiThreeFlash or \GPTFiveTwo}
	\label{tab:cross_model_ci_component}
	\small

	\begin{subtable}[b]{\textwidth}
		\centering
		\caption{Results by component and model}
        \label{tab:cross_model_ci}
		\small
		\begin{tabular}{l|l|rrrr|ccc}
			\toprule
			\textbf{Component} & \textbf{Model} & \textbf{TP} & \textbf{FP} & \textbf{TN} & \textbf{FN} & \textbf{Precision} & \textbf{Recall} & \textbf{F2} \\
			\midrule
			\multirow{3}{*}{\goal}
			& \OFourMini            & 105 & 32  & 2138 & 5  & 76.6\% & 95.5\% & 91.0\% \\
			& \GPTFiveTwo           & 107 & 78  & 2092 & 3  & 57.8\% & 97.3\% & 85.6\% \\
			& \GeminiThreeFlash     & 108 & 123 & 2047 & 2  & 46.8\% & 98.2\% & 80.5\% \\
			\midrule

			\multirow{3}{*}{\need}
			& \OFourMini            & 51 & 79  & 2145 & 5 & 39.2\% & 91.1\% & 72.0\% \\
			& \GPTFiveTwo           & 52 & 123 & 2101 & 4 & 29.7\% & 92.9\% & 65.2\% \\
			& \GeminiThreeFlash     & 53 & 221 & 2003 & 3 & 19.3\% & 94.6\% & 53.2\% \\
			\midrule

			\multirow{3}{*}{\alternative}
			& \OFourMini            & 8 & 44 & 2226 & 2 & 15.4\% & 80.0\% & 43.5\% \\
			& \GPTFiveTwo           & 5 & 23 & 2247 & 5 & 17.9\% & 50.0\% & 36.8\% \\
			& \GeminiThreeFlash     & 6 & 58 & 2212 & 4 & 9.4\%  & 60.0\% & 28.9\% \\
			\midrule

			\multirow{3}{*}{Overall}
			& \OFourMini            & 164 & 155 & 6509 & 12 & 51.4\% & 93.2\% & 80.2\% \\
			& \GPTFiveTwo           & 164 & 224 & 6440 & 12 & 42.3\% & 93.2\% & 75.1\% \\
			& \GeminiThreeFlash     & 167 & 402 & 6262 & 9  & 29.4\% & 94.9\% & 65.6\% \\
			\bottomrule
		\end{tabular}%
	\end{subtable}

	\begin{subtable}[b]{\textwidth}
		\centering
		\caption{Relative Improvement (RI) between models}
        \label{tab:cross_model_ci_ri}
		\small
		\begin{tabular}{c|c|ccc}
			\toprule
			\textbf{Comparison} & \textbf{Component} & \textbf{Precision} & \textbf{Recall} & \textbf{F2} \\ \midrule

			\multirow{4}{*}{\makecell{\OFourMini\\vs\\\GPTFiveTwo}}
			& \goal        & 32.5\%  & -1.9\% & 6.3\%  \\
			& \need        & 32.0\%  & -1.9\% & 10.5\% \\
			& \alternative & -13.9\% & 60.0\% & 18.3\% \\
			\cline{2-5}
			& \textbf{Overall} & \textbf{21.6\%} & \textbf{0.0\%} & \textbf{6.8\%} \\ \midrule

			\multirow{4}{*}{\makecell{\OFourMini\\vs\\\GeminiThreeFlash}}
			& \goal        & 63.9\%  & -2.8\% & 13.1\% \\
			& \need        & 102.8\% & -3.8\% & 35.4\% \\
			& \alternative & 64.0\%  & 33.3\% & 50.7\% \\
			\cline{2-5}
			& \textbf{Overall} & \textbf{75.2\%} & \textbf{-1.8\%} & \textbf{22.2\%} \\ \midrule
		\end{tabular}%
	\end{subtable}
\end{table*}

Table~\ref{tab:cross_model_ci_component} shows the rationale identification results of different models with the prompts developed for \OFourMini. For the three models, \goal\ is consistently the easiest component to identify, followed by \need, while \alternative\ remains the most difficult one. All models achieve high recall for \goal\ and \need, indicating that rationale sentences for these components are broadly detectable across LLMs. Interestingly, overall, \OFourMini outperforms \GPTFiveTwo by 6.8\% and \GeminiThreeFlash by 22.2\% relative improvement on F2 Score (see \Cref{tab:cross_model_ci_ri}). Although their recall remains very similar for \goal and \need, the main difference in performance arises from precision. Component-wise, all the instances except for \alternative, where \GPTFiveTwo's precision is 13\% higher than \OFourMini, \OFourMini achieves at least 32\% 
higher precision than other models. Notably, for \need, \OFourMini shows a 102\% higher precision boost compared to \GeminiThreeFlash. This result shows that prompt development on a relatively lightweight model like \OFourMini can help achieve higher performance compared to a more powerful model like \GPTFiveTwo and \GeminiThreeFlash.

One possible explanation for these results is that the rationale identification task rewards careful rule-following more than general reasoning ability. Rationale identification requires interpretation, but the model must still follow the annotation rules and avoid inferring rationale when the evidence is not explicit. A stronger reasoning model such as \GPTFiveTwo may sometimes over-interpret contextual or implementation-related sentences as implicit rationale, which can increase false positives. In contrast, \OFourMini may have benefited more from the reasoning-based few-shot prompt because it followed the provided examples and decision boundaries more conservatively. This can explain why \OFourMini achieved higher precision and F2-score even though \GPTFiveTwo and \GeminiThreeFlash are generally considered stronger reasoning models.

\begin{table*}[t]
\centering
\small
\caption{Rationale identification performance of different prompts executed on different models. CI-RFS is the prompt developed for \OFourMini, while CI-RFS-G3F and CI-RFS-G5.2 are the prompts developed for \GeminiThreeFlash and \GPTFiveTwo, respectively}
\label{tab:cross_model_ci_component_optimized}
\small
\begin{tabular}{l|l|l|cccc|ccc}
\toprule
\textbf{Component} & \textbf{Model} & \textbf{Prompt} & \textbf{TP} & \textbf{FP} & \textbf{TN} & \textbf{FN} & \textbf{Precision} & \textbf{Recall} & \textbf{F2 Score} \\
\midrule

\multirow{7}{*}{\goal}
& \multirow{2}{*}{\GeminiThreeFlash}
& CI-RFS & 108 & 123 & 2047 & 2 & 46.8\% & 98.2\% & 80.5\% \\
& & CI-RFS-G3F & 109 & \textbf{117} & 2053 & 1 & \textbf{48.2\%} & 99.1\% & \textbf{81.8\%} \\
\cline{2-10}
& \multirow{2}{*}{\GPTFiveTwo}
& CI-RFS & 107 & 78 & 2092 & 3 & 57.8\% & 97.3\% & 85.6\% \\
& & CI-RFS-G5.2 & 106 & \textbf{47} & 2123 & 4 & \textbf{69.3\%} & 96.4\% & \textbf{89.4\%} \\
\cline{2-10}
& \multirow{3}{*}{\OFourMini}
& CI-RFS & 105 & 32 & 2138 & 5 & 76.6\% & 95.5\% & \textbf{91.0\%} \\
& & CI-RFS-G5.2 & 99 & \textbf{19} & 2151 & 11 & \textbf{83.9\%} & 90.0\% & 88.7\% \\
& & CI-RFS-G3F & 101 & 23 & 2147 & 9 & 81.5\% & 91.8\% & 89.5\% \\
\midrule

\multirow{7}{*}{\need}
& \multirow{2}{*}{\GeminiThreeFlash}
& CI-RFS & 53 & 221 & 2003 & 3 & 19.3\% & 94.6\% & 53.2\% \\
& & CI-RFS-G3F & 51 & \textbf{184} & 2040 & 5 & \textbf{21.7\%} & 91.1\% & \textbf{55.6\%} \\
\cline{2-10}
& \multirow{2}{*}{\GPTFiveTwo}
& CI-RFS & 52 & 123 & 2101 & 4 & 29.7\% & 92.9\% & 65.2\% \\
& & CI-RFS-G5.2 & 50 & \textbf{86} & 2138 & 6 & \textbf{36.8\%} & 89.3\% & \textbf{69.4\%} \\
\cline{2-10}
& \multirow{3}{*}{\OFourMini}
& CI-RFS & 51 & 79 & 2145 & 5 & 39.2\% & 91.1\% & \textbf{72.0\%} \\
& & CI-RFS-G5.2 & 46 & \textbf{64} & 2160 & 10 & \textbf{41.8\%} & 82.1\% & 68.9\% \\
& & CI-RFS-G3F & 47 & 69 & 2155 & 9 & 40.5\% & 83.9\% & 69.1\% \\
\midrule

\multirow{7}{*}{\alternative}
& \multirow{2}{*}{\GeminiThreeFlash}
& CI-RFS & 6 & 58 & 2212 & 4 & 9.4\% & 60.0\% & 28.9\% \\
& & CI-RFS-G3F & 6 & \textbf{38} & 2232 & 4 & \textbf{13.6\%} & 60.0\% & \textbf{35.7\%} \\
\cline{2-10}
& \multirow{2}{*}{\GPTFiveTwo}
& CI-RFS & 5 & 23 & 2247 & 5 & 17.9\% & 50.0\% & 36.8\% \\
& & CI-RFS-G5.2 & 5 & \textbf{13} & 2257 & 5 & \textbf{27.8\%} & 50.0\% & \textbf{43.1\%} \\
\cline{2-10}
& \multirow{3}{*}{\OFourMini}
& CI-RFS & 8 & 44 & 2226 & 2 & 15.4\% & 80.0\% & 43.5\% \\
& & CI-RFS-G5.2 & 7 & \textbf{21} & 2249 & 3 & \textbf{25.0\%} & 70.0\% & \textbf{51.5\%} \\
& & CI-RFS-G3F & 8 & 30 & 2240 & 2 & 21.1\% & 80.0\% & 51.3\% \\
\midrule

\multirow{7}{*}{Overall}
& \multirow{2}{*}{\GeminiThreeFlash}
& CI-RFS & 167 & 402 & 6262 & 9 & 29.4\% & 94.9\% & 65.6\% \\
& & CI-RFS-G3F & 166 & \textbf{339} & 6325 & 10 & \textbf{32.9\%} & 94.3\% & \textbf{68.7\%} \\
\cline{2-10}
& \multirow{2}{*}{\GPTFiveTwo}
& CI-RFS & 164 & 224 & 6440 & 12 & 42.3\% & 93.2\% & 75.1\% \\
& & CI-RFS-G5.2 & 161 & \textbf{146} & 6518 & 15 & \textbf{52.4\%} & 91.5\% & \textbf{79.6\%} \\
\cline{2-10}
& \multirow{3}{*}{\OFourMini}
& CI-RFS & 164 & 155 & 6509 & 12 & 51.4\% & 93.2\% & \textbf{80.2\%} \\
& & CI-RFS-G5.2 & 152 & \textbf{104} & 6560 & 24 & \textbf{59.4\%} & 86.4\% & 79.2\% \\
& & CI-RFS-G3F & 156 & 122 & 6542 & 20 & 56.1\% & 88.6\% & 79.4\% \\
\midrule
\end{tabular}%
\end{table*}

To assess whether model-specific prompt adaptation affects performance, we adapted the CI-RFS prompt for the other two models, which resulted in \textit{CI-RFS-G5.2} for \GPTFiveTwo and \textit{CI-RFS-G3F} for \GeminiThreeFlash. Across the three versions, we retained the rationale taxonomy, target commit diff, artifact sentences, output format, and two few-shot exemplar commits. For \GPTFiveTwo, following OpenAI's Prompt Optimizer guidelines~\cite{OpenAiPromptOptimizer}, we reorganized the instructions into explicit annotation constraints, exclusion rules, and stricter output requirements, while emphasizing conservative labeling and grounding predictions in the commit diff. For \GeminiThreeFlash, following Gemini's Prompting Strategies guidelines~\cite{GeminiPromptingStrategies}, we separated contextual analysis, relevance filtering, and multi-labeling into an explicit execution logic section. All three prompt versions are available in our replication package~\cite{package}.

In \Cref{tab:cross_model_ci_component_optimized}, we report the component
identification results obtained by executing each model-specific prompt on all
three models.
To evaluate the benefit of model-specific prompting, we compared each tailored
prompt with the CI-RFS prompt (developed for \OFourMini). All comparisons were
performed on \OFourMini, which is the model used in \tool.
The table reveals that the performance of each model in component identification increased consistently in F2 score. Every model performs best on their own prompt compared to the prompt developed for \OFourMini. For \OFourMini, its performance decreased when it was evaluated with CI-RFS-G3F and CI-RFS-G5.2. We analyzed the source of this boost, and we found that when prompts are adapted for each model, their true positives remained similar but their false positives were reduced significantly. For example, when the \GPTFiveTwo model was executed with the CI-RFS prompt (developed for \OFourMini) the overall FPs were 224, but when the same model was run with the CI-RFS-G5.2 prompt (developed for \GPTFiveTwo) the overall FPs decreased to 146. Similarly, when the \GeminiThreeFlash model was executed with CI-RFS, the FPs were 402, but with the CI-RFS-G3F prompt, the model produced 339 FPs overall.

Across prompts, the true positives remained relatively stable, while the false positives decreased substantially. This suggests that model-specific prompt development mainly helped each model become more selective about assigning positive labels. In other words, the model-specific prompts did not help the models identify many new rationale sentences; instead, they helped reduce over-identification. This also explains the slight recall drops in some cases: stricter prompts remove false positives, but may also filter out a small number of true positives.

For rationale generation, we applied the \tools generation prompt, {CG-FS}, to the three models and manually evaluated the generated summaries using the same two dimensions described earlier: information coverage (IC) and extra information (EI). We executed the prompt on identified rationale sentences from the associated artifacts to assess the impact of the model on rationale generation. \Cref{tab:cross_model_cg_component} reports the component-wise average ratings. The three models show broadly similar behavior across components. For \goal, all three models produce strong summaries with very similar IC scores. For \need, the models again behave similarly, recovering most of the ground truth information. In both \goal and \need components, \GPTFiveTwo shows a slightly higher IC score, followed by \GeminiThreeFlash. For \alternative, all three models obtain identical average scores. Overall, all three models show good information coverage (3.8 - 3.9) which means they are able to recover most of the information from ground truth data. However, the \GeminiThreeFlash model included little extra information on \goal, \OFourMini model presented little extra information on the \need component. Taken together, these results suggest that rationale generation quality is relatively stable across the tested models, with no major shifts in behavior compared to the performance differences observed for rationale identification.

\begin{table*}[t]
	\centering
	\caption{Rationale generation results obtained by each model}
	\label{tab:cross_model_cg_component}

    \begin{subtable}[b]{\textwidth}
		\centering
		\caption{Results by component and model}
		\label{tab:cross_model_cg_results}
		\small
		\begin{tabular}{c|c|c|c|c}
			\toprule
			\makecell[c]{\textbf{Component}}
			&
			\makecell[c]{\textbf{Model}}
			&
			\makecell[c]{\textbf{Information}\\\textbf{Coverage (IC)}}
			&
			\makecell[c]{\textbf{Extra}\\\textbf{Information (EI)}}
			&
			\makecell[c]{\textbf{F2}}
			\\ \midrule

			\multirow{3}{*}{\goal}
				& \OFourMini        & 4.5 & 3.7 & 4.3 \\
				& \GPTFiveTwo       & \textbf{4.6} & 3.5 & \textbf{4.4} \\
				& \GeminiThreeFlash & \textbf{4.6} & \textbf{3.9} & \textbf{4.4} \\ \midrule

			\multirow{3}{*}{\need}
				& \OFourMini        & 3.5 & \textbf{3.2} & 3.4 \\
				& \GPTFiveTwo       & \textbf{3.6} & 3.0 & \textbf{3.5} \\
				& \GeminiThreeFlash & 3.5 & 3.0 & 3.4 \\ \midrule

			\multirow{3}{*}{\alternative}
				& \OFourMini        & \textbf{2.1} & \textbf{1.5} & \textbf{2.0} \\
				& \GPTFiveTwo       & \textbf{2.1} & \textbf{1.5} & \textbf{2.0} \\
				& \GeminiThreeFlash & \textbf{2.1} & \textbf{1.5} & \textbf{2.0} \\ \midrule

			\multirow{3}{*}{Overall}
				& \OFourMini        & 3.8 & 3.2 & \textbf{3.7} \\
				& \GPTFiveTwo       & \textbf{3.9} & 3.1 & \textbf{3.7} \\
				& \GeminiThreeFlash & \textbf{3.9} & \textbf{3.3} & \textbf{3.7} \\ \bottomrule
		\end{tabular}%
	\end{subtable}

	\begin{subtable}[b]{\textwidth}
		\centering
		\vspace{0.3cm}
		\caption{Relative Improvement (RI) between models}
		\label{tab:cross_model_cg_ri}
		\small
		\begin{tabular}{c|c|c|c|c}
			\toprule
			\makecell[c]{\textbf{Comparison}}
			&
			\makecell[c]{\textbf{Component}}
			&
			\makecell[c]{\textbf{Information}\\\textbf{Coverage (IC)}}
			&
			\makecell[c]{\textbf{Extra}\\\textbf{Information (EI)}}
			&
			\makecell[c]{\textbf{F2}}
			\\ \midrule

			\multirow{4}{*}{\makecell{\OFourMini\\vs\\\GPTFiveTwo}}
				& \goal        & -2.2\% & 5.1\% & -0.5\% \\
				& \need        & -5.2\% & 7.1\% & -2.6\% \\
				& \alternative & 0.0\%  & 0.0\% & 0.0\% \\ \cline{2-5}
				& \textbf{Overall}
				& \textbf{-3.1\%}
				& \textbf{5.6\%}
				& \textbf{-1.1\%}
				\\ \midrule

			\multirow{4}{*}{\makecell{\OFourMini\\vs\\\GeminiThreeFlash}}
				& \goal        & -1.3\% & -4.6\% & -2.1\% \\
				& \need        & -2.0\% & 5.0\%  & -0.5\% \\
				& \alternative & 0.0\%  & 0.0\%  & 0.0\% \\ \cline{2-5}
				& \textbf{Overall}
				& \textbf{-1.6\%}
				& \textbf{-1.2\%}
				& \textbf{-1.5\%}
				\\ \bottomrule
		\end{tabular}%
	\end{subtable}
\end{table*}

It is worth mentioning that two annotators independently rated the generated rationale components, obtaining substantial to near-perfect overall inter-rater agreement across models. Specifically, we obtained a Cohen’s $\kappa$ of 0.98, 0.91, and 0.91 for \OFourMini, \GPTFiveTwo, and \GeminiThreeFlash, respectively.

Overall, the results are broadly consistent across models. For rationale identification, all three models exhibit the same qualitative trend: \goal\ is easiest to identify, followed by \need, while \alternative\ remains the most difficult, but \OFourMini is clearly more precise and therefore more accurate overall. For rationale generation, by contrast, the three models produce broadly similar ratings across all components, suggesting that summary generation is more stable across models than sentence-level rationale identification. These results suggest that \tools overall feasibility does not depend on a single LLM, although the choice of model affects rationale identification accuracy, with \OFourMini providing the best results among the tested models. Among
the evaluated models, \OFourMini achieves the best overall performance while
also being the smallest model, making it well suited for future
real-world deployments of \tool.

\section{Evaluating \tool's Perceived Usefulness}
\label{sec:user_study}

We conducted a user study to understand whether Java programmers perceive \tool's generated rationale summaries as useful and effort-saving when understanding unfamiliar code changes. We also sought feedback on the software engineering tasks that could benefit from generated rationale and on the features and integration mechanisms programmers would like to see in a production version of \tool.

We focus on a scenario in which developers need to understand code changes that are unfamiliar to them. Such situations commonly arise during maintenance, onboarding, and debugging, where developers must determine why a change was made despite having little prior knowledge of the change itself. This scenario is particularly relevant because recovering the rationale behind unfamiliar changes is one of the primary goals of \tool.

With this in mind, our study addresses the following research questions (RQs):

\begin{enumerate}[start=7, label=\textbf{RQ$_\arabic*$:}, ref=\textbf{RQ$_\arabic*$}, itemindent=0cm,leftmargin=1cm]
	
    \item \label{rq:usefulness}{\textit{How useful and effort-saving do Java programmers perceive \tool's generated rationale summaries to be?}}
	
    \item \label{rq:feedback}{\textit{What tasks, features, and integration opportunities do Java programmers envision for \tool?}}

\end{enumerate}

\ref{rq:usefulness} investigates whether programmers perceive the generated
rationale summaries as helpful for understanding unfamiliar code changes and
reducing the effort required to recover the code change rationale. \ref{rq:feedback} identifies how generated rationale summaries may fit into existing development workflows and informs future improvements to \tool.
\looseness=-1

\subsection{Study Design}

Our study consisted of three phases. First, participants completed a short demographic questionnaire. Second, they evaluated rationale summaries generated by \tool for three commits from a single software project. Finally, they provided feedback regarding use cases, desired features, and workflow integration opportunities for \tool.  The study was administered as an online survey through Qualtrics~\cite{Qualtrics2026} and was designed to take approximately 30 minutes. 

The remainder of this subsection describes each phase in detail. Our replication package~\cite{package} provides the study questionnaire, commits, and additional materials needed to replicate and validate the study.
The study protocol, including participant recruitment procedures, questionnaires, and evaluation tasks, was approved by the Institutional Review Board (IRB) of our university.

\subsubsection{Participant Recruitment\\}

We recruited Java programmers with diverse software development backgrounds and experience levels. Participants were invited through our professional network via email and direct messages. 
We specifically sought individuals with Java development experience because the study required participants to inspect and reason about Java code changes.

We contacted 26 programmers, of whom 18 responded, and 15 completed the study. Following data validation, three participants were excluded from the analysis, resulting in a final set of 12 participants (see \Cref{sec:user_study_results} for details). No compensation was provided to participants.

\begin{table}[t]
	\centering
	\small
	\caption{Distribution of the 18 commits used in the study}
	\label{tab:commit-distribution}
	\setlength{\tabcolsep}{6pt}
	\renewcommand{\arraystretch}{0.95}
	\begin{tabular}{@{}l|l|c@{}}
		\toprule
		\textbf{Dimension} & \textbf{Category} & \textbf{\#Commits (\%)} \\
		\midrule
		
		\multirow{5}{*}{Project}
		& JUnit4      & 6 (33.3\%) \\
		& Spring Boot & 3 (16.7\%) \\
		& Dubbo       & 3 (16.7\%) \\
		& OkHttp      & 3 (16.7\%) \\
		& Retrofit    & 3 (16.7\%) \\
		\midrule
		
		
		\multirow{3}{*}{Rationale components}
		& 1 component  & 2 (11.1\%) \\
		& 2 components & 10 (55.6\%) \\
		& 3 components & 6 (33.3\%) \\
		\midrule
		
		\multirow{2}{*}{Code diff size}
		& Short ($\leq$ 15 LOC) & 13 (72.2\%) \\
		& Long ($>$ 15 LOC)     & 5 (27.8\%) \\
		
		\bottomrule
	\end{tabular}
\end{table}

\subsubsection{Commit Selection Process\\}

We selected 18 commits from the evaluation dataset described in \Cref{subsec:dev_data}. These commits were organized into six groups of three commits each. Each group contained commits from a single project. 
Each participant evaluated one group consisting of three commits from a single project. This allowed participants to remain within a consistent project context throughout the study and reduced context switching between different projects. 

The six groups collectively covered all five studied projects: Spring Boot~\cite{springboot}, Dubbo~\cite{dubbo}, OkHttp~\cite{okhttp}, Retrofit~\cite{retrofit}, and JUnit4~\cite{junit4}. Three commits were selected from each project, with an additional three commits selected from JUnit4, resulting in a total of 18 commits. Groups were assigned to participants so that each commit group, hence each commit, was evaluated by multiple (two to three) participants whenever possible. 

We selected three commits per participant to keep study duration within approximately 30 minutes while allowing exposure to multiple rationale summaries. The selected commits were chosen to provide diversity in code diff size and coverage of rationale components. As shown in \Cref{tab:commit-distribution}, the final set contains commits with one, two, and three generated rationale components, as well as short and long code diffs. We used the rationale summaries produced by \tool during the evaluation phase, ensuring that participants assessed the same type of output analyzed in \ref{rq:tool-eval-gen}. 

\begin{figure}[t]
    \centering
    \includegraphics[width=\columnwidth]{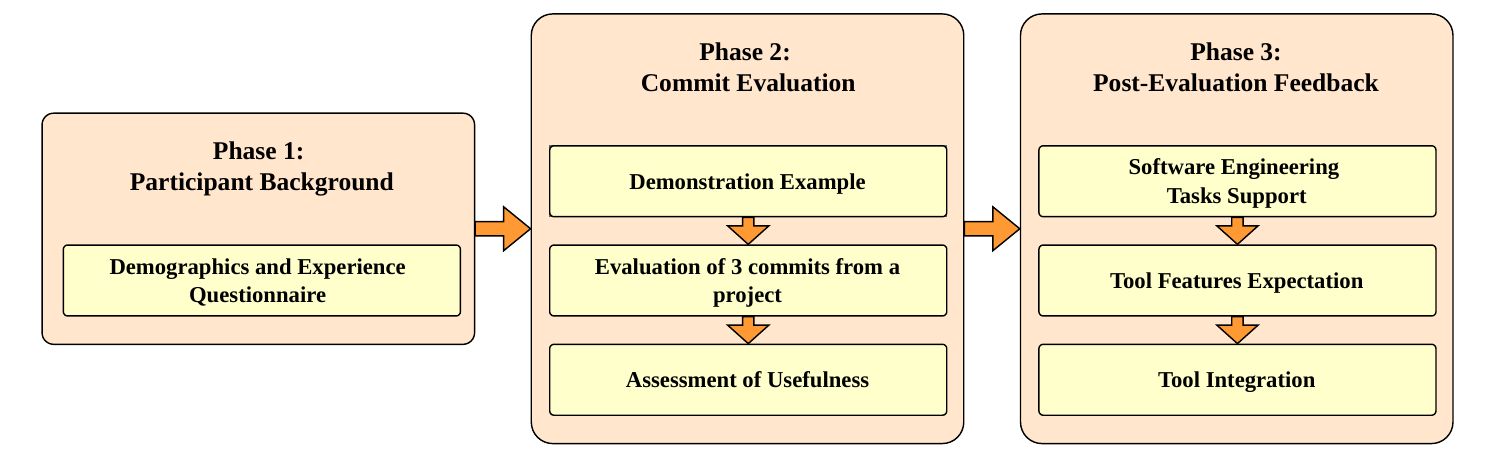}
    \caption{Overview of the user study procedure}
    \label{fig:commit_questions}
\end{figure}

\subsubsection{Study Procedure\\}

\Cref{fig:commit_questions} provides an overview of the study procedure followed by the participants.

In the first phase, participants completed a background questionnaire capturing their programming experience, Java development experience, software development context (\eg proprietary or open source), application domains, and primary software engineering role (\eg programmer, tester, or software architect).

In the evaluation phase, participants were first presented with definitions of code change rationale and the three rationale components that \tool generates: \goal, \need, and \alternative. They were also provided with a complete example consisting of a commit message, code diff, and generated rationale summaries to familiarize them with the study format and the types of rationale they would evaluate.

Participants then evaluated the three commits assigned to them. For each commit, they were presented with the original commit message, the corresponding code diff, and the rationale summaries generated by \tool.
Participants were instructed to inspect this information and then answer four questions about the usefulness of the rationale summaries.
Two questions asked participants to rate, using 5-point Likert scales, the extent to which:
\begin{enumerate}
\item the generated rationale components helped them understand the code change rationale; and
\item the generated rationale components could save effort when understanding why the code change was made.
\end{enumerate}
The remaining two questions asked the participants to provide open-ended explanations for their ratings.

Effort saving was defined as the participant's perceived reduction in effort required to understand why the change was made when using the generated rationale summaries, compared to relying on the commit message and code diff alone. Participants were explicitly instructed to judge whether summaries would reduce the time and cognitive effort required to understand the rationale behind the code change. Consequently, the effort-saving ratings reflect participants' perceptions of potential effort reduction rather than direct measurements of task performance. We discuss the implications of this design choice and the associated threats to validity in \Cref{sec:threats}.

After evaluating all three commits, participants completed a post-study questionnaire selecting software engineering tasks that could benefit from generated rationale summaries from a predefined list, and identifying desired features for \tool and preferred integration mechanisms via open-ended responses.

\subsection{Results and Analysis}
\label{sec:user_study_results}

We report participants' demographics, their perceptions of the usefulness and effort-saving value of \tool's generated rationale summaries, and their feedback regarding potential applications, features, and integration opportunities.

\subsubsection{Participant Exclusions\\}

Of the 15 participants who completed the study, three responses were excluded prior to analysis. Two participants completed the study in under 10 minutes, substantially shorter than the expected completion time of approximately 30 minutes, and provided extremely brief open-ended responses with little detail, suggesting limited engagement with the evaluation tasks. One additional response was excluded because its open-ended answers appeared to be generated by a generative AI model rather than reflecting the participant's own assessment. As such, the response contained generic explanations repeated across commits and failed to reference any commit-specific details. The final analysis therefore includes commit evaluations from 12 participants.

\subsubsection{Demographics\\}

The 12 participants included programmers with diverse levels of software development experience. Seven participants reported 1--4 years of programming experience, four reported 5--9 years, and one reported 10--19 years. Java experience followed a similar distribution, with seven participants reporting 1--4 years of Java experience and two reporting 5--9 years. Regarding experience in software development and maintenance in industry, nine participants reported prior experience, with seven having 1--4 years of experience and two with less than 1 year.
\looseness=-1

Participants also reported experience developing software in a variety of domains, including education and research (5 participants), healthcare (4), productivity software (2), fiduciary/banking systems (1), and rental services (2). 
Seven participants were affiliated with academic or research institutions, and five worked in industry. Overall, these demographics suggest that the study captured perspectives from programmers with varying levels of experience, application domains, organizational contexts, and software engineering responsibilities.

\begin{table}[t]
    \small
    \centering
    \caption{Participants' ratings of the usefulness and effort-saving value of the generated rationale summaries (N = 35 ratings; 12 participants evaluated 2--3 commits each)}
    \label{tab:rationale-distribution}
    \begin{tabular}{l|c|c}
        \toprule
        \textbf{Rating} & \textbf{Usefulness} & \textbf{Effort saving} \\
        \midrule
        Strongly disagree              & 1 (2.9\%)  & 2 (5.7\%)  \\
        Somewhat disagree              & 3 (8.6\%)  & 3 (8.6\%)  \\
        Neither agree nor disagree     & 2 (5.7\%)  & 0 (0.0\%)  \\
        Somewhat agree                 & 10 (28.6\%) & 8 (22.9\%) \\
        Strongly agree                 & 19 (54.3\%) & 22 (62.9\%) \\
        \bottomrule
    \end{tabular}
\end{table}

\subsubsection{\ref{rq:usefulness}: Perceived Usefulness and Effort-Saving of \tool's Generated Rationale Summaries\\}

Table~\ref{tab:rationale-distribution} summarizes participants' ratings of perceived usefulness and effort saving. The 12 participants provided 35 valid commit evaluations, resulting in 35 usefulness ratings and 35 effort-saving ratings. One additional evaluation was excluded from the analysis due to a survey configuration error that displayed rationale summaries for a different commit than the one being evaluated.

For usefulness, 29 of the 35 ratings (82.9\%) indicated that participants perceived the generated rationale summaries as helpful to understand the rationale behind the code change. Effort-saving ratings followed a similar pattern: 30 of 35 ratings (85.7\%) indicated agreement that the generated rationale summaries could reduce the effort required to understand why the change was made.

Our analysis of open-ended responses provides supporting evidence for these results:

\paragraph{\underline{Understanding the Motivation Behind a Change.}} Participants reported that the rationale summaries were most valuable when they provided context that was not immediately apparent from the commit message or code diff, particularly information explaining the \need behind a change. For example, P5 noted that the generated rationale helped them understand ``\textit{not just the changes made, but also the goal behind the changes and why the developers need the improvement in the code.}'' Similarly, P10 explained that ``\textit{The commit message alone only told me that this change was trying to `restore' something,}'' whereas the rationale summary ``\textit{gave me insight into the need for this change and the context surrounding the problem.}''

\paragraph{\underline{Reducing the Effort Needed to Recover Rationale.}} Participants also perceived the rationale summaries as effort-saving because they believed the summaries reduced the need to manually inspect code and infer developer intent. P2 explained that the ``\textit{natural language description helps to save time rather than interpret the code change,}'' while P3 noted that the rationale ``\textit{could save multiple minutes of having to read through the code to determine the goal and need.}'' In reference to commit \bluelink{https://github.com/square/retrofit/commit/a6025429e62a58d107b8d9385e68c4a058ac6615}{ a602542} from project \texttt{Retrofit}~\cite{retrofit}, which restored fine-grained logging of HTTP response bodies by splitting large responses into smaller chunks before logging them, P11 provided a concrete example, explaining that ``\textit{Without the answer, if I really had to, I would need to spend more time to understand why chunking is needed.}''.

\paragraph{\underline{Dependence on Rationale Quality.}} Participants also identified situations where the generated rationale was less useful. In particular, summaries were perceived as less helpful when they closely resembled the commit message, lacked specificity, or did not align with the code change. For example, P3 noted that one rationale ``\textit{just summarized the commit message and code comments, but included less detail},'' while P12 observed that ``\textit{the rationale did not match the commit.}'', which happened because the LLM misclassified sentences as rationale. These responses suggest that participants' perceptions of usefulness depend on the quality and grounding of the generated rationale.

Overall, participants generally perceived \tool's rationale summaries as useful for understanding unfamiliar code changes and believed they could reduce the effort required to recover and understand code change rationale.

\begin{table}[t]
	\centering
    \small
	\caption{Software engineering tasks for which participants find rationales useful}
	\label{tab:tasks}
	\begin{tabular}{@{}l|c|c@{}}
		\toprule
		\textbf{Task} & \textbf{\#Participants} & \textbf{Proportion} \\
		\midrule
		Understanding design/implementation/features & 9 & 75.0\% \\
		Code reviews & 9 & 75.0\% \\
		Documentation (design/implementation) & 9 & 75.0\% \\
		Teaching / onboarding & 8 & 66.7\% \\
		Diagnosing / locating / correcting bugs & 6 & 50.0\% \\
		Refactoring & 5 & 41.7\% \\
		Designing test cases & 5 & 41.7\% \\
		Reproducing bugs & 5 & 41.7\% \\
		Reusing code for new functionality & 5 & 41.7\% \\
		\bottomrule
	\end{tabular}
\end{table}

\subsubsection{\ref{rq:feedback}: Tasks, Features, and Integration Opportunities for \tool\\}

\paragraph{\underline{Tasks Supported by Generated Rationales.}}
Table~\ref{tab:tasks} summarizes the software engineering tasks participants believed could benefit from generated rationale summaries. The most frequently selected tasks were understanding code design and implementation, code review, and documentation, each selected by nine participants (75\%). Teaching and onboarding developers was selected by eight participants (66.7\%).

Participants frequently reported that rationale summaries helped them understand implementation decisions. For example, P11 explained: ``\textit{The statement in GOAL is more informative and helps me understand better.}'' Several participants also connected rationale to debugging and maintenance activities. For example, P10 stated that ``\textit{Rationales would be particularly helpful in cases where there’s a need to go back and understand what developers were thinking when making a change.}'' Such situations frequently occur during debugging, regression investigation, and maintenance tasks, where developers must reason about the intent behind earlier modifications before devising a solution and making further changes.
\looseness=-1

These responses suggest that participants view generated rationale summaries as potentially useful for supporting understanding of individual code changes, but also as a resource supporting a broad range of maintenance, documentation, review, and onboarding activities.

\paragraph{\underline{\tool Features and Integration Opportunities.}}

The feature requested most frequently was traceability. Three participants wanted to know where the generated rationale originated. For example, P3 stated that ``\textit{the tool should cite where its information comes from to mitigate hallucinations,}'' while P9 suggested that rationale summaries should be ``\textit{traceable (linked to the artifact).}''

This request aligns with \tool's architecture. \tool's rationale identification module already records which artifact sentences were classified as \goal, \need, or \alternative and retains links to their original artifacts. Consequently, a user-facing implementation of \tool could expose these traceability links, allowing developers to inspect the underlying evidence supporting each generated rationale summary.

Participants also requested support for interactive exploration of rationale. For example, P3 suggested allowing developers to ask follow-up questions about generated rationales. 
This feedback suggests that participants view rationale summaries not only as static documentation but also as a starting point for further exploration of design decisions and implementation trade-offs.
Others requested rationale search capabilities (P10: ``\textit{I would also want to be able to search through code rationales for my project, either literally or semantically}.''), support for impact analysis (P1: ``[The tool should] notify if the modifications have a side effect on the other parts of the project''), and code quality insights such as policy checking (P7: ``\textit{A comprehensive evaluation whether the changed code conforms with all the code design principles}'') or code smell detection (P6: ``\textit{I think a feature that can detect code smells will help a lot}'').

Integration preferences largely centered on existing development environments. Four participants preferred IDE integration, four participants preferred integration into pull request or code review platforms, two participants suggested integration into version control systems, and two participants suggested integrating the tool as a web extension. These responses indicate that developers would prefer rationale support to be embedded within tools they already use rather than exposed through standalone applications.

\RQ{\textbf{\ref{rq:usefulness} and \ref{rq:feedback} Findings:} Participants generally perceived \tool's generated rationale summaries as useful and effort-saving for understanding unfamiliar code changes. They identified code understanding, code review, documentation, and onboarding as the primary tasks that could benefit from generated rationale. Participants most frequently requested traceability support linking rationale summaries back to their source artifacts and expressed a preference for integrating \tool into existing development environments such as IDEs, code review systems, and version control platforms.}


\section{Threats to Validity}
\label{sec:threats}

We discuss the factors that affect the validity of our findings.

\subsection{Construct Validity}
We defined the rationale components based on the prior rationale model~\cite{Safwan:FSE19} and refined them through our qualitative study. This resulted in operational definitions for \goal, \need, and \alternative, which introduced validity threat concerns for the rationale component taxonomy. To reduce ambiguity, annotators used a shared codebook with definitions, examples, and decision rules, which was iteratively refined during annotation. We followed a multi-coder methodology involving iterative coding sessions, discussion, and consensus to resolve disagreements. Across all annotation tasks, we observed high agreement according to Cohen's $\kappa$, providing evidence for the quality of the process and resulting data.

Another threat concern of our study is the creation of ground truth sentence labels. Identifying whether an artifact sentence expresses rationale is subjective, especially when rationale is implicit or spread across multiple artifacts. We mitigated this threat through independent annotation, multiple annotation iterations, reconciliation meetings, and consensus-based resolution of disagreements. We also measured inter-rater agreement to assess the reliability of the labels, obtaining high agreement results.

A third threat concerns the ground truth rationale summaries used to evaluate generation. These summaries could be incomplete, overly interpretive, or biased by the annotators' understanding of the change. To mitigate this, one annotator synthesized the component summaries from the commit diff, commit message, and labeled artifact sentences, and a second annotator reviewed them for completeness, correctness, and accuracy with respect to the available artifacts. Disagreements were resolved through discussion until consensus was reached.

The process and metrics we used to evaluate rationale classification and generation also pose threats to construct validity. For rationale identification, we used standard classification metrics because the task is formulated as multi-label classification. We used F2 because missing rationale information is more harmful for our rationale generation task than including some additional candidate sentences. For rationale generation, automatic text similarity metrics alone would not capture whether a summary preserved the correct rationale meaning. Therefore, we used human ratings of information coverage and extra information, with explicit scoring criteria, independent ratings, and reconciliation to reduce evaluator bias. Overall, we spent 15 person-hours manually evaluating rationale summaries across several stages of the study, including prompt development, \tool evaluation, and analysis with multiple models. We also spent an additional 8 person-hours reconciling cases in which annotators disagreed on the evaluated rationales. 

Another threat concerns how we measured usefulness and effort saving in the user study. Since the study measured participants' perceptions rather than task performance, these results should be interpreted as evidence of perceived usefulness and perceived effort savings, rather than direct evidence that \tool improves comprehension or reduces actual task completion time. Evaluating comprehension accuracy, task performance, and time savings in controlled tasks remains future work.

\subsection{Internal Validity}
The selection of commits and associated artifacts poses a threat to internal validity. To mitigate this, we ensured that the sampled commits used for analysis were diverse. The heuristic-based approach used to retrieve rationale artifacts associated to each commit could influence the results. We tried to mitigate this threat by combining multiple retrieval strategies, including explicit-reference matching, GitHub API queries, and cross-artifact searches. Two annotators independently assessed the relevance of the retrieved issues and pull requests, resolved disagreements through discussion, and excluded irrelevant artifacts from the analysis. Nevertheless, some relevant artifacts may have remained unidentified.

\tool prompt development followed an iterative, data-driven methodology: failed cases were systematically analyzed, and prompts were refined based on evaluation metrics. To control for LLM inconsistency, we executed the developed prompts multiple times and found consistent results.
\looseness=-1

Regarding the user study, novelty bias may have influenced participants' ratings, as participants could react positively to seeing an automatically generated rationale summary. We did not directly measure this effect, so the perceived usefulness results should be interpreted with caution. Still, participants' open-ended explanations often pointed to concrete reasons for their ratings, such as whether the summaries added context, clarified the \need, or aligned with the code change, which helps contextualize their ratings beyond the numeric responses.

A further threat is that participants evaluated only the condition containing \tool's generated rationale summaries. We did not include a control condition in which participants assessed the same commits using only the commit message and code diff. Because participants only reported perceived benefits, we cannot conclude that the summaries improved actual comprehension or reduced effort. Establishing these effects would require a controlled comparison with and without rationale summaries.

Another threat is the choice of the LLM, but we executed an experiment to assess the effect of the LLM using two additional recent LLMs, \GPTFiveTwo and \GeminiThreeFlash. The results showed that model choice affects rationale identification performance, especially precision, but the same overall trend held across models: \goal was the easiest component to identify, followed by \need, while \alternative remained the most difficult. For rationale generation, the models produced broadly similar ratings across components.


\subsection{External Validity}
Our findings may not fully generalize beyond the selected Java projects, commits, artifact types, LLMs, and participants. We included projects from different domains, commits of varying sizes and complexities, multiple artifact types, and participants with diverse development backgrounds. The study remains limited to five open source Java projects and three LLMs, so the results may differ for other programming languages, software ecosystems, artifact sources, or models.

The user study findings are also limited by the small convenience sample of 12 participants. Individuals accessible to the researchers or interested in code rationale and LLM-based tools may have been more likely to participate, potentially making the sample more receptive to the proposed approach. In addition, seven participants were affiliated with academic or research institutions, while five worked in industry. Although they represented different experience levels, domains, and development roles, the sample may not reflect the broader population of professional developers. We therefore interpret the user study results as exploratory evidence of perceived usefulness rather than broadly generalizable findings.


\section{Related Work}
\label{sec:related_work}

We discuss prior work on rationale modeling and management in software engineering, focusing on four complementary areas: (i) rationale models and taxonomies, (ii) identification of rationale in software artifacts, (iii) approaches for capturing and documenting rationale during development, and (iv) the role of rationale in commit message generation.

\subsection{Design Rationale Models and Taxonomies}

Software rationale captures the reasoning and justification underlying decisions made during software development and evolution. Such decisions occur at different levels of abstraction, ranging from requirements and architectural choices to the implementation and maintenance of individual code changes~\cite{Sutcliffe1995,Lamsweerde2001RE,Tang2007,Gilson2011rationale,dutoit2006rationale,Zannier:IST07}. These levels form a continuum rather than strictly separate categories: a low-level code change may implement an architectural decision, address a defect introduced by that decision, or implement a higher-level requirement. Our work focuses on rationale associated with concrete code changes while recognizing that this rationale may be connected to higher-level design and architectural decisions.

Prior work has proposed several models for representing rationale and decision-making~\cite{Kunz1970,MacLean1991,PottsICSE1988,LeeICSE1991DR,BurgeJSS2008rationale,Anton1996,Kaiya2002,tang2006survey}. IBIS~\cite{Kunz1970}, for example, organizes rationale around issues, positions, and supporting or opposing arguments. QOC~\cite{MacLean1991} represents the questions being addressed, the options considered, and the criteria used to compare those options. Potts and Bruns~\cite{PottsICSE1988} modeled rationale through design issues, alternatives, and justifications, while Lee \etal~\cite{LeeICSE1991DR} proposed the Decision Representation Language, which connects issues, alternatives, goals, claims, and supporting or opposing relationships in a structured decision graph. RATSpeak~\cite{BurgeJSS2008rationale} extended this representation with requirements, assumptions, and a richer argument ontology. Tang \etal~\cite{tang2006survey} developed an empirical taxonomy of architectural rationale containing categories such as assumptions, constraints, trade-offs, costs, and benefits. These models primarily support the representation of rationale surrounding design and architectural decisions.

More recent work has adapted rationale models to information contained in development artifacts~\cite{Kleebaum:REFSQ20,Jiuang2024ASE,al2022developers,Li:EASE20,BhatECSA17,DhaouadiASE2022rationale}. Kleebaum \etal~\cite{Kleebaum:REFSQ20}, for example, represented rationale from issue trackers and version control systems as decision problems, alternatives, decisions, and pro/con arguments connected in a knowledge graph. DRMiner~\cite{Jiuang2024ASE} modeled solutions discussed in issue logs together with their supporting arguments. Dhaouadi \etal~\cite{DhaouadiASE2022rationale} proposed a decision rationale graph that represents decisions, their rationale, and supporting facts extracted from commit messages.

Our work adopts the taxonomy of Safwan \etal~\cite{Safwan:FSE19,al2022developers} because it specifically models the fine-grained information developers seek when understanding code changes. The taxonomy decomposes code change rationale into 15 components, including a change's \goal, \need, \benefit, \sideEffect, and  \alternative. Safwan \etal developed the taxonomy from developers' reported information needs; however, they did not examine how these components are actually documented and distributed across commit-related artifacts. We build on their taxonomy by studying whether and where its components appear across multiple software artifact types and by supporting the automated identification and synthesis of the three most frequently documented components: \goal, \need, and \alternative.

\subsection{Rationale Identification in Software Artifacts}

Rationale is often not recorded consistently in formal models or dedicated decision documents~\cite{Alkadhi:SANER18,BurgeJSS2008rationale}. Instead, it is expressed informally and inconsistently across artifacts such as commit messages, issue discussions, code comments, emails, meeting notes, chat messages, and wikis~\cite{soria2024characterizing,Alkadhi:SANER18,Lopez2012}. Recovering rationale therefore requires developers to search, navigate, and connect fragments distributed across heterogeneous sources. Empirical studies confirm that these artifacts contain relevant decision information. Bi \etal~\cite{bi2021JSSarchitecture}, for example, found that architectural rationale was frequently communicated in the mailing lists of two open source projects. Soria \etal~\cite{soria2024characterizing} similarly found that software decisions and their underlying reasoning were discussed during meetings but were only partially preserved in meeting documentation.

Prior studies have investigated automated and semi-automated methods for identifying rationale from individual artifact types~\cite{UllahSpringer2023,Lopez2012,Alkadhi:MSR17,RogersSpringer2015,Sharma:ICSE21,LesterSpringer2018,Rani2021JSSCOMMENTTYPE,dhaouadi2025automated,Rogers2012}. TREx~\cite{Lopez2012} used rule-based natural-language processing to identify ontology-based rationale elements in emails, meeting notes, and wikis. Alkadhi \etal~\cite{Alkadhi:MSR17} applied Na\"ive Bayes and SVM classifiers to development-team chat messages to identify rationale elements such as issues, alternatives, arguments, and decisions. Rogers \etal~\cite{RogersSpringer2015} combined NLP and machine learning techniques to classify rationale elements in Chrome bug reports. Sharma \etal~\cite{Sharma:ICSE21} developed heuristics based on textual patterns, temporal proximity, author roles, message structure, and decision-related expressions to identify rationale in email archives. Rani \etal~\cite{Rani2021JSSCOMMENTTYPE} and Pascarella \etal~\cite{pascarella2019classifying} used machine learning classifiers to identify rationale-oriented code comments. More recently, Dhaouadi \etal~\cite{dhaouadi2025automated} combined ML/DL classification with an LLM-based extraction step to recover decision rationale pairs from commit messages.

These approaches demonstrate that rationale can be identified in diverse software artifacts, but they typically focus on one artifact type or communication channel and use rationale schemas designed for the corresponding source or task. To the best of our knowledge, no prior approach combines the following three capabilities: identifying fine-grained code change rationale components, jointly analyzing the multiple artifact types associated with a commit, and using LLM-based cross-artifact reasoning for sentence-level identification and synthesis. Our approach examines six artifact types associated with each commit: commit messages, issue reports, pull requests, code reviews, Javadocs, and code comments, and identifies sentences expressing \goal, \need, and \alternative in relation to the actual code change. This formulation requires not only detecting rationale content, but also determining whether the information is relevant to the target commit and which fine-grained component it expresses.

\subsection{Rationale Capture, Documentation, and Synthesis}

Because rationale is often incomplete, scattered, and weakly connected to the corresponding software elements, prior work has proposed tools for capturing and organizing it during development~\cite{Mannan:FSE20,Ko:ICSE07,Kleebaum:SEKE19,Alkadhi:MSR17,FarenhorstIJCIS2007,Kleebaum:WRCSE18,Manteuffel:JSS16,Mehrpour:VLHCC19,BabarSHARK2007}. PAKME~\cite{BabarSHARK2007}, for example, supports architectural decision making by organizing design options, patterns, and their associated rationale. Decision Architect~\cite{Manteuffel:JSS16} allows architects to document architecture decisions and their rationale through multiple modeling viewpoints. EAGLE~\cite{FarenhorstIJCIS2007} combines a repository of architectural knowledge with project-specific discussion and user notification features. ACTIVE DOCUMENTATION~\cite{Mehrpour:VLHCC19} allows developers to document design decisions as rules and rationale, then checks whether the implementation conforms to those rules.

Reviews of rationale management tools have highlighted the burden of manually documenting and maintaining rationale~\cite{Capilla:JSS16}. Consequently, more recent approaches have explored semi-automated and automated recovery from existing development artifacts~\cite{Bhat:ICSA19,Kleebaum2021,Jiuang2024ASE,Li:FSE2024,Tian:ICSE2022,Rastkar:ICSE13,casillo2025towards,zhou2025using}. ADeX~\cite{Bhat:ICSA19}, for example, extracts architectural design decisions from textual sources such as issue reports, meeting minutes, chats, and commits to support architectural knowledge management. DRMiner~\cite{Jiuang2024ASE} recovers design solutions and supporting arguments from Jira issue discussions. Zhou \etal~\cite{zhou2025using} used LLMs to generate architectural rationale by providing an architecture-related problem and its corresponding decision as context. These approaches focus primarily on architecture or design-level decisions, whereas our work focuses on rationale directly associated with concrete code changes and their surrounding implementation artifacts.
\looseness=-1

The work most closely related to our rationale generation task includes the approaches of Rastkar \etal~\cite{Rastkar:ICSE13} and Casillo \etal~\cite{casillo2025towards}. Rastkar \etal proposed a multi-document summarization technique that combines information from commit messages, bug reports, tasks, user stories, and feature requests to generate a general explanation of why a change was made. Their work demonstrates the value of synthesizing information across documents, but it does not distinguish among fine-grained rationale components or first identify the sentences supporting each component.

Casillo \etal~\cite{casillo2025towards} formulated rationale generation as a translation task from code changes to natural-language rationale. They mined commits whose messages contained explicit rationale and fine-tuned a CodeT5+ model to generate rationale from the corresponding code changes. This formulation learns a mapping from code differences to rationale available in commit messages. However, rationale is frequently absent or incomplete in commit messages and instead is recorded in issues, pull requests, reviews, or code documentation, as we found in our study. Our approach therefore formulates the task as cross-artifact identification followed by synthesis: it first retrieves explicit rationale evidence from the artifacts associated with a commit and then generates separate summaries for \goal, \need, and \alternative. This design preserves component distinctions and enables each generated summary to be traced to the artifact sentences from which it was synthesized.

Overall, prior work has supported rationale capture, recovery, or summarization, but generally produces architectural decision representations, decision-rationale pairs, or a single general rationale summary. In contrast, our work generates concise natural-language summaries for individual fine-grained code change rationale components by integrating evidence identified across multiple heterogeneous artifacts.

\subsection{Commit Message Generation with Code Change Rationale}

Commit messages are an important form of code change documentation because they typically describe what changed and may also explain why the change was needed~\cite{Tian:ICSE2022,li2023commit}. Earlier commit message generation approaches, including CommitGen~\cite{JiangASE2017}, NMT~\cite{Loyola2017}, PtrGNCMsg~\cite{LiuMSR2019}, NNGen~\cite{LiuASE2018}, and ATOM~\cite{LiuTSE2020}, generated or retrieved messages primarily from code diffs and structural code information. Their main objective was to summarize the implemented modification rather than to recover the reasoning behind it.

More recent approaches have attempted to include rationale information in generated commit messages~\cite{Li:FSE2024,XueTSE2024,wu2025empirical,Mandli2025}. ERICommitter~\cite{XueTSE2024} uses in-context examples with informative commit messages to help an LLM infer \textit{why} information from a target code diff. COMET~\cite{Mandli2025} uses the what/why labeling method introduced by Tian \etal~\cite{Tian:ICSE2022} to select informative training messages and generates commit messages from code changes and their surrounding code context. OMG~\cite{Li:FSE2024} prompts GPT-4 with a code diff and additional context, including issue reports, pull requests, class and method summaries, and commit type, to generate rationale-aware commit messages.

These approaches are related to our work because both seek to capture the motivation behind code changes. However, their primary goal is different. Existing approaches generate a single commit message that summarizes what changed and briefly explains why. In contrast, our approach recovers explicit rationale components from software development artifacts. Specifically, we identify and summarize a change's \goal, motivating \need, and the \alternative solutions considered. Unlike commit message generation, which produces a single summary, our pipeline first identifies the artifact sentences supporting each rationale component and then generates a summary for each component, making the generated rationale traceable to specific evidence in the artifacts.



\section{Conclusions and Future Work}
\label{sec:conclusions}

We analyzed 63 commits from five Java projects to examine how software artifacts capture code change rationale components. We found evidence of seven rationale components, with \goal, \need, and \alternative being the most common. Our results show that rationale is unevenly distributed across artifacts: \cm and \prs mainly document \goal, while \need is more often captured in \prs and \issues. We also found that rationale beyond \goal is often scattered across multiple artifacts, making it difficult for developers to reconstruct why a change was made by inspecting a single artifact (\eg a commit message). These findings highlight the need for automated support to identify and synthesize rationale information from multiple development artifacts.

To address this need, we developed \tool, an LLM-powered approach that identifies rationale sentences across artifacts and generates concise summaries of a commit's \goal, \need, and \alternative. Our prompt development results show that reasoning-based few-shot prompting improves rationale identification. In the evaluation on 50 commits, \tool outperformed the prompting baseline overall and accurately identified \goal sentences, although performance was lower for \need and \alternative. This suggests that \goal is easier to detect because it is more explicit in artifacts, whereas \need and \alternative often require more contextual reasoning and are expressed in more varied ways.

For rationale generation, \tool produced strong \goal summaries that closely matched the ground truth. However, \need and \alternative summaries were less reliable than \goal, mainly because the false positive sentences from the identification step sometimes led the generator to produce components that were absent from the ground truth. This finding suggests that improving the precision of rationale identification, especially for \need and \alternative, is important for improving rationale generation. Our cross-model analysis further shows that \tool\ generalizes across the evaluated LLMs. While model choice affects rationale identification accuracy, all evaluated models produce similarly high-quality rationale summaries when given the same identified rationale sentences.

Finally, our user study with 12 Java programmers showed that participants generally perceived \tool's generated rationale summaries as useful for understanding change rationale. Participants especially valued summaries that explained the \need behind a change and provided context not obvious from the commit message or code diff alone. They also indicated that generated rationales could support code reviews, documentation, teaching, debugging, and understanding unfamiliar code. At the same time, participants highlighted the need for traceability, concise summaries, and integration into development workflows such as IDEs, code review platforms, and version control systems.

Future work should improve \tool in several directions. First, more diverse few-shot exemplars and decision rules could help reduce false positives for \need and \alternative. Second, future versions of \tool could support interactive rationale exploration, where developers ask follow-up questions, provide missing rationale context, or correct generated summaries. Third, future work should investigate personalized and task-specific rationale generation. Different developers may need different levels of detail depending on their role, project familiarity, and task. Finally, larger evaluations with more projects, programming languages, more professional developers, and real development tasks (such as identifying tangled commits or debugging regressions) are needed to assess how \tool affects developers' decision-making and productivity in practice.




\bibliographystyle{ACM-Reference-Format}
\bibliography{references}

\end{document}